\documentclass[oldversion]{aa}
\usepackage{times}
\usepackage{amssymb}
\usepackage{mathptmx}
\usepackage{natbib}
\usepackage{pdflandscape}
\usepackage{rotating}
\usepackage{longtable}
\usepackage[]{fontenc}
\usepackage[latin1]{inputenc}
\usepackage[]{graphicx}

\begin{document}

\title{The SINFONI survey of powerful radio galaxies at
  z$\sim$2: Jet-driven AGN feedback during the Quasar
  Era \thanks{Based on observations collected at the Very Large
    Telescope of ESO. Program IDs 070.A-0545, 070.A-0229, 076.A-0684,
    079.A-0617, 081.A-0468, 381.A-0541, 082.A-0825, 083.A-0445.}} 
\author{N.~P.~H.~Nesvadba\inst{1,2}, C.~De~Breuck\inst{3},
  M.~D.~Lehnert\inst{4}, P.~N.~Best\inst{5}, and C.~Collet\inst{1}}
\institute{ Institut d'Astrophysique Spatiale, Centre Universitaire
  d'Orsay, Bat.~121, 91405 Orsay, France
\and 
email: nicole.nesvadba@ias.u-psud.fr 
\and 
European Southern Observatory, Karl-Schwarzschild Strasse, Garching
bei M\"unchen, Germany
\and
Institut d'Astrophysique de Paris, CNRS \& Universit\'e Pierre et
Marie Curie, 98bis, bd Arago, 75014 Paris, France
\and
SUPA, Institute for Astronomy, Royal Observatory of Edinburgh,
Blackford Hill, Edinburgh EH9 3HJ, UK
 } 
\titlerunning{Imaging spectroscopy of HzRGs}
\authorrunning{Nesvadba et al.}  \date{Received / Accepted }

\abstract{
We present VLT/SINFONI imaging spectroscopy of the rest-frame optical
emission lines of warm ionized gas in 33 powerful radio galaxies at
redshifts z$\gtrsim$2, which are excellent sites to study the
interplay of rapidly accreting active galactic nuclei and the
interstellar medium of the host galaxy in the very late formation
stages of massive galaxies. Our targets span two orders of magnitude
in radio size (2$-$400~kpc) and kinetic jet energy (a few $10^{46}-$
almost $10^{48}$ erg s$^{-1}$). All sources have complex gas
kinematics with broad line widths up to $\sim$1300 km s$^{-1}$. About
half have bipolar velocity fields with offsets up to 1500 km s$^{-1}$
and are consistent with global back-to-back outflows. The others have
complex velocity distributions, often with multiple abrupt velocity
jumps far from the nucleus of the galaxy, and are not associated with a
major merger in any obvious way. We present several empirical
constraints that show why gas kinematics and radio jets seem to be physically
related in all galaxies of the sample. The kinetic energy in the gas
from large scale bulk and local outflow or turbulent motion
corresponds to a few $10^{-3}$ to $10^{-2}$ of the kinetic energy output
of the radio jet. In galaxies with radio jet power $\ga 10^{47}$ erg
s$^{-1}$, the kinetic energy in global back-to-back outflows dominates
the total energy budget of the gas, suggesting that bulk motion of
outflowing gas encompasses the global interstellar medium. This might
be facilitated by the strong gas turbulence, as suggested by recent
analytical work. We compare our findings with recent hydrodynamic
simulations, and discuss the potential consequences for the subsequent
evolution of massive galaxies at high redshift. Compared with recent
models of metal enrichment in high-z AGN hosts, we find that the
gas-phase metallicities in our galaxies are lower than in most low-z
AGN, but nonetheless solar or even super-solar, suggesting that the
ISM we see in these galaxies is very similar to the gas from which
massive low-redshift galaxies formed most of their gas. This further
highlights that we are seeing these galaxies near the end of their
active formation phase.}


\maketitle
\section{Introduction}
\label{sec:introduction}

Substantial observational and theoretical progress in the last decade
has left little doubt that the supermassive black holes that seem to
be nearly ubiquitous in the nuclei of galaxies play a significant role
in shaping the properties of their host galaxies. \citet[][]{silk98}
recognized that the energy output of active galactic nuclei, in spite
of their short lifetimes of only a few $10^{7}$ yrs, is sufficient to
unbind most of the interstellar gas even in very massive, gas-rich
host galaxies, if this energy can be efficiently injected into the
interstellar medium. The details of this mechanism are still
relatively poorly understood. Observations have provided evidence that
outflows can be seen in galaxies with bolometrically luminous AGN as
well as AGN that are dominated by the radio jets produced by narrow
beams of relativistic, synchrotron emitting particles.

Of particular interest is the role of powerful AGN in the early
evolution of massive galaxies at high redshift. Observational
properties of massive low-redshift galaxies as well as cosmological
simulations suggest that star formation in high-z galaxies was
prematurely truncated before star formation could exhaust the
available gas supply. Immense winds driven by the energy released from
AGN are now the truncation mechanism which is preferred
by most cosmological models of galaxy evolution. Detailed models of
how radio jets drive such winds have made impressive progress in the
last years. For example, they predict that prior to forming well
collimated jets upon breakout from the ISM, the jet will very
effectively deposit its momentum and energy during a ``flood and
channel'' phase, as it permeates the ISM along low-density channels, in
agreement with observations \citep[][]{sutherland07, wagner11,
  wagner12}.

Powerful radio galaxies (HzRGs) at high redshift (z$\ga$2) are ideal sites to
study the late formation stages of massive galaxies in the early
Universe. They have high stellar \citep[e.g.,][]{debreuck03,
seymour07, debreuck10} and dynamical masses \citep[][]{villar03,
nesvadba07b}, and high star formation rates of up to 1000 M$_{\odot}$
yr$^{-1}$ \citep[][]{archibald01, reuland04, drouart14}, with implied
formation times of few 100~Myr. They host luminous, obscured quasars
with bolometric luminosities of few $\times 10^{45-46}$ erg s$^{-1}$
\citep[e.g.,][]{carilli02, overzier05, drouart14}, and have powerful
radio jets \citep[e.g.,][]{carilli97,pentericci00}, indicating that
they are the host galaxies of some of the most powerful active
galactic nuclei. Their black hole masses fall near the upper end of
the mass function of supermassive black holes in nearby galaxies
\citep[][]{nesvadba11a}, and scale with the mass of their host
galaxies in a fairly similar way as nearby galaxies which fall onto
the local black-hole bulge mass relationship, suggesting they must be
near the end of their active formation period. \citet{drouart14}
argued that the black holes of HzRGs will outgrow the plausible mass
range for supermassive black holes even in very massive galaxies, if
their growth continues for more than a few $10^7$ yrs, further
highlighting that we are observing these sources at an
outstanding moment of their evolution. The same is suggested by their
high stellar masses, which exceed the amounts of remaining molecular
gas by factors of 10 or more \citep[][]{emonts14}, limiting their
potential future growth in stellar mass.

HzRGs are often surrounded by extended nebulosities of warm ionized
gas \citep[e.g.,][]{villar03, nesvadba08} with sizes of up to about
60~kpc, and irregular gas kinematics, with velocity offsets and line
FWHMs of up to 1000~km s$^{-1}$, respectively. These velocities are
above the escape velocity from the gravitational potential of massive
galaxies, suggesting this gas is outflowing \citep[][]{villar03,
  nesvadba06a, nesvadba08}. These structures are in most cases
elongated along an axis that is approximately aligned with the axis of
the radio jet, and have sizes that are smaller than the jet size, and
dynamical times comparable to the radio-jet lifetime. This has
previously been considered evidence that these are outflows of warm
ionized gas that has been entrained by the expanding cocoon of hot,
shocked gas inflated by the radio jet \citep[][]{nesvadba06a,
 nesvadba08}.

Here we present an analysis of 33 powerful radio galaxies at redshifts
z$\ge$2 with new SINFONI observations, revisiting the kinematic and
energetic signatures of AGN feedback as previously described for
individual sources by \citet{nesvadba06a, nesvadba07b, nesvadba08} and
for a small sample of lower-power radio galaxies by
\citet{collet14b}. We confirm the basic previous results, and
significantly expand them, in particular by studying the trends
between radio power and various gas properties. We find a good overall
qualitative agreement with hydrodynamic jet models, but significant
differences when compared with more detailed predictions.

The paper is organized as follows: In \S\ref{sec:observations} we
describe our sample and observations before outlining our analysis
methods in \S\ref{sec:methodology}. These include in particular how we
constructed the kinematic maps and subtracted the broad-line emission
from the nucleus in the subset of sources which show signatures of direct AGN
light. In \S\ref{ssec:intspec} we discuss the integrated spectral
properties of our sample (the detailed properties of individual
sources are listed in the appendix \ref{sec:individualobjects}), like
emission-line diagnostics, electron densities, gas masses and
extinction, before describing the results from spatially resolved maps
in \S\ref{sec:spatiallyresolvedproperties}. This includes continuum
maps, [OIII]$\lambda$5007 morphologies, maps of H$\alpha$/[OIII] line
ratios, and basic morphological properties of the emission-line
regions. In \S\ref{sec:kinematics} we discuss the maps of relative
velocities and line widths, the radial dependence of the gas
kinematics and surface brightness, and search for signatures of
rotationally or pressure-supported gravitational motion. In
\S\ref{sec:spatiallyresolvedproperties} and \ref{sec:kinematics} we
study the relationship between the gas and radio jet properties, like
position angles, sizes, axis ratios, and the gas kinematics. We also
discuss the implications of these relationships in the context of AGN
feedback, study the energetics of the jets and gas flows, the
efficiency of the energy transfer from the jet to the gas, and compare
with predictions from hydrodynamic models. In \S\ref{sec:feedback} we
interpret our results in the context of feedback, and turn to the
implications for galaxy evolution in \S\ref{sec:galaxyevolution},
before summarizing our results in \S\ref{sec:summary}.  

Throughout the
paper we adopt a flat H$_0$= 70 km s$^{-1}$ Mpc$^{-1}$ concordance
cosmology with $\Omega_M=0.3$ and $\Omega_{\rm \Lambda}=$0.7.

\section{Observations and sample selection}
\label{sec:observations}

\subsection{Sample selection}
\label{ssec:sample}
\begin{figure*}
\centering
\includegraphics[width=0.48\textwidth]{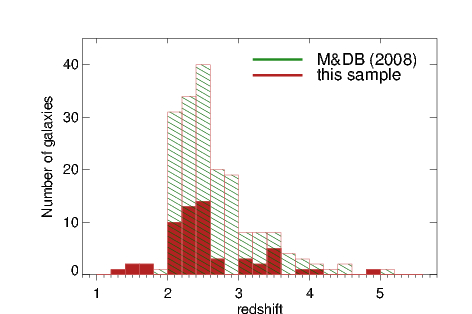}
\includegraphics[width=0.48\textwidth]{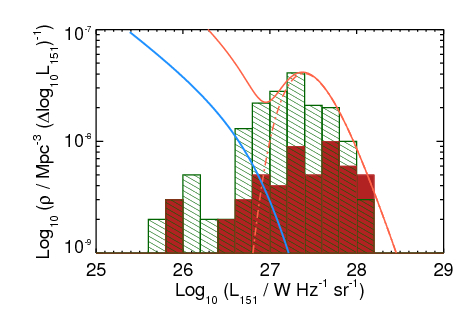}
\caption{Distribution of our targets in redshift
  ({\it left}) and radio power ({\it right}). In both panels, the red
  solid and dark green hatched histograms show our sample and the
  parent sample of all known powerful radio galaxies at z$\ge$2 from
  \citet{miley08}, respectively. The gap between z$=$2.6 and z$=$2.9
  in the left panel is because no bright optical emission lines fall
  into the near-infrared atmospheric windows at these redshifts. The
  light blue and light red lines in the right panel show the
  luminosity function of powerful radio galaxies at redshifts z$=0$
  and z$=2$ from \citet{willott01}, respectively. The dot-dashed light
  red line highlights their high-power 'high-z' population \citep[for
    details see][]{willott01}.}
\label{fig:sample}
\end{figure*}

In total, we observed 49 radio galaxies at z$\ge$2 in a suite of
observing programs with SINFONI between 2009 and 2012. This
corresponds to 32\% of the known radio galaxies at z$\gtrsim$2 in the
compilation of \citet{miley08}. The redshift distribution of our
sample is shown in Fig.~\ref{fig:sample}. All sources are between redshifts 
z$=$1.4$-$4.8, and most between z$=$2.0 and 3.6. We
note a small redshift gap between z$\sim$2.6 and z$\sim$3.0, where no
bright line emission falls into the near-infrared atmospheric
windows. Sixteen of these sources have already been discussed in our
previous publications \citep[][]{nesvadba06a, nesvadba07b, nesvadba08,
  collet14a, collet14b}.

Since these targets were selected in a similar way and observed with
very similar setups, they form a homogeneous sample with the 33 new targets
presented here. For six targets where nuclear broad-line emission was
observed in H$\alpha$, \citet{nesvadba11a} presented an analysis of
their black-hole properties, but not their extended line
emission. They are therefore included in our present study.

The targets are taken from a number of different catalogs, including
the Molonglo Reference Catalog \citep[][]{large81}, the 3$^{{\rm
    rd}}$, 4$^{{\rm th}}$, and 5$^{{\rm th}}$ Cambridge Catalogs
\citep[][]{laing83,pilkington65,pearson75}, the Parkes
\citep[][]{wright90}, Parkes-MIT-NRAO \citep[PMN~][]{griffith93} and
Texas \citep[TXS,][]{douglas96} surveys, the catalog of
Ultra-Steep-Spectrum radio sources \citep[USS;][]{debreuck00}, and the
Sydney University Molonglo Sky Survey \citep[SUMSS][]{mauch03}. Our
sample is therefore not complete in a statistical sense, however,
given the rarity of powerful high-redshift radio galaxies, it would
not be practical to collect a significant number of sources from a
statistically complete sample. Moreover, we wish to study trends in
gas properties as a function of the radio size and radio power, so
that a uniform sampling of radio properties is more important than
matching the radio luminosity function. For a population study this
would of course be different.  In a general sense, these galaxies were
selected for their steep spectral indices from low-frequency radio
surveys. All had known spectroscopic redshifts, typically from
rest-frame UV observations \citep[most redshifts were taken
  from][]{debreuck00}, which implies that our sample is biased towards
galaxies with bright UV or optical line emission. With the exception
of TXS~2355$-$003 \citep[already discussed in][]{collet14a}, we
confirmed the previous redshifts of all galaxies. TXS~2355$-$003 lies
at z$=$1.49, not at z$=$2.49, as previously estimated by
\citet{debreuck00}.

Our choice of targets was guided by our aim to investigate how the
radio properties of the AGN in our sample affect the warm ionized gas
in their host galaxies. This makes it necessary to sample a range in
radio properties as uniformly as possible rather than to match the
radio luminosity function, as would be the case for a population
study. In Fig.~\ref{fig:sample} we compare the distribution of radio
power of our galaxies with the radio luminosity function of
\citet{willott01} of very powerful high-z radio sources. Our sample is
very naturally biased towards galaxies with bright emission lines,
corresponding to the ``high-excitation line'' mode of radio galaxies,
or the ``Quasar mode'' of AGN feedback models, which in turn
correspond to the high-power population of \citet{willott01}. The
figure illustrates that we sample a range in radio power from
$10^{26.3}$ to $10^{29.3}$ W~Hz$^{-1}$, covering the entire range of
about three orders of magnitude that is characteristic for this
population, which, according to \citet{willott01}, fades quickly
towards lower redshift. Our faintest radio sources are only factors of
a few stronger than implied by the radio flux in the most intense
starbursts \citep[][]{collet14b}, which is likely dominated by star
formation \citep[few$\times 10^{25}$ W Hz$^{-1}$, e.g.,][]{barger14},
whereas our most luminous sources are amongst the most powerful radio
sources known in the Universe. For comparison, typical powerful radio
galaxies at low redshift have few$\times 10^{25-26}$ W Hz$^{-1}$
\citep[e.g.,][]{wall85, tadhunter93}.

\subsection{VLT/SINFONI observations and data reduction}
\label{ssec:obsdatred}

Observations were carried out with the near-infrared imaging
spectrograph SINFONI on UT4 of the Very Large Telescope of ESO between
September 2009 and September 2012 under program IDs 079.A-0617,
081.A-0468, 381.A-0541, 082.A-0825, and 083.A-0445. SINFONI is an
image slicer which operates in the J, H, and K-band in a contiguous
8\arcsec$\times$8\arcsec\ field of view at a pixel scale of
125~mas$\times$250~mas.  The spatial resolution of most of our data is
set by the size of the seeing disk, between 0.6\arcsec\ and
1.5\arcsec. The size of the seeing disk for individual galaxies
  in shown as a black ellipse in the lower left corner of each line
  map in Figs.~\ref{fig:maps} to \ref{fig:maps5}. At z$\sim$2,
1.2\arcsec\ corresponds to 10~kpc. This is comparable to a typical
galaxy scale, so that our data allow us to infer for all targets
whether the gas is confined to within the AGN host galaxy or extends
beyond.  MRC~2104$-$242, MG~2037$-$0011, and NVSS~J2135$-$3337
fortuitously have nearby bright stars that served as natural guide
stars for adaptive-optics assisted observations. For those galaxies,
we obtained high-resolution observations with a spatial resolution of
about 0.4\arcsec\ ($\sim$3 kpc). The spatial resolution for these
sources is limited by the pixel size. We used the largest pixel size
also for the adaptive-optics data, because we wanted to reach the
highest possible sensitivity to the lowest surface-brightness gas in
our sources.

\begin{table*}
\centering
\begin{tabular}{lccccccccc}
\hline
Source              & RA(J2000)  & Dec (J2000)  & redshift & DL     & DA    & Band  &  ToT.\\
                    &            &              &          &  [Gpc] & [Gpc] &       &  [s] \\
\hline
MRC~0114$-$211      & 01:16:51.4 & $-$20:52:07  & 1.41     & 10.1 & 8.4   & HK    &  4800 \\
BRL~0128$-$264      & 01:30:27.9 & $-$26:09:58  & 2.35     & 18.9 & 8.2   & HK    &  7500 \\
MRC~0156$-$252      & 01:58:33.6 & $-$24:59:31  & 2.02     & 15.7 & 8.4   & HK    &  13800 \\
USS0211$-$122       & 02:14:17.4 & $-$11:58:47  & 2.34     & 18.8 & 8.2   & HK    &  10500 \\
MRC~0251$-$273       & 02:53:16.7 & $-$27:09:13  & 3.16     & 27.1 & 7.6   & HK    &  11400 \\
RC~J0311$+$0507     & 03:11:47.9 & $+$05:08:01  & 4.80     & 44.5 & 6.4   & HK    &  5400 \\
MP~J0340$-$6507     & 03:40:44.6 & $-$65:07:12  & 2.29     & 18.3 & 8.2   & HK    &  12600\\
PKS~0529$-$549      & 05:30:25.4 & $-$54:54:22  & 2.58     & 21.2 & 8.0   & HK    &  21600 \\
5C~7.269            & 08:28:38.8 & $+$25:28:27  & 2.22     & 17.6 & 8.3   & HK    &  7200 \\
MRC~1017$-$220      & 10:19:49.0 & $-$22:19:58  & 1.77     & 13.4 & 8.5   & HK    &  11100 \\
TN~J1112$-$2948     & 11:12:23.9 & $-$29:48:07  & 3.09     & 26.3 & 7.6   & HK    &  22500 \\
TXS~1113$-$178      & 11:16:14.7 & $-$18:06:23  & 2.24     & 17.8 & 8.2   & HK    &  1800\\
3C~257              & 11:23:09.4 & $+$05:30:18  & 2.46     & 20.0 & 8.1   & HK    &  8100 \\
USS~1243$+$036      & 12:45:38.4 & $+$03:23:21  & 3.57     & 31.3 & 7.3   & HK    &  12000 \\
MG~1251$+$1104      & 12:51:00.0 & $+$11:04:22  & 2.32     & 18.6 & 8.2   & HK    &  2400 \\
MRC~1324$-$262      & 13:26:54.7 & $-$26:31:42  & 2.28     & 18.2 & 8.2   & HK    &  2400 \\
TN~J1338$-$1942     & 13:38:26.0 & $-$19:42:31  & 4.11     & 37.0 & 6.9   & J,HK  &  14100, 10500 \\
USS~1410$-$001      & 14:13:15.1 & $-$00:23:00  & 2.36     & 19.0 & 8.2   & HK    &  10500 \\
MRC~1558$-$003      & 16:01:17.3 & $-$00:28:48  & 2.53     & 20.7 & 8.1   & HK    &  21300 \\
USS~1707$+$105      & 17:01:06.5 & $+$10:31:06  & 2.35     & 18.9 & 8.2   & HK    &  10500 \\
3C~362              & 17:47:07.0 & $+$18:21:10  & 2.28     & 18.2 & 8.2   & HK    &  7200 \\
MP~1758$-$6738      & 17:58:51.3 & $-$67:37:28  & 2.03     & 15.8 & 8.4   & HK    &  7200 \\
TN~J2007$-$1316     & 20:07:53.2 & $-$13:16:44  & 3.84     & 34.1 & 7.1   & HK    &  9600 \\
MRC~2025$-$218      & 20:27:59.5 & $-$21:40:57  & 2.63     & 21.7 & 8.0   & HK    &  18600 \\
MG~2037$-$0011      & 20:37:13.4 & $-$00:10:59  & 1.51     & 11.0 & 8.5   & J,H   &  5400, 4050\\
MRC~2048$-$272      & 20:51:03.6 & $-$27:03:03  & 2.06     & 16.1 & 8.3   & HK    &  13200 \\
MRC~2104$-$242      & 21:06:58.2 & $-$24:05:11  & 2.49     & 20.3 & 8.1   & H,K   &  8100,10800\\
4C~23.56            & 21:07:14.8 & $+$23:31:45  & 2.48     & 20.2 & 8.1   & HK    &  1500\\
NVSS~J2135$-$3337   & 21:35:10.5 & $-$33:37:04  & 2.52     & 20.6 & 8.1   & K     &  18600 \\
MG~2144$+$1928      & 21:44:07.5 & $+$19:29:15  & 3.59     & 31.5 & 7.3   & HK    &  16500 \\
MRC~2224$-$273      & 22:27:43.2 & $-$27:05:02  & 1.68     & 17.0 & 8.3   & HK    &  2700 \\
TN~J2254$+$1857     & 22:54:53.7 & $+$18:57:04  & 2.15     & 17.0 & 8.3   & H,K   &  3600,10800 \\
MG~2308$+$0336      & 23:08:25.2 & $+$03:37:03  & 2.46     & 20.0 & 8.1   & HK    &  1800\\
TXS~2353$-$003      & 23:55:35.9 & $-$00:02:48  & 2.59     & 21.3 & 8.0   & HK    &  8700\\
\hline
\end{tabular}
\caption{
Our sample. Column (1) -- Source ID. Column (2) -- Right
ascension. Column (3) -- Declination. Column (4) -- Luminosity
distance in Gpc.  Column (5) -- kpc per arcsec.. Column (6) --
Observing band. Column (7) -- On-source observing time in seconds.}
\label{tab:sample}
\end{table*}

On-source observing times were between 30 and 375 minutes. The range
in observing times is large because data obtained in service mode
(most of the sample) were not always completed. As a consequence of
this and varying spectral bands and observing conditions, the depth of
our observations ranges between rms$=$0.8 and $2 \times 10^{-17}$ erg s$^{-1}$ cm$^{-2}$
\AA$^{-1}$ arcsec$^{-2}$, for a circular aperature with
1\arcsec\ diameter and after smoothing spatially by 3$\times$3 pixels,
akin to our data cubes prior to analysis. In spite of this, we
detected line emission from all galaxies at least in one band,
including those with the shortest exposure times. With very short
exposures we may however have missed faint, extended line emission in
some cases. Observing parameters and exposure times of individual
sources are listed in Table~\ref{tab:sample}. We also point out in our
descriptions of individual sources in
Appendix~\ref{sec:individualobjects} which sources might be affected.

Most data were obtained with the H$+$K grating which covers the H and
K band simultaneously at R$\sim$1500 (corresponding to FWHM$\approx$200 km
s$^{-1}$) and is sufficient to resolve all emission lines spectrally
in our targets. Some galaxies, in particular those observed
during the visitor run 079.A-617, were observed in the H and K-band
individually. Where appropriate, for example owing to the redshift of
the source, we also observed in the J band. The spectral resolving
power in the J, H, and K band is R$\sim$2000, R$\sim$3000, and
R$\sim$4000, respectively. Table~\ref{tab:sample} lists all bands in
which each target was observed. The bands covered for each source can
also be found in the detailed descriptions of individual sources in
Appendix~\ref{sec:individualobjects}.

\subsection{New and archival VLA observations}
\label{ssec:vlaobs}

We also collected a set of new VLA continuum imaging for parts of our
sample through program AN~0129, to increase the number of targets with
arcsecond resolution radio morphologies. C. Carilli kindly shared his
radio imaging of 17 sources with us, which had previously been
discussed by \citet{carilli97} and \citet{pentericci00}, and we
obtained the cm imaging of another eight sources from the VLA
archive. Tab.~\ref{tab:radioobs} lists the sets of the new and
archival radio images used for this study.

Observations of our proprietary program AN~0129 were carried out in
one observing session with 25 antennae on 2007 July 5, during the
upgrade to the EVLA, and under rather unstable conditions. This made
it unfortunately impossible to measure polarizations robustly in two
bands and obtain rotation measures, as we had initially
planned. However, the total power measurements were not affected. 

New data were obtained with the A-array (BnA for targets at
Dec$<-20^\circ$) at one or both of the frequencies 4.885~GHz and
8.485~GHz, limited by the constraints of observing time.  We observed
each source several times at different hour angles, covering typically
about 4-5~hrs per source. On-source integration times were between
1260 and 1310~s per source in the 8.4~GHz band, and between 5100 and
5480~s per source in the 4.8~GHz band, respectively. The resulting rms
is between 17 and 86~$\mu$Jy in the final maps. 

\begin{table*}
\begin{tabular}{lrrlllllll}
\hline
\hline
Source       & RA(J2000) & Dec(J2000) & freq. &  rms           & Beam     & PA   & Program & Reference \\
             &           &            &     [GHz]     & [$\mu$Jy/bm]  & [arcsec,arcsec] & [deg.] &         & \\
\hline
MRC0114$-$211   & 01:16:51.40  &  20:52:06.8  &  23.3      & 362  & 0.1, 0.1 & -11  & AJ206 & \\
BRL~0128$-$264  & 01:30:27.82  & -26:09:57.0  &  4.8       & 28   & 1.0, 0.8 & -75  & AN129 & \tablefootmark{a}\\
MRC~0156$-$252  & 01:59:04.40  & -25:28:38.9  &  8.2       & 33   & 0.9, 0.4 & -5   & AC374 & \citet{carilli97} \\
USS~0211$-$122  & 02:14:17.40  & -11:58:46.0  &  8.2       & 28   & 0.4, 0.3 & -5   & AC374 & \citet{carilli97} \\
MRC~0251$-$273  & 02:53:16.74  & -27:09:13.0  &  8.5       & 13   & 0.9, 0.3 & 13.  & AD520 & \\
RC~0311$+$0507  & 03:09:09.86  &  04:56:48.3  &  4.9       & 287  & 0.7, 0.6 & 1.0  & AH167 & \citet{parijskij14} \\
MRC~0406$-$244  & 04:08:51.50  & -24:18:16.0  &  4.9       & 22   & 1.1, 0.7 & 89   & AN129 & \tablefootmark{a}\\
PKS~0529$-$549  & 05:30:25.40  & -54:54:23.1  & 18.5       & 54   & 1.0, 0.8 & 43   & ATCA  & \citet{Broderick07b} \\ 
5C~07.269       & 08:28:38.97  &  25:28:27.2  &  8.4       & 146  & 0.2, 0.2 &  35  & AB808 &  De Breuck \\
TXS~0828$+$193  & 08:30:53.40  &  19:13:16.0  &  4.9       & 21   & 0.6, 0.4 & -69  & AN129 &  \tablefootmark{a}\\
3C~257          & 11:17:69.32  &  06:03:14.0  &  8.4       & 13   & 0.4, 0.2 & 43   & AV165 &  \\
TN~J1112$-$2948 & 11:12:23.94  & -29:48:07.0  &  4.8       & 130  & 4.0, 1.3 & -36  & ADA000& DeBreuck   \\
TXS~1113$-$178  & 11:16:14.68  & -18:06:23.6  &  4.7       & 57   & 0.8, 0.4 & -2   & AC374 & \citet{carilli97} \\
USS1243$+$036   & 12:45:38.39  &  03:23:21.0  &  8.4       & 25   & 0.3, 0.2 & 48   & AM336 & \\
MRC~1324$-$262  & 13:26:54.66  & -26:31:41.4  &  8.2       & 32   & 0.5, 0.3 & 12   & AC374 & \citet{carilli97} \\
TN~J1338$-$1942 & 13:38:26.06  & -19:42:30.1  &  8.2       & 12   & 0.4, 0.2 & 1    & AP360 & \citet{carilli97} \\
USS~1410$-$001  & 14:13:15.15  & -00:23:00.8  &  4.7       & 43   & 0.6, 0.5 & 17   & AC374 & \citet{carilli97} \\ 
MRC~1558$-$003  & 16:01:17.43  & -00:28:46.4  &  8.5       & 14   & 0.3, 0.3 & 22   & AP360 & \citet{carilli97}\\
USS~1707$+$105  & 17:10:06.85  &  10:31:09.0  &  4.8       & 38   & 0.5, 0.4 & -24  & AD520 & \\
MP~1747$+$182   & 17:47:07.00  &  18:21:10.8  &  4.7       & 13   & 0.5, 0.5 & 13   & AP360 & \citet{carilli97}\\ 
TN~J2007$-$1316 & 20:07:53.26  & -13:16:43.6  &  8.5       & 37   & 0.4, 0.2 & 16   & AD520 & DeBreuck  \\  
MRC~2025$-$218  & 20:27:59.49  & -21:40:56.9  &  4.8       & 56   & 2.9, 1.2 & -30  & AN129 & \tablefootmark{a}\\
MG~2037$-$0011  & 20:37:13.40  & -00:10:58.0  &  8.4       & 22   & 0.3, 0.2 &  36  & AN129 & \tablefootmark{a}\\
MRC~2104$-$242  & 21:06:58.10  & -24:05:11.0  &  4.8       & 28   & 1.6, 0.6 &  56  & AN129 & \tablefootmark{a} \\  
4C~23.56        & 21:07:14.28  &  23:31:41.2  &  8.4       & 25   & 0.2, 0.2 & -34  & AC379 &   \\
NVSSJ2135$-$3337& 21:35:10.48  & -33:37:04.4  &  4.8       & 28   & 1.3, 1.0 &  30  & AN129 & \tablefootmark{a}\\
MG~2144$+$1928  & 21:44:07.52  &  19:29:14.8  &  8.4       & 66   & 0.2, 0.2 &  -8  & AS446 & \\
TN~J2254$+$1857 & 22:54:53.71  &  18:57:04.6  &  4.9       & 17   & 0.4, 0.4 & -73  & AN129 & \tablefootmark{a}\\
MRC~2308$+$0336 & 23:03:19.62  &  03:04:25.1  &  5.0       & 36   & 0.5, 0.4 &  40  & AB375 &  \\
TXS~2353$-$003  & 23:55:35.90  & -00:02:48.0  &  4.8       & 19   & 0.4, 0.4 & -21  & AN129 & \tablefootmark{a}\\
TXS~2353$-$003  & 23:55:35.90  & -00:02:48.0  &  8.4       & 17   & 0.3, 0.3 &  40  & AN129 & \tablefootmark{a}\\
\hline
\hline
\end{tabular}
\caption{List of new and archival VLA observations complementing this study.\label{tab:radioobs} }
\tablefoottext{a}{This study.}
\end{table*}

3C~48 was used as a primary calibrator to set the flux scale, and
secondary calibrators located within $\approx 10^\circ$ of the target
source were observed every $20-30$ min for phase calibration. These
data were all reduced using standard techniques within {\sc
 AIPS}. Sources were self-calibrated first in phase, and then in
amplitude and phase, to improve the final image quality. Final images
were made using a ROBUST$=0$ antenna weighting as a compromise between
angular resolution and sensitivity.

Already existing, non-proprietary VLA observations of our sources were
obtained from the VLA archive and reduced in a similar manner. The
archival data of MG~2308$+$0336 did not have a primary flux
calibrator, so we were forced to bootstrap the flux density from the
secondary calibrator. As a result, the uncertainty in radio flux is
likely high compared to the other sources, about 50\%. Given that our
primary use of these data is to constrain the radio morphology and the
kinetic power, which introduces astrophysical uncertainties of factors
of a few, we do not consider this an important limitation of our
analysis.

\section{Methodology}
\label{sec:methodology}

\subsection{Integrated spectra}
\label{sssec:intspectra}

We show integrated spectra for each source in
Figs.~\ref{fig:intspec}. These spectra were
obtained by summing over all spatial pixels where the signal-to-noise
ratio of the [OIII]$\lambda$5007 line exceeded 3$\sigma$. We extracted
other line fluxes from the same area. Except for H$\alpha$ in some
cases, [OIII]$\lambda$5007 was typically more extended than other
emission lines.  Before adding the spectrum in a given spatial pixel,
we corrected for the velocity shift in this pixel relative to the
nucleus. Our integrated line profiles therefore give a
luminosity-weighted measure of the intrinsic line widths of the
extended emission-line gas, which is not affected by the resolved
velocity offsets. This also helps to maximize the signal-to-noise
ratios of the fainter lines.

\subsection{Removal of nuclear broad line emission}
\label{ssec:blrremoval}

Six of the galaxies presented here have bright broad H$\alpha$ line
emission from the AGN \citep[][]{nesvadba11a}, a seventh with lower
radio power was found by \citet{collet14b}. To analyze the extended
emission-line regions of these galaxies, it was necessary to remove
the nuclear component first. This is not an easy task, because the
broad and narrow-line profiles are often not very well approximated by
Gaussian line profiles, and therefore, fitting the spectra of a galaxy
with multiple Gaussians corresponding to the broad and narrow lines
often leads to significant residuals. Several approaches have been
proposed in the literature \citep[e.g.,][]{christensen06, canodiaz12},
but all leave large residuals, which is why we developed another
method.

Rather than simply approximating the full BLR profile with a Gaussian,
we only use a Gaussian fit to interpolate the observed spectrum at the
wavelengths covered by the narrow emission lines, and construct a
hybrid BLR profile in which the intrinsic line profile along the wings
is not changed. We subtract this hybrid BLR line profile from all
spatial pixels contaminated by BLR emission, after scaling the flux of
the nuclear component by the flux expected from the point spread
function as measured from the AGN continuum. Through this scaling, the
size of the seeing disk as measured during the observation itself is
automatically taken into account. For a more detailed description and
the performance of these fits see \citet{colletthesis} and Nesvadba et
al. (2017, in prep.).

\subsection{Map construction}
\label{ssec:mapconstruction} 

We constructed emission-line maps of these galaxies in very similar
ways to the SINFONI samples previously analyzed by \citet{nesvadba06a,
  nesvadba08, nesvadba11a}, \citet{collet14a, collet14b}, and Nesvadba
et al. (2016, in prep.). Namely, we obtained integrated spectra from
box apertures in each galaxy, covering 3$\times$3 pixels
(0.4\arcsec$\times$0.4\arcsec). This is less than the size of the
seeing disk, and corresponds to the spatial resolution that can be
obtained with adaptive-optics assisted observations and a pixel scale
of 125~mas$\times$250~mas (\S\ref{ssec:obsdatred}). This
therefore minimizes pixel-to-pixel noise without introducing
artificial beam-smearing effects. For galaxies where
[OIII]$\lambda\lambda$4959,5007 is observed, we fitted the
[OIII]$\lambda\lambda$4959,5007 lines first, which probe the same gas,
and have a fixed flux ratio $R_{\rm 4959,5007}=1/3$. For most galaxies
we use the [OIII]$\lambda$5007 maps to measure the morphology and
kinematics of the warm ionized gas. In \S\ref{ssec:oiiihacomparison}
we will present our arguments why this gives representative results
for the warm ionized gas overall. H$\alpha$ is bright enough to allow
for independent fits, but is too broad and heavily blended with
[NII]$\lambda\lambda$6548,6583 in most galaxies to provide good
kinematic constraints. Other lines, including H$\beta$, are too faint
to be well mapped over large radii. Generally speaking,
single-component Gaussians were sufficient to obtain good
fits. Exceptions are mentioned in the individual target descriptions
in Appendix~A.

To monitor the goodness-of-fit and evaluate the need for multiple
emission-line components, we also constructed maps of the reduced
$\chi^2$, as well as cubes of the fit residuals. Fit residuals are
usually below a few percent of the fitted line flux, and median
reduced $\chi^2$ are between 0.7 and 5, depending on the degree of
contamination with night-sky lines to the emission lines. A notable
exception is USS~0211$-$122, which we will further discuss below in
\S\ref{ssec:induss0211}.

Our galaxies are too faint in the continuum to study their spectral
energy distribution, however, we were able to localize the peak of the
continuum emission in 23 targets.  We constructed line-free continuum
images by averaging along the spectral direction of our data cubes
after masking the brightest night-sky lines and the line emission from
our targets. Continuum morphologies, where detected, are shown as
contours in Figs.~\ref{fig:maps} to~\ref{fig:maps5}. Imperfections in
the night-sky subtraction and flat fielding, and night-sky line
residuals make it unfortunately very difficult to obtain a good sky
subtraction from SINFONI cubes, which makes continuum flux
measurements very challenging. Typical uncertainties are about
0.3$-$0.5~mag, so that measurements through broad-band filters are
more reliable.

\section{Integrated spectral properties of the sample}
\label{ssec:intspec}

\subsection{Integrated spectra}
\label{sssec:results.intspec}

In Fig.~\ref{fig:intspec} we show the integrated spectra of our
targets. Wavelengths between 1.8~$\mu$m and 2.0~$\mu$m, i.e., between
the H and K~bands, and where the atmospheric transmission is very low,
are clipped. For galaxies at redshifts z$\sim 2.0-2.6$, the
[OIII]$\lambda\lambda$4959,5007 doublet and H$\beta$ fall into the H,
and [OI]$\lambda$6300, H$\alpha$, [NII]$\lambda\lambda$6548,6583, and
[SII]$\lambda\lambda$6716,6731 fall into the K-band. For galaxies at
$z\sim 3.0-3.8$, [OII]$\lambda$3727 falls into the H-band and H$\beta$
and [OIIII]$\lambda\lambda$4959,5007 fall into the
K-band. [OI]$\lambda$6300 has been detected in six sources. The
[SII]$\lambda\lambda$6716,6731 doublet has been detected in 13
sources, and the doublets have been spectrally resolved into
individual components in two galaxies. In eleven sources, the two
lines of the doublet are blended due to large intrinsic widths and the
relatively low spectral resolving power of R$=$1500. H$\alpha$ and
[NII]$\lambda$6548,6583 are also blended in most sources. Nonetheless,
visual inspection of integrated spectra already shows that the
[NII]$\lambda$6583/H$\alpha$ ratios are generally below unity, and
often below 0.5, unlike in AGN host galaxies at low redshift. We will
come back to this point in \S\ref{ssec:diagnostics}
and~\S\ref{ssec:metallicity}.

\subsection{Diagnostic diagrams and gas-phase metallicities}
\label{ssec:diagnostics}

The bright, rest-frame optical emission lines contain a multitude of
constraints on the gas properties, including electron temperatures and
densities, ionization parameter, extinction, and gas-phase metal
abundance. Taken together, combinations of these lines provide
observationally very convenient diagnostics of the gas heating
mechanism by young stellar populations or active galactic nuclei
\citep[e.g.,][]{veilleux87, kewley01, kauffmann03, kewley06}.

Identifying HII regions and AGN narrow-line regions in galaxies by
means of their bright optical line ratios is a very successful
tool for galaxy evolution studies at low redshift, where the
relative uniformness of most HII and AGN narrow-line regions results
in fairly narrow sequences in diagnostic diagrams of, e.g., the
[NII]/H$\alpha$ and [OIII]/H$\beta$ line ratios
\citep[``BPT-diagrams''][]{veilleux87, kewley01, kewley06}.

\begin{figure}
\centering
\includegraphics[width=0.5\textwidth]{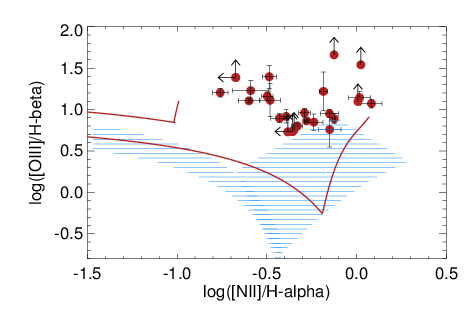}
\caption{
\label{fig:metallicity} BPT
 diagram of our sources with two of the models of \citet{kewley13a}
 shown for z$=$2.5. Red circles show the 25 radio galaxies from the
 present analysis and those of \citet{collet14b}, which have
 measurements of all four lines or sensitive upper limits (on H$\beta$
 or [NII]$\lambda$6583). We also mark the range of line ratios
 expected for low-redshift galaxies with ordinary star formation
 properties and high-metallicity AGN narrow-line regions (light blue
 hatched region). The red lines show the 'high-redshift' starburst
 region of \citet{kewley13a} and the AGN branch with low-metallicity
 narrow-line regions (their scenario 4)}.
\end{figure}

In Fig.~\ref{fig:metallicity} we show where our galaxies fall relative
to the [NII]/H$\alpha$ vs. [OIII]/H$\beta$ diagram. Given that our
sources host very luminous obscured quasars
\citep[e.g.,][]{overzier05, drouart14}, it may not be surprising that
our galaxies fall outside the sequence of HII regions, and into the
part of the diagram that is characteristic of AGN photoionization
\citep[see also][]{villar97, villar03, humphrey09}. However, our
galaxies do not fall onto the classical AGN branch from low-redshift
studies either, but towards higher [OIII]$\lambda$5007 fluxes for a
given H$\beta$ flux and [NII]/H$\alpha$ ratio.

\citet{kewley13a} and \citet{kewley13b} predicted such a shift for gas
photoionized by powerful AGN, when imposing a redshift evolution of
narrow-line metallicity as modeled by \citet{dave10}, which scales
with the cosmic history of star formation. They distinguish between
two models of metal-poor narrow-line regions, either in combination
with a low-redshift star-forming branch, or with a star-forming branch
shifted towards lower [NII]/H$\alpha$ values for a given
[OIII]/H$\beta$ ratio, as found in many actively star-forming galaxies
at z$\sim$2 \citep[][]{erb06b, lehnert09, nmfs09, steidel14}. 

The [NII]/H$\alpha$ ratios of our galaxies fall into the range
predicted by the low-metallicity models of \citet{kewley13a}, however,
the [OIII]/H$\beta$ ratios exceed this range by few 0.1~dex. 
Our data are not sensitive to distinguishing
whether the star-forming ISM in our galaxies may be more akin to
low-redshift conditions, or the more extreme environs in high-z
starburst galaxies. But given the high star formation rates
implied by the far-infrared continuum in at least some of our sources
\citep[][]{drouart14}, and because the offset of the star formation
branch in these galaxies appears to be most likely due to a higher ionization
parameter in these galaxies, we suspect that the latter will likely be
the case. We will discuss this point further in \S\ref{ssec:metallicity}

\subsection{Electron densities}
\label{sssec:densities} 

Amongst the bright rest-frame optical lines, the
[OII]$\lambda\lambda$3726,3729 and [SII]$\lambda\lambda$6716,6731
doublets are particularly interesting, because they provide
constraints on the electron density of the warm ionized gas for
densities between about 100 and $10^5$ cm$^{-3}$
\citep[e.g.,][]{osterbrock89}. Unfortunately, both components of these
doublets are very near in wavelength to each other, which makes it
challenging to isolate the two lines in most of our targets, given
their broad intrinsic widths of typically $\sim$800 km s$^{-1}$
(Appendix~\ref{sec:individualobjects} gives the FWHM for individual
targets), moderate spectral resolution of R$=$1500 provided by the
H$+$K grating, presence of night-sky line residuals, and relatively
faint line fluxes.

In spite of this, we were able to identify and resolve the
[OII]$\lambda\lambda$3726,3729 doublet in RC~J0311$+$0507
(\S\ref{ssec:indrcs0311}) with a line ratio of F(3727)/F(3729)=0.96,
which suggests an electron density of 350~cm$^{-3}$ for the best-fit
value of the line ratio of F(3727)/F(3729)=0.96 \citep[see also
  Table~\ref{tab:emlines}][]{osterbrock89}. To our knowledge, this is
the first estimate of the electron density in an AGN host galaxy at
z$\ge$4. In USS~0211-122 at z$=$2.34, we find a
[SII]$\lambda\lambda$6716,6731 ratio of 1.3, corresponding to an
electron density of 370~cm$^{-3}$ (Table~\ref{tab:emlines}). In
MRC~0114$-$211 at z$=$1.4, we find a line ratio of 1.25 between the
two [SII] doublet lines, which corresponds to a best-fit value of
$n_e=310$~cm$^{-1}$.  \citet{nesvadba06a} and \citet{nesvadba08}
previously found electron densities of 390~cm$^{-3}$ and
500~cm$^{-3}$, in MRC~1138$-$262 and MRC~0406-242, respectively, and
\citet{collet14b} found $n_e = 500$~cm$^{-3}$ and 750~cm$^{-3}$ for
two radio galaxies with somewhat lower radio power at similar
redshifts, NVSS~J210626-314003, and NVSS~J012932-385433. As a fiducial
value for the whole sample, we will adopt n$_e=$500~cm$^{-3}$ in the
present analysis.

\subsection{Extinction and ionized gas masses}
\label{sssec:gasmasses}

H$^+$ is the most abundant species of warm ionized gas and can therefore
be used to estimate the total mass of warm ionized gas, if extinction
and electron density (\S\ref{sssec:densities}) are known. We have
observed both lines, H$\alpha$ and H$\beta$, in  15 galaxies, and follow
\citet{osterbrock89} and \citet{dopita03} in estimating the extinction from the
observed line ratios and Balmer decrements. We assume a Galactic
extinction law and a Balmer decrement of H$\alpha$/H$\beta=2.88$.  We
find a large range of (nominally) $A_V=$0.0~mag to 4.7~mag. Results
for individual galaxies are found in Table~\ref{tab:gasmasses}. For
H$\alpha$, this corresponds to correction factors between 1 and 13. We
note that we do not correct the H$\beta$ line for potential underlying
stellar absorption. As previously discussed by \citet{nesvadba08}, the
expected absorption line equivalent widths of few \AA\ are very small
compared to the large emission-line equivalent widths of H$\beta$,
which makes such a correction unnecessary.

We can derive a mass of warm ionized gas mass from the H$\alpha$ line
flux by setting

\begin{equation}
M_{\rm WIM,H\alpha}=\frac{L_{\rm H\alpha}}{h\ \nu_{\rm H\alpha}\alpha_{\rm eff}^{\rm H\alpha}} = 3.3\times 10^8 L_{\rm H\alpha,43}n_{\rm e,100}^{-1} {\rm M_{\odot}},
\label{eqn:wimmass}
\end{equation}

where L$_{\rm H\alpha}$ is the H$\alpha$ luminosity, $h$ the Planck constant,
$\nu_{\rm H\alpha}$ the rest-frame frequency of the H$\alpha$ line,
$\alpha_{\rm eff}^{\rm H\alpha}$ the effective H$\alpha$ recombination
coefficient, and $n_{\rm e,100}$ the electron density in units of 100
cm$^{-3}$. 

For the galaxies at z$\sim$3.0$-$3.6, where H$\beta$ falls into the
K-band, and H$\alpha$ outside the atmospheric windows, we can use an
analogous estimate for the warm ionized gas mass based on the H$\beta$
luminosity:

\begin{equation}
M_{\rm WIM,H\beta}=\frac{L_{\rm H\beta}}{h\ \nu_{\rm H\beta}\alpha_{\rm eff}^{\rm H\beta}} = 9.5\times 10^8 L_{\rm H\beta,43}n_{\rm e,1oo}^{-1} {\rm M_{\odot}}.
\end{equation}

However, we will treat the warm ionized gas mass estimates from
H$\beta$ as lower limits. The reason is the higher dust attenuation in
the rest-frame V compared to the R-band, and the small number of
galaxies with good detections of Balmer lines higher than H$\beta$. We
only have a good detection of H$\gamma$ for PKS~0529$-$549, for which
we also have H$\alpha$ and H$\beta$ measured. For redshifts
$z\sim3.1-3.6$, H$\gamma$ falls inbetween the H and the K band, which
makes it unobservable for most of the galaxies where we have only
H$\beta$ measured.

The estimates provided here are a factor~3 lower than those previously
given by \citet{nesvadba06a, nesvadba08}, due to a missing factor 3 in
the previous estimates. Given the high masses of warm ionized gas
found in that study, and large systematic uncertainties of factors of
a few, this difference has however no impact on the scientific
arguments presented previously. We list the new estimates of the warm
ionized gas masses in these galaxies in Table~\ref{tab:gasmasses}.

With a fiducial electron density n$_e=$500~cm$^{-3}$ and the
extinctions given in Table~\ref{tab:gasmasses} we find warm ionized
gas masses of $2\times 10^8-5\times 10^9$ M$_{\odot}$. 
These masses are much greater than the masses of warm ionized gas
typically found in star-forming galaxies or radio-quiet quasar host
galaxies at similar redshifts. For example, using the H$\alpha$
luminosities measured by \citet[][]{forster09} and \citet{lehnert09},
we find warm ionized gas masses of few times $10^7$~M$_{\odot}$ or
less in typical UV/optical and sub-mm selected high-z star-forming
galaxies.

For eight of our HzRGs we can compare the warm ionized and the
molecular gas masses directly. \citet{emonts14} measured masses
between $3.6$ and $7\times 10^{10}$ M$_{\odot}$ of molecular gas from
CO(1-0) observations at ATCA in five of our sources, and placed upper
limits between $2$ and $3\times 10^{10}$ M$_{\odot}$ for another three
(Table~\ref{tab:gasmasses}). In TXS~0828$+$193, \citet{nesvadba09a}
found an upper limit of $2\times 10^{10}$ M$_{\odot}$ in warm
molecular gas, which has $1.7\times 10^{10}$ M$_{\odot}$ of warm
ionized gas \citep[][but using equation
  \ref{eqn:wimmass}]{nesvadba08b}. Overall, we find fractions of warm
ionized to molecular gas masses between 2 and 14\% when considering
galaxies with CO detections, and between $\ge$ 20\% and 80\%
when using upper limits on the molecular gas mass in galaxies without
CO detections. In comparison, the galaxies of \citet{tacconi10},
which have CO observations from PdBI and are also included in the
sample of \citet{nmfs09} with SINFONI imaging spectroscopy,
have ionized-to-molecular mass ratios of few $\times 10^{-4}$.
\citet[][]{hopkins10} showed analytically that blastwaves as driven,
e.g., by jets, can deform, ablate, or even destruct clouds, which also
lowers their self-shielding capabilities significantly. They suggest
that between about 20\% and 100\% of the gas can become photoionized
by the AGN within few times $10^7$ yrs, broadly consistent with our
results.

\section{Spatially resolved properties}
\label{sec:spatiallyresolvedproperties}

\subsection{Continuum morphologies}
\label{ssec:continuum} 
The rest-frame optical continuum morphologies, where detected, are
shown as contours in Fig.~\ref{fig:maps} to \ref{fig:maps5}. 
  These morphologies are not contaminated by line emission. They were
  obtained by collapsing over line-free wavelength ranges over the
  full available spectral bandwidth, i.e., the H and K-band in case of
  data taken with the HK grating. We detect
the continuum in 23 sources. Most sources have only a single,
unresolved continuum source. Exceptions include MRC~0114-211, which
has another nearby continuum emitter seen in projection just outside
the emission-line region (see \S\ref{ssec:indmrc0114} for details).
3C~257 and MP~0340$-$6507 have both an extended continuum.  This could
either indicate the presence of two partially blended, and presumably
interacting sources, as expected in the merger scenario, or extended
dust in a single extended source, perhaps dust illumiated by the AGN
or extended star-forming regions \citep[e.g.,][]{cimatti97, vernet01,
  hatch08}. \citet{debreuck10} suggest that extended dust emission
could also indicate binary AGN. In MRC~2048$-$272, we find a second
continuum source at a projected distance of 3\arcsec, which is however
at a very different redshift, z$=$1.52 (the radio galaxy is at
z$=$2.06). This intervening galaxy is also the source of the
far-infrared emission reported by \citet{drouart14}.

Overall, we have therefore little evidence for on-going merging
activity in our galaxies. Of course, with a field-of-view of
8\arcsec$\times$8\arcsec, corresponding to 64~kpc$\times$64~kpc at
z$\sim$2, we are not able to probe the more extended environment where
companion galaxies have been found for several of our sources in
previous studies \citep[e.g.,][]{chambers96, lefevre96, kurk04,
  venemans07, hatch11, koyama13, hayashi12, wylezalek13, cooke14,
  collet14a}. With a seeing of typically 1\arcsec, corresponding to
8~kpc at z$\sim$2, we are not sensitive to advanced major mergers that
are approaching coalescence, and we might also miss low-mass or
strongly obscured sources \citep[][]{ivison08}. However, even
relatively distant, early-stage major mergers have been proposed to
trigger phases of rapid star formation and galaxy growth in massive
high-z galaxies \citep[][]{hopkins07}, which does not seem to be the
case for most of our sources.

\subsection{[OIII] morphologies and surface luminosities}
\label{ssec:oiiimorphologies}
The morphologies of the warm ionized gas in our sources are shown in
Figs.~\ref{fig:maps} to ~\ref{fig:maps5}. Most are very
irregular. Isophotal sizes of the bright [OIII]$\lambda$5007 line
emission down to the 3$\sigma$ depths of our data sets are given in
Table~\ref{tab:eelrproperties} along the minor and major axis,
respectively. Error bars were derived from the positional accuracy of
the point spread function at a signal-to-noise ratio of S/N$=$3,
corresponding to the surface-brightness cutoff in our maps, FWHM
[2\ S/N]$^{-1}$. We multiply by a factor $\sqrt{2}$ to take into
account that we need to measure two positions to obtain the size of a
major or minor axis, respectively. Sizes range from the resolution
limit (5$-$10~kpc, depending on the seeing at the time of observation
of each source, see Table~\ref{tab:sample}) to 69~kpc$\times$23~kpc
for our largest source, MRC~0156$-$252.

\begin{figure}
\centering
\includegraphics[width=0.5\textwidth]{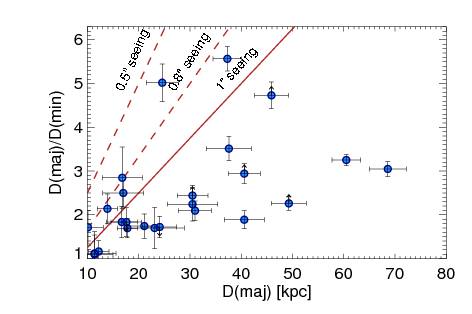}
\caption{The ratio between major and minor axis of the extended
  emission-line regions in our 27 galaxies with well resolved line
  emission (Table~\ref{tab:eelrproperties}), as a function of major
  axis length. Typical ratios are between 2 and 4. More extended
  regions are also more elongated.  
\label{fig:gassizevsgasratio}}
\end{figure}

In Fig.~\ref{fig:gassizevsgasratio} we show the ratios of the major
and minor axis in the 27 galaxies with well extended line emission
(Table~\ref{tab:eelrproperties}) as a function of the size of the
major axis. The red dashed and solid lines show the axis ratios than
can be measured as a function of the size of the major axis during
seeing of 0.5\arcsec, 0.8\arcsec, and 1.0\arcsec, respectively. We
find a total range between $2-4$, with most ratios between 2 and
3. This range is consistent with the ratios predicted by hydrodynamic
``cocoon'' models of radio jets, which predict values between 2 and 4
\citep[e.g.,][]{gaibler07, krause03, wagner12}.

Most galaxies have a single extended emission line region, which is
centered on the continuum peak. The peak in [OIII] surface brightness
is however not always found at the geometric center of the emission
line region. Examples of slightly off-center peaks include
BLR~0128$-$264, or MP~1758$-$6738, where we find offsets of 4.4~kpc
and 6.3~kpc, respectively, seen in projection on the sky. Likewise,
the emission-line and continuum peak do not always coincide. For
example, in USS~1410$-$001 or 3C~257, continuum and emission-line
surface brightness peaks are offset from each other by about 10~kpc.
Three galaxies (TN~J1112$-$2948, MG~2144$+$1928, and USS1243$+$036)
have multiple bright emission-line regions, which are not connected to
each other, at least at the surface brightness levels of about $1-3\times 
10^{-17}$ erg s$^{-1}$ cm$^{-2}$ arcsec$^{-2}$ that we reach with our
data. All three are roughly aligned with the radio jet axis
(\S\ref{sssec:positionangles}).

\subsection{Alignment with the radio axis} 
\label{sssec:positionangles}
\begin{figure}
\centering
\includegraphics[width=0.5\textwidth]{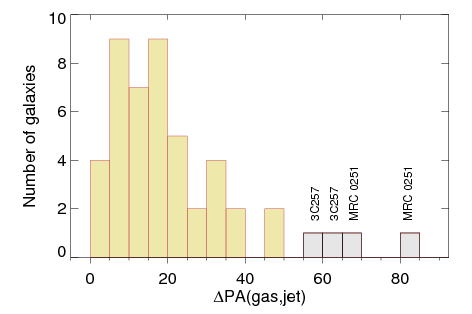}
\caption{Offsets in position angle between
  the major axis of the extended emission-line gas on each side of the
  nucleus and the associated radio jet for galaxies where jet and gas
  are well spatially resolved. See \S\ref{sssec:positionangles} for details.}
\label{fig:PAdistribution}
\end{figure}

Alignments between radio jet and extended continuum or emission-line
regions are very characteristic of high-redshift radio galaxies
\citep[e.g.,][]{mccarthy91, cimatti97}, whereas galaxies with
emission-line regions that are perpendicular to the radio jet axis are
commonly found in the more nearby Universe. \citet{baum92} found that
the radio galaxies they classify as 'rotators' have disks that are
orthogonal to the radio jet axis within about 20$^\circ$ \citep[see
  also ][]{deruiter02}. In Fig.~\ref{fig:PAdistribution} we show a
histogram with the offsets in position angle between the extended
emission line regions and the radio jets. Because the emission-line
morphologies in many galaxies are irregular (and in some cases also
the radio morphologies and jet directions), we estimated position
angles for each side of the nebulae from the nucleus, and also for
each individual radio jet. We include all 23 galaxies with well
extended emission-line regions and radio jets from the present work,
as well as the previously published sources MRC~0406$-$244,
TXS~0828$+$193, TN~J0205$+$2242 and MRC~0316$-$257 from
\citet{nesvadba08} and \citet{nesvadba07b}. We also include
MG~0340$-$65, for which \citet{mcconnell12} list a position angle
PA$=59^\circ$, and USS~1707$+$105, using the radio morphology shown by
\citet{hatch11} and the position of the outer
cloud as reference for the gas in the northern part of the galaxy.

Fig.~\ref{fig:PAdistribution} shows that jet and emission-line gas are
well aligned in the great majority of cases. Offsets are $<30^\circ$,
consistent with the measurement uncertainties (in particular in the
lowest-surface brightness areas which are often decisive to determine
the major axis of the line-emitting gas, see Fig.~\ref{fig:maps} to
\ref{fig:maps4}), jittering of the jet or perhaps jet precession
\citep[e.g.,][]{navaz14}, and the broad lateral sizes of many
emission-line regions in our galaxies perpendicular to the jet
direction. Each of these effects can produce offsets in position angle
of a few degrees. The axis ratios of typical jet cocoons in
hydrodynamic simulations of radio galaxies
\citep[e.g.,][]{gaibler09,wagner12} suggest maximal misalignments of
$20-30^\circ$ as we find in our data. More pronounced offsets can be
produced by jet deflection on dense clouds \citep[as is not uncommon
  in HzRGs][]{vanbreugel98}. In all these cases, however, we can
safely say that the radio source intercepts a significant part of the
emission-line region.

A few sources, however, have offsets that are $>30^\circ$, and deserve
a more detailed discussion. The largest offsets, $>50^\circ$, are
found for MRC~0251$-$273 and 3C~257, which have both small double
radio sources embedded within the ISM of their host galaxy
(\S\ref{sec:feedback}). Another eight jets show offsets
between 30$^\circ$ and 50$^\circ$, namely, the two jets in 4C~23.56,
the south-western jet in MRC~1558$-$003, and the northern jet in
USS~1410$-$001, whereas the second jet is better aligned with the
gas. The radio source is much larger than the size of the
emission-line region in all these cases, and the axis towards the
radio hotspots intercepts at least parts of the gas. In two of these
cases, the gas kinematics are perturbed along the jet direction: In
MRC~1558$-$003, faint extended radio emission extends from the radio
core towards the south-western jet through a narrow funnel-like
structure with high gas velocities ($+500$ km s$^{-1}$). In
USS~1410$-$001, the highest velocities in the northern emission-line
cloud are also found along the direction towards the radio
hotspot. The jets in both galaxies therefore seem to have been
deflected through jet-cloud interactions. In USS~1410$-$001, this
interaction must have occurred under a grazing angle.

We discuss two more outliers with very large radio sources and large
offsets in position angle, $\ge 60^\circ$, NVSS~J210626$-$314003 and
TXS~2353$-$003 in detail in \citet{collet14a}. Unlike MRC~0251$-$273
and 3C~257, which have equally large offsets, but radio jets that are
still embedded within the ISM of their host galaxy, these galaxies
have very extended radio sources $>$100 kpc. We also do not see any
gas associated with the direction of the radio jets in the two
\citeauthor{collet14a} sources, suggesting that the jet and gas are
not interacting in a direct way.

\subsection{Jet and emission-line region sizes and axis ratios}
\label{ssec:gassizesandaxisratios}

\begin{figure}
  \includegraphics[width=0.5\textwidth]{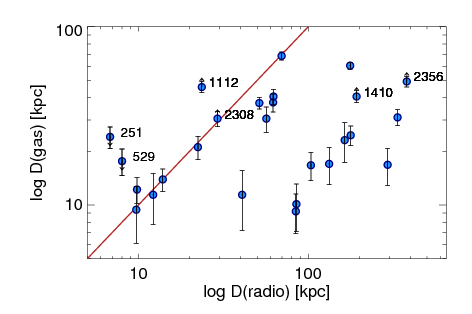}
\caption{Size of the bright emission-line regions as a function of the
  largest angular scale of the radio jet in the 27 sources with well
  extended emission-line regions. We label sources with upper and
  lower limits, respectively. In MRC~0251$-$273 and PKS~0529$-$549, the
  jet has not yet broken out of the ISM, see
  \S\ref{sec:feedback}}
\label{fig:radiogassize}
\end{figure}

In Fig.~\ref{fig:radiogassize} we compare the sizes of the
emission-line regions with those of the radio jets. The red line shows
the one-to-one relationship between the two. The two sizes are not
correlated, however, in most sources, the emission-line gas extends
over radii that are smaller than the radio jet size. Eight galaxies
have line emission that extends over similar radii as the radio
jets. Three galaxies have line emission that is significantly more
extended than the jet size. This includes two galaxies where we do not
spatially resolve the line emission in our SINFONI data, and which are
shown as upper limits in Fig.~\ref{fig:radiogassize}. Line emission in
these galaxies extends over sizes $<$10~kpc, typical sizes of high-z
galaxies without radio jets and similarly deep or even deeper SINFONI
observations \citep[e.g.,][]{forster09}. In galaxies with compact
radio sources, the onset of the jet activity could be too recent for
the jet cocoon to have encompassed the large-scale interstellar medium
of the host galaxy \citep[e.g.,][]{owsianik98,murgia99}.

In another 18 galaxies with radio sizes between 15~kpc and 200~kpc,
and gas sizes between 10~kpc and 70~kpc, the radio jets have already
broken out of their host galaxy, with radio sizes that exceed the size
of the emission-line regions by up to about an order of
magnitude. Although we find galaxies that have much larger radio
sources than emission-line regions for all gas sizes, we do not find
any galaxy with emission-line regions greater than 70~kpc. 

The SINFONI field-of-view of nominally
8\arcsec$\times$8\arcsec\ covers very similar ranges at z$\sim$2,
64~kpc$\times$64~kpc, but this is not the reason for this cut-off. Our
dither strategy (\S\ref{sec:observations}) required us to obtain
small mosaics, which cover about twice as large a size along the radio
jet axis, and are significantly larger than the sizes of the
emission-line regions in most cases.

In four galaxies, however, the line emission does extend to near the
edge of the data cubes. These galaxies are 4C~23.56, TN~J1112$-$2948,
USS~1410$-$001, and MG~2308$+$0336. These galaxies are formally shown
as lower limits in Fig.~\ref{fig:radiogassize}, but we do not expect
that their emission-line sizes exceed the SINFONI field-of-view. In
USS~1410$-$001, \citet{villar03} find from optical longslit
spectroscopy that bright Ly$\alpha$ emission extends out to about
10\arcsec\ along the radio jet axis, comparable to the size we measure
in [OIII]$\lambda$5007. \citet{tanaka11} obtained H$\alpha$
narrow-band imaging of 4C~23.56 down to a flux limit of $7.5\times
10^{-17}$ erg s$^{-1}$ cm$^{-2}$, finding a size of about 7\arcsec\ for 
the radio galaxy along the radio jet axis (Tanaka, private communication). 
 MG~2308$+$0336 has a radio size that is comparable to the
SINFONI field-of-view (right panel of Fig.~\ref{fig:maps5}). For
TN~J1112$-$2948, we obtained a second pointing around the northern
bubble, but did not find any bright nearby line emission.

\begin{figure}
\includegraphics[width=0.5\textwidth]{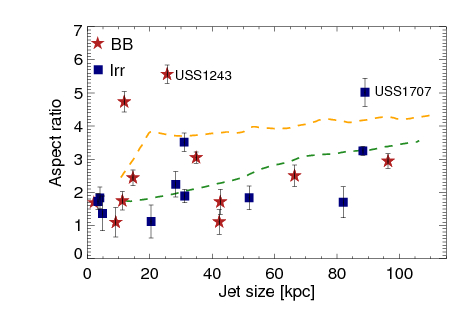}
\caption{Aspect ratio (size along the major/minor axis) of the
  emission-line region as a function of the jet size. Red stars and
  blue boxes show sources with bipolar velocity fields and more
  complex kinematics, respectively. The orange and green lines are for
  density parameters of 0.1 and 10$^{-3}$ between jet and ambient
  medium, respectively, and are taken from \citet{gaibler09}. We only
  show the 22 sources from Table~\ref{tab:eelrproperties} with well
  resolved emission-line regions, which also have jet sizes LAS/2
  $\le$ 120 kpc, the size range modeled by \citet{gaibler09}.}
\label{fig:radiovsgasratio}
\end{figure}

The elongation of the line emission is another interesting quantity,
that can also be compared with expectations from jet cocoon models. In
Fig.~\ref{fig:radiovsgasratio} we show the ratios of major to minor
axis of the emission-line regions as a function of the size of the
radio jets, finding a mild trend towards more elongated emission-line
regions with increasing radio jet size. Correlation probabilities
implied by the Spearman's rank test and Kendall's tau are 0.02 and
0.03, respectively, i.e., a random distribution is excluded at
confidence levels of 0.98 and 0.97, respectively. An increasing
elongation as observed here is expected in cocoon models of light
jets, because the bow shock should expand faster than the blastwave as
the cocoon grows larger \citep[][]{krause03}. These models also
predict a range of factors 2$-$4 between the major and minor axis of
the emission-line gas, in agreement with our data 
(\S\ref{ssec:oiiimorphologies} and Fig.~\ref{fig:gassizevsgasratio}).

In Fig.~\ref{fig:radiovsgasratio} we also make a more specific
comparison with the model predictions of \citet{gaibler09}. The yellow
and green lines show density parameters (i.e., the density ratio
between jet and ambient gas) of 0.1 and 10$^{-3}$, respectly. Values
for both densities are in good correspondence to the overall data set,
in particular when taking into account that the density parameter in
high-redshift galaxies is likely to be smaller \citep[e.g.,][suggest
  $10^{-4}$ for Cyg~A and high-z radio galaxies]{krause03}. We do not
find significant differences between galaxies with regular
back-to-back and irregular kinematics.

\subsection{Comparison of [OIII]$\lambda$5007 and H$\alpha$ morphologies}
\label{ssec:oiiihacomparison}

The [OIII]$\lambda\lambda$4959,5007 lines are observationally very
convenient to study the morphology and kinematics of the warm ionized
gas in HzRGs. They are the brightest emission lines in the rest-frame
optical spectra of most powerful HzRGs, and they do not suffer
blending with other nearby lines, except in the few galaxies with
broad nuclear emission lines. However, the broad component of H$\beta$
is spectrally well offset from the [OIII]$\lambda\lambda$4959,5007
lines, faint, and are a few times broader than [OIII], which allows
us to clearly distinguish it from the [OIII]$\lambda\lambda$4959,5007
doublet \citep[][]{nesvadba11a}. Having two lines probing the same gas
is also convenient to keep track of uncertainties due to telluric
features.

However, O$^{++}$ is a minor constituent of the gas, and the lines are
more sensitive to temperature, metallicity, and ionization parameter
than the mass of warm ionized gas \citep[e.g.,][]{ferland03}. We
should therefore verify that that the [OIII]$\lambda$5007 morphology
and kinematics is indeed an adequate representative of the overall
distribution and kinematics of the warm ionized gas.

\begin{figure*}
\centering
\includegraphics[width=0.32\textwidth]{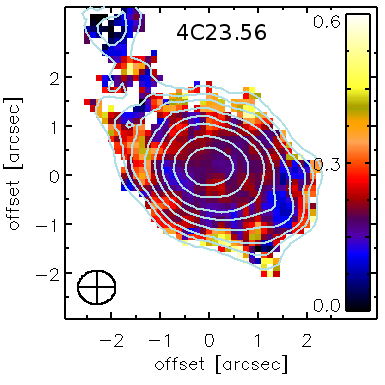}
\includegraphics[width=0.32\textwidth]{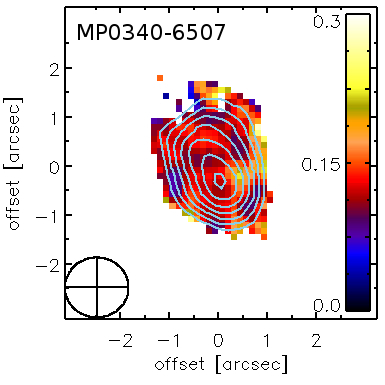}
\includegraphics[width=0.32\textwidth]{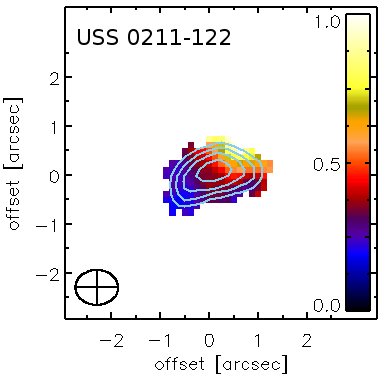}
\includegraphics[width=0.32\textwidth]{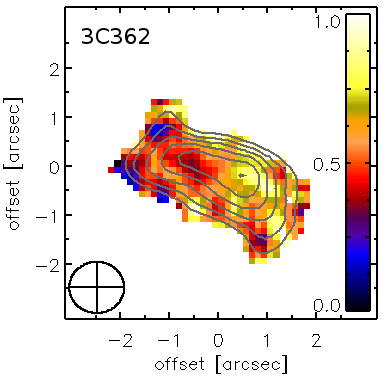}
\includegraphics[width=0.32\textwidth]{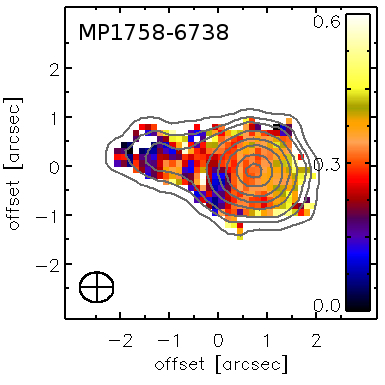}
\includegraphics[width=0.32\textwidth]{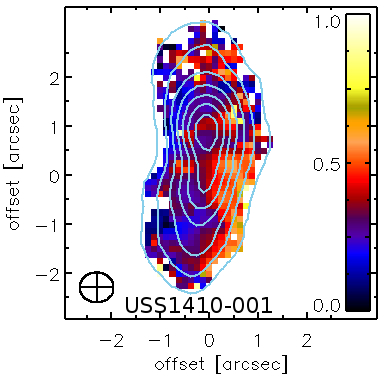}
\includegraphics[width=0.32\textwidth]{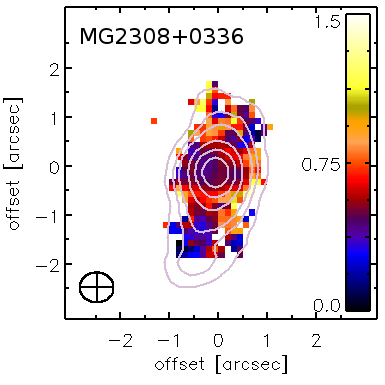}
\includegraphics[width=0.32\textwidth]{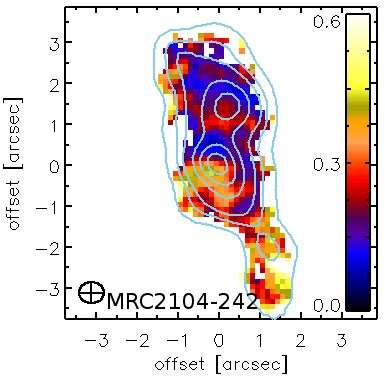}
\includegraphics[width=0.32\textwidth]{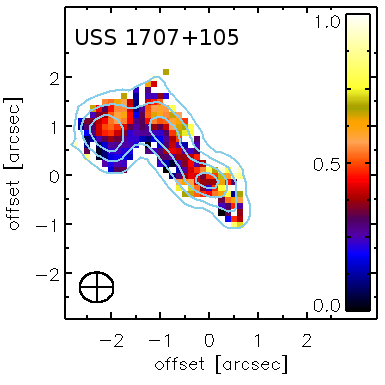}
\caption{
Ratios of H$\alpha$ to
  [OIII]$\lambda$5007 for nine galaxies with extended emission-line
  regions and at redshifts where we cover both lines with our SINFONI
  data cubes. Contours show the [OIII]$\lambda$5007 morphology. The
  ellipse in the lower left corner illustrates the size of the seeing
  disk. \label{fig:hao3comparison}}
\end{figure*}

\begin{figure}
\centering
\includegraphics[width=0.45\textwidth]{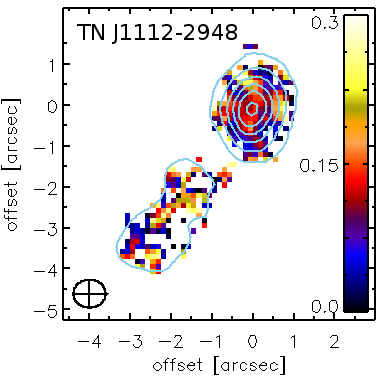}
\caption{
Ratios of H$\beta$ and [OIII]$\lambda$5007 fluxes in TN~J1112$-$2948 at
  z=3.09. \label{fig:hbo3comparisontnj1112}.}
\end{figure}

Several of our galaxies are bright and extended enough to allow for a
detailed comparison between the morphologies of [OIII]$\lambda$5007
and H$\alpha$. In Fig.~\ref{fig:hao3comparison} we show maps of the
H$\alpha$-to-[OIII]$\lambda$5007 ratios in the nine galaxies where
such a comparison is possible. In a 10$^{\rm th}$ galaxy,
TN~J1112$-$2948, H$\beta$ is bright enough for a similar analysis
(Fig.~\ref{fig:hbo3comparisontnj1112}). Line ratios change within
factors of typically 2-4 in individual galaxies, but the overall
morphologies of the emission-line regions traced in either line are
very similar in all galaxies. Both lines are emitted from similar
environments, although the detailed local gas conditions (or
extinction) may change.

To isolate H$\alpha$ and [NII]$\lambda\lambda$6548,6583, we had to
impose the same kinematic properties for H$\alpha$ and the
[NII]$\lambda\lambda$6548,6583 lines which we had previously measured
for [OIII]$\lambda$5007. The height of each Gaussian was however a free
parameter in our fits, and so these maps would have shown less
extended H$\alpha$ or H$\beta$ emission-line regions if such a difference to
[OIII] was present. Visual inspection of the data cubes ensured that
we are not missing any H$\alpha$ emission (blended with [NII]) that is
more extended than [OIII]. 

Performing a similar analysis for the velocity offsets and line widths
is more difficult because the blending of the H$\alpha$ and [NII]
lines makes it more challenging to probe the line wings and to account
for spectral shifts.  However, the small residuals in the integrated
spectra of H$\alpha$ and [NII] after our constrained line fits are
small (of-order a few percent), which gives us confidence that the
overall kinematic properties measured in both lines are generally
comparable. Unfortunately, the extended H$\beta$ emission is too faint
in these galaxies to produce resolved extinction maps \citep[but
  see][for extinction maps of two other HzRGs with comparable
  properties to our sources.]{nesvadba08}

\subsection{Emission-line surface brightnesses}
\label{ssec:totallineflux}
\begin{figure}
\centering
\includegraphics[width=0.48\textwidth]{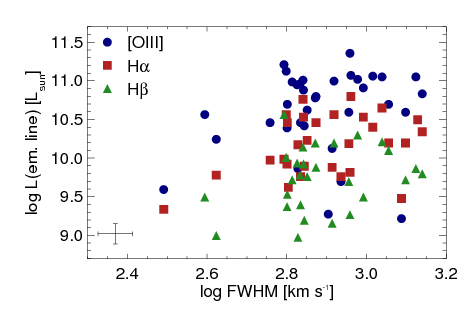}
\caption{Emission-line luminosity as a function of FWHM
  line width for [OIII]$\lambda$5007, H$\alpha$, and H$\beta$. All
  three lines show trends of increased luminosity with increasing line
  width. 
\label{fig:linelumis}}
\end{figure}

We also investigated whether trends exist between the emission-line
surface luminosity of [OIII]$\lambda$5007 and H$\alpha$ and various
other parameters. We prefer to investigate surface luminosities
instead of surface brightness to minimize the impact of cosmological 
surface-brightness dimming for our results, which scales as $(1+z)^4$ for
spectrally integrated line fluxes, corresponding to factors 33
to 1132 for our sources.

We used the line maps extracted from our SINFONI data cubes and
extracted the highest emission-line surface luminosities, the average
surface luminosity, and the most common surface luminosities, i.e.,
the peak of the histograms of the [OIII]$\lambda$5007 surface
luminosities extracted from each individual pixel. We also calculated
the sum of the line emission in all spatial pixels to derive the total
line flux. We did not find any obvious trends between surface
luminosity and the size or axis ratio of the emission line regions.

\begin{table*}
\centering
\begin{tabular}{lcccccccc}
\hline
Source & LAS      & LAS   & $\log{P_{\rm 500}}$  & $\log{P_{\rm 1.4}}$ & $\log{P_{\rm 151}}$ & $\log{E_{\rm mech,W}}$ & $\log{E_{\rm mech,C}}$ \\
       & [arcsec] & [kpc] & [W Hz$^{-1}$]   & [W Hz$^{-1}$]   &  [W Hz$^{-1}$] & [erg s$^{-1}$]     & [erg s$^{-1}$] \\
\hline
MRC~0114-211  & 0.7  & 5.9   & 28.8$\pm$0.1 & 28.6$\pm$0.1 & 29.1$\pm$0.1 &  47.0$\pm$0.3 & 47.4$\pm$0.3 \\
TN~J0121+1320 & 0.3  & 2.2   & 28.5$\pm$0.1 & 28.0$\pm$0.1 & 29.1$\pm$0.1 &  47.0$\pm$0.3 & 47.0$\pm$0.3 \\
BRL~0128-264  & 35.8 & 293.6 & 29.1$\pm$0.1 & 28.8$\pm$0.1 & 29.7$\pm$0.1 &  47.5$\pm$0.3 & 46.6 $\pm$0.3\\
MRC~0156-252  & 8.3  & 69.7  & 28.5$\pm$0.1 & 28.0$\pm$0.1 & 28.9$\pm$0.1 &  46.8$\pm$0.3 & 47.0$\pm$0.3 \\
TN~J0205+2242 & 2.7  & 19.7  & 28.5$\pm$0.1 & 28.0$\pm$0.1 & 29.1$\pm$0.1 &  47.0$\pm$0.3 & 47.0$\pm$0.3 \\
USS~0211-122  & 16.2 & 132.8 & 28.5$\pm$0.1 & 28.0$\pm$0.1 & 29.0$\pm$0.1 &  46.9$\pm$0.3 & 47.0$\pm$0.3 \\
MRC~0251-273  & 0.9  & 6.8   & 28.5$\pm$0.1 & 28.3$\pm$0.1 & 29.0$\pm$0.1 &  46.9$\pm$0.3 & 47.2$\pm$0.3 \\
RC~J0311+0507 & 2.8  & 18.5  & 29.5$\pm$0.1 & 29.1$\pm$0.1 & 30.0$\pm$0.1 &  47.7$\pm$0.3 & 47.8$\pm$0.3 \\
MRC~0316-257  & 7.6  & 57.8  & 29.0$\pm$0.1 & 28.5$\pm$0.1 & 29.5$\pm$0.1 &  47.3$\pm$0.3 & 47.4$\pm$0.3 \\
MP~J0340-6507 & 20.0 & 164.0 & 28.8$\pm$0.1 & 28.3$\pm$0.1 & 29.5$\pm$0.1 &  47.3$\pm$0.3 & 47.2$\pm$0.3 \\
MRC~0406-244  & 7.3  & 59.1  & 29.0$\pm$0.1 & 28.6$\pm$0.1 & 29.7$\pm$0.1 &  47.5$\pm$0.3 & 47.5$\pm$0.3 \\
PKS~0529-549  & 1.0  & 8.0   & 29.2$\pm$0.1 & 28.6$\pm$0.1 & 29.8$\pm$0.1 &  47.6$\pm$0.3 & 47.5$\pm$0.3 \\
TXS~0828+193  & 20.0 & 160.  & 28.4$\pm$0.1 & 27.9$\pm$0.1 & 29.0$\pm$0.1 &  46.9$\pm$0.3 & 46.9$\pm$0.3 \\
5C~7.269      & 0.5  & 4.2   & 27.8$\pm$0.1 & 27.4$\pm$0.1 & 28.3$\pm$0.1 &  46.3$\pm$0.3 & 46.6$\pm$0.3 \\
MRC~1017-220  & 8.3  & 70.6  & 28.1$\pm$0.1 & 27.9$\pm$0.1 & 28.2$\pm$0.1 &  46.2$\pm$0.3 & 46.9$\pm$0.3 \\
TN~J1112-2948 & 3.1  & 23.6  & 28.8$\pm$0.1 & 28.2$\pm$0.1 & 29.5$\pm$0.1 &  47.3$\pm$0.3 & 47.2$\pm$0.3 \\
TXS~1113-178  & 10.3 & 84.5  & 28.5$\pm$0.1 & 28.1$\pm$0.1 & 29.0$\pm$0.1 &  46.9$\pm$0.3 & 47.1$\pm$0.3 \\
3C~257        & 12.8 & 103.7 & 29.2$\pm$0.1 & 29.1$\pm$0.1 & 29.4$\pm$0.1 &  47.2$\pm$0.3 & 47.8$\pm$0.3 \\
MRC~1138-262  & 11.4 & 94.6  & 29.1$\pm$0.1 & 28.6$\pm$0.1 & 29.7$\pm$0.1 &  47.5$\pm$0.3 & 47.5$\pm$0.3 \\
USS~1243+036  & 7.0  & 51.1  & 29.2$\pm$0.1 & 28.7$\pm$0.1 & 29.9$\pm$0.1 &  47.7$\pm$0.3 & 47.5$\pm$0.3 \\
MG~1251+1104  & 1.2  & 9.8   & 28.4$\pm$0.1 & 28.0$\pm$0.1 & 28.9$\pm$0.1 &  46.8$\pm$0.3 & 47.0$\pm$0.3 \\
MRC~1324-262  & 1.7  & 13.9  & 28.5$\pm$0.1 & 28.1$\pm$0.1 & 28.9$\pm$0.1 &  46.8$\pm$0.3 & 47.1$\pm$0.3 \\
TN~J1338-1942 & 1.4  & 9.7   & 28.7$\pm$0.1 & 28.3$\pm$0.1 & 29.3$\pm$0.1 &  47.1$\pm$0.3 & 47.2$\pm$0.3 \\
USS~1410-001  & 23.5 & 192.7 & 28.4$\pm$0.1 & 28.0$\pm$0.1 & 29.0$\pm$0.1 &  46.9$\pm$0.3 & 47.0$\pm$0.3 \\
MRC~1558-003  & 7.7  & 62.4  & 28.8$\pm$0.1 & 28.4$\pm$0.1 & 29.5$\pm$0.1 &  47.3$\pm$0.3 & 47.3$\pm$0.3 \\
USS~1707+105  & 21.7 & 177.9 & 28.6$\pm$0.1 & 28.2$\pm$0.1 & 29.1$\pm$0.1 &  47.0$\pm$0.3 & 47.2$\pm$0.3 \\
3C~362        & 6.9  & 56.6  & 28.9$\pm$0.1 & 28.6$\pm$0.1 & 29.4$\pm$0.1 &  47.2$\pm$0.3 & 47.5$\pm$0.3 \\
MP~1758-6738  & 40   & 336.  & 29.0$\pm$0.1 & 28.0$\pm$0.1 & 28.5$\pm$0.1 &  47.3$\pm$0.3 & 47.0$\pm$0.3 \\ 
TN~J2007-1316 & 12   & 85.2  & 29.1$\pm$0.1 & 28.5$\pm$0.1 & 29.9$\pm$0.1 &  47.7$\pm$0.3 & 47.4$\pm$0.3 \\
MRC~2025-218  & 5.1  & 40.8  & 28.7$\pm$0.1 & 28.3$\pm$0.1 & 29.3$\pm$0.1 &  47.1$\pm$0.3 & 47.2$\pm$0.3 \\
MG~2037-0011  & 0.2  & 1.7   & 27.9$\pm$0.1 & 27.6$\pm$0.1 & 28.2$\pm$0.1 &  46.7$\pm$0.3 & 46.2$\pm$0.3 \\
MRC~2048-272  & 6.7  & 55.6  & 28.7$\pm$0.1 & 28.2$\pm$0.1 & 29.4$\pm$0.1 &  47.2$\pm$0.3 & 47.2$\pm$0.3 \\
MRC~2104-242  & 21.8 & 176.6 & 28.8$\pm$0.1 & 28.4$\pm$0.1 & 29.5$\pm$0.1 &  47.3$\pm$0.3 & 47.3$\pm$0.3 \\
4C~23.56      & 47.0 & 380.7 & 28.9$\pm$0.1 & 28.4$\pm$0.1 & 29.5$\pm$0.1 &  47.3$\pm$0.3 & 47.3$\pm$0.3 \\
NVSS~J2135$-$3337 & 0.5 & 4.1& 27.0$\pm$0.1 & 27.2$\pm$0.1 & 28.5$\pm$0.1 &  46.5$\pm$0.3 & 46.4$\pm$0.3 \\
MG~2144+1928  & 8.5  & 62.0  & 29.1$\pm$0.1 & 28.6$\pm$0.1 & 29.6$\pm$0.1 &  47.4$\pm$0.3 & 47.5$\pm$0.3 \\
MRC~2224-273  & 0.4  & 3.3   & 27.8$\pm$0.1 & 27.6$\pm$0.1 & 27.9$\pm$0.1 &  45.9$\pm$0.3 & 46.7$\pm$0.3 \\
TN~J2254+1857 & 2.7  & 22.4  & 27.8$\pm$0.1 & 27.2$\pm$0.1 & 28.5$\pm$0.1 &  46.5$\pm$0.3 & 46.4$\pm$0.3 \\
MG~2308+0336  & 3.6  & 29.2  & 28.5$\pm$0.1 & 28.3$\pm$0.1 & 28.9$\pm$0.1 &  46.8$\pm$0.3 & 47.2$\pm$0.3 \\
TXS~2353-003  & 38.8 & 310.4 & 28.8$\pm$0.1 & 28.3$\pm$0.1 & 29.3$\pm$0.1 &  47.1$\pm$0.3 & 47.2$\pm$0.3 \\
\hline
\end{tabular}
\caption{Radio size, radio power and kinetic jet energy estimates.
  Errors are the measurement uncertainties, except for the kinetic
  power where we use the scatter in the relationships we used to
  derive our estimates. We do not provide uncertainties for the
  Largest Angular Scale, because these are strongly dominated by the
  dynamic range, resolution, and frequency of the data set used, which
  in our case is rather heterogeneous. For all but the very compact
  sources, these uncertainties are in the range of few percent.
\label{tab:radiopower}}
\end{table*}

\section{Gas kinematics}
\label{sec:kinematics}

\subsection{Velocity patterns and total velocity offsets}
\label{sssec:velocities} 
The central panels of Fig.~\ref{fig:maps} to \ref{fig:maps5} show the
maps of relative velocities in the extended emission-line regions of
our sources. Total velocity offsets range from 100~km~s$^{-1}$ to
2000~km~s$^{-1}$. The lowest value is measured in TN~J2007$-$1316,
which is only marginally resolved and where the observed velocity
offset is probably significantly lowered by beam smearing effects. The
median velocity offset in all targets is 680~km~s$^{-1}$.

To obtain robust results in spite of occasional velocity spikes in low
signal-to-noise pixels towards the periphery of our sources, we did
not consider isolated pixels with very high velocity offsets near the
edge of our targets for our velocity estimates. Finding such values in
isolated pixels is inconsistent with the smoothing expected from
oversampling the seeing disk like we do in our data sets (a typical
seeing disk with FWHM$\sim0.8-1$\arcsec\ is sampled with pixels of
0.125\arcsec$\times$0.125\arcsec). Typical uncertainties of the
velocity offsets are between 30 and 50~km~s$^{-1}$. Velocity offsets
for individual sources are listed in Table~\ref{tab:eelrproperties}.

Inspection of Figs.~\ref{fig:maps} to \ref{fig:maps5} shows the
immense variety of velocity patterns in our galaxies. We broadly
distinguish between sources with regular velocity fields, where
monotonic, large-scale velocity gradients dominate, that are
approximately aligned along the major axis of the emission-line
region. Galaxies with irregular velocity fields show multiple velocity
minima and maxima. Good examples for regular velocity fields are,
e.g., MRC~0156$-$252 or 4C~23.56. Irregular fields are found,e.g., in
USS~1707$+$105, or MRC~2025$-$218.

Table~\ref{tab:eelrproperties} lists our
classification as either bipolar, i.e., 'back-to-back', monotonically
rising or falling gradients ('BB') or irregular velocity fields
('Irr'). We label compact sources with 'C', where we cannot exclude that the
compactness of the line emission is smearing out any putative velocity
offset. In total, we find 17 sources that are dominated by monotonic
large-scale gradients. Fifteen have more obvious irregular velocity
components (see Figs.~\ref{fig:maps} to \ref{fig:maps5}). In three
galaxies with compact line emission we do not see any velocity
offsets, and attribute this to the compactness of the source.

The right panels of Figs.~\ref{fig:maps} to \ref{fig:maps5} also show
the gas morphology and kinematics with the radio morphology overlaid
as contours. Most radio maps were obtained at observed frequencies
between 1.4~GHz and 8.5~GHz, with typical resolutions of 1\arcsec\ or
better. Observation parameters for individual sources can be found in
Table~\ref{tab:radioobs}. 

Several sources have extended radio morphologies associated with the
emission-line regions, which enable a more detailed comparison. Two
radio galaxies, MRC~0251$-$271 and 3C~257 have small double radio
sources which have either not yet broken out of the surrounding
interstellar gas, or which are seen under particularly small angles to
the line of sight. In MG~2308$+$0336, the radio source is just
adjacent to the emission line region, suggesting that the jet has
apparentely just broken out of the cocoon of hot gas. The velocity
offsets and line widths are greatest in the periphery of the
emission-line region and adjacent to the radio lobes. In
USS1243$+$036, both bubbles, although not directly connected to the
central region of the galaxy, are associated with radio emission. In
the southern bubble, the jet escapes from the gas cloud at an
intermediate radius, and is associated with a region where the gas
shows a velocity jump and increase in line width. In PKS~0529$-$549,
we see two knots of radio emission well within the extended emission
line region. Both radio knots are associated with regions of high
velocity offsets and broad line widths.

\subsection{Line widths} 
\label{sssec:linewidths}

\begin{figure}
\centering
\includegraphics[width=0.48\textwidth]{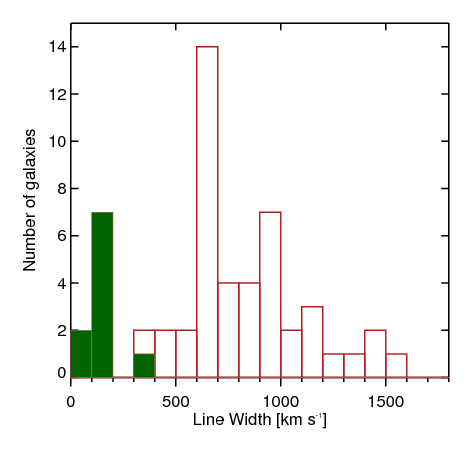}
\caption{Comparison of the FWHM line widths in our targets (empty red
  histogram) and the galaxies of \citet{buitrago13} at z$\sim$1.5,
  which have comparablly high stellar masses, but not prominent AGN
  (green filled histogram). 
\label{fig:fwhmmasssel}}
\end{figure}

FWHM (full-width-at-half-maximum) line widths are shown in the right
panel of Figs.~\ref{fig:maps} to \ref{fig:maps5}, and are between
450~km s$^{-1}$ and 3200 km~s$^{-1}$. This is significantly higher
than the line widths found by \citet{buitrago13} in a sample of
mass-selected galaxies at z$\sim$1.5 with stellar masses of few
$10^{11}$ M$_{\odot}$, comparable to the stellar masses in our sources
\citep[][]{seymour07,debreuck10}, suggesting that another mechanism
than gravity maintains the FWHM in our sources as high as observed. In
Fig.~\ref{fig:fwhmmasssel} we show a comparison of the line widths in
the \citet{buitrago13} sample and for our galaxies.

All sources have irregular FWHM maps. Some of the galaxies with
well extended emission-line gas, e.g., MRC~2025$-$218, show large
jumps in FWHM within about the size of a seeing disk of up to $\sim
800$~km s$^{-1}$. FWHMs in galaxies with two rather symmetric bubbles
can have very different FWHMs on either side. Examples are
USS~1410$-$001, where gas in the northern lobe reaches
FWHM~$\sim$900~km s$^{-1}$, compared to only 450~km s$^{-1}$ in the southern
lobe. In 3C~257, we find a FWHM gradient that runs nearly
perpendicular to the major axis of the gas, with FWHMs increasing from
500~km~s$^{-1}$ to 1300~km~s$^{-1}$ from the north-east to the
south-west. Not always are regions of broad FWHMs and high velocity
offsets associated with each other. For example in TN~J1338$-$192, the
gas with the lowest line widths (FWHM$\sim 500$~km s$^{-1}$) is also
at the highest redshift ($+500$ km s$^{-1}$ relative to the average).
Some galaxies, e.g., MRC~$2025-218$, have multiple regions of very
broad line emission, however, 4C~23.56 has uniformly broad lines
(FWHM$\sim$800 km s$^{-1}$) that only become more narrow in the very
periphery of the emission-line region. 

The maximal FWHM is generally
not associated with the central regions of the host galaxies,
as probed by the position of the continuum peak. This is only the case
for BLR~0128$-$264, TXS~1113$-$178, MG~1251$+$1104, MRC~2025$-$218,
NVSS~J2135$-$3337, and MRC~2224$-$273. In, e.g., MG~2308$+$0336, the
gas near the nucleus is in fact the gas with the lowest FWHM: FWHMs
are $\sim 1200$ km s$^{-1}$ in the southern periphery adjacent to the
radio lobe, and only FWHM$\sim$500 km s$^{-1}$ near the center.

Even galaxies with regular, clearly bipolar velocity fields have
irregular FWHM distributions. The broadest FWHMs are not generally
associated with the largest velocity gradients in these galaxies,
which rules out beam smearing as primary cause. Moreover, the FWHMs
are generally comparable to, if not larger than, the large-scale velocity
gradients in our sources, further limiting the potential effects of
beam-smearing (a notable exception is however BLR~0128$-$264). This of
course applies only to the large-scale velocity gradients that we
spatially resolve. Local FWHM maxima could trace gas flows on scales
unresolved in our data, which can encompass regions of several kpc.

Given this great complexity, it is obviously not straight-forward to
define one characteristic FWHM for each galaxy, which can then be
compared with other sources. We have therefore considered several
quantities:

Firstly, probably the easiest way to obtain a luminosity-weighted global
estimate is by integrating the data cube over all pixels where line
emission is detected after removing the local offsets measured from
the velocity maps. Since most of our FWHM maps are derived from
[OIII]$\lambda$5007, which has a line flux that is more strongly
determined by the ionization parameter than gas mass
\citep[e.g.,][]{ferland03}, this is in essence an ionization-weighted
FWHM. FWHMs obtained in this way are listed for each galaxy and line in
Table~\ref{tab:emlines}.
Secondly, we also list the median FWHM in each source in
Table~\ref{tab:eelrproperties} to give an estimate of the 'typical'
FWHM in each galaxy that is measured in most pixels. This may serve
as an approximation to the projected, geometrically weighted
FWHM. Thirdly, in the same table we also list the maximal FWHMs found
in smaller regions of each galaxy, but still over apertures of several
pixels to avoid bias by noisy pixels.

\begin{figure}
\includegraphics[width=0.48\textwidth]{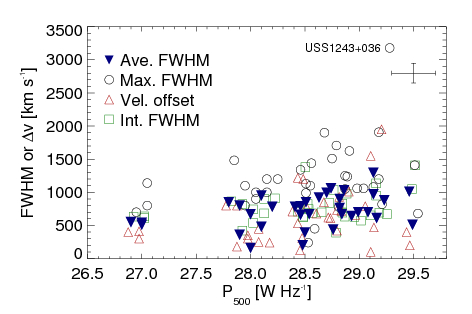}
\caption{FWHM and velocity offsets as a function of the jet power at
  500~MHz in the 25 galaxies with extended emission-line regions from
  our sample. We also add 9 targets from \citet{collet14b} to enhance
  the number of galaxies with radio power below $10^{28.5}$ erg
  s$^{-1}$.}\label{fig:fwhmemech}
\end{figure}

The FWHM line widths are not correlated with the emission-line
luminosities. In Fig.~\ref{fig:linelumis} we show this on the example
of the luminosity-weighted integrated widths (the first method). The
figure includes all 35, 29, and 28 galaxies with [OIII]$\lambda$5007,
H$\alpha$, and H$\beta$ where these lines have been measured in our
sample.
Spearman rank test and Kendall's $\tau$ rule
out non-significant correlations only at probabilities $0.27, 0.14$,
and $0.27$ for [OIII]$\lambda$5007, H$\alpha$, and H$\beta$,
respectively.

In Fig.~\ref{fig:fwhmemech} we show the velocity offsets, average,
maximum, and integrated line widths as a function of the jet power at
500~MHz in the rest frame (the kinematic properties and the jet power
of individual galaxies are listed in Tab.~\ref{tab:eelrproperties} and
\ref{tab:radiopower}, respectively). All quantities increase with
increasing radio power, although the scatter is large. Spearman rank
tests and Kendall's $\tau$ exclude non-significant correlations at
confidence levels of 0.1 and better. The tightest relationship is
between radio power and average FWHM, with confidence levels of 0.02
and 0.015 for Kendall's $\tau$ and the Spearman rank test,
respectively.

We also investigated whether the properties of the gas and radio
emission in the two main classes 'BB' and `Irr' depended on whether a
given galaxy falls into one class or another. We did not find
significant trends with radio size, radio power, or jet kinetic
energy. A Kolmogorov-Smirnov test suggests probabilities of between
0.2 and 0.9 that the two classes are drawn from the same overall
distribution of these parameters amongst our full sample.  Likewise,
we did not find significant differences in the sizes of emission-line
regions and kinematic parameters (velocity offsets, average and
maximal Gaussian line widths, i.e., FWHM/2.355). Probabilities that
both subsets are distinguishable in other parameters, including FWHM
and velocity offsets alone, are between 0.1 (for velocity offsets) and
0.9 (for radio power), which is not significant.  The only quantity
where we do see a significant difference between the two classes is
the ratio between bulk and unordered velocities, which we will discuss
in the next subsection.

\subsection{Ratio of velocity gradients and line widths}
\label{sssec:voversigma}

\begin{figure}
\centering
\includegraphics[width=0.45\textwidth]{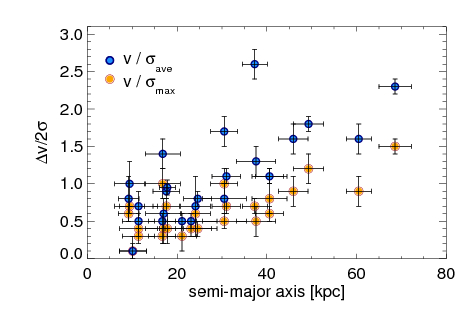}\\
\includegraphics[width=0.45\textwidth]{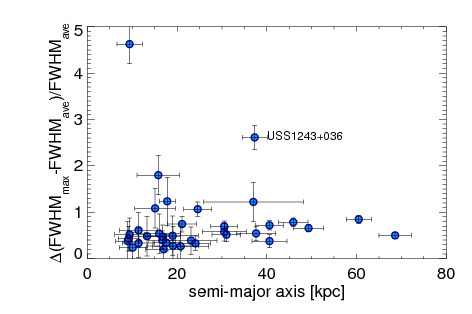}
\caption{The ratio of bulk to turbulent velocity
  (parametrized as $\Delta v/ 2\sigma$, see
  \S\ref{ssec:gaskinematics}) in the 25 sources with extended gas in
  Table~\ref{tab:eelrproperties} scales with the size of the
  emission-line region (top panel). We plot $\Delta v/2$ instead of
  $\Delta v$, because this gives an approximation to the circular
  velocity in case of rotationally dominated motion. The offset
  between maximal and average FWHM does however not depend on the
  source size (bottom panel). \label{fig:vsigma}}
\end{figure}

The ratios between velocity gradient and line widths provide
interesting constraints on the importance of ordered bulk motion and
unordered, possibly turbulent motion of the gas. In galaxies
considered to be dominated by large-scale rotation, such estimates are
usually derived by using the large-scale velocity gradient and the
central velocity dispersion \citep[][]{forster09}. However, given the
more complex velocity fields of our sources, the situation is less
straight-forward in this case. We therefore list two values in
Table~\ref{tab:eelrproperties}. Firstly, we state the v/2$\sigma_{\rm
  ave}$, using the Gaussian line width derived from the average FWHM
(i.e., $\sigma_{\rm ave}=$FWHM$_{\rm ave}/$2.355) in the maps, and
secondly, the $\Delta$v/2$\sigma_{\rm max}$ using the Gaussian width
corresponding to the broadest FWHM in each source. $\Delta$v is the
total velocity gradient, and we divide by an additional factor 2
  to obtain values that are comparable to those found for other
  samples of high-z galaxies, where v approximates the circular
  velocity. We consider these two quantities representations of the
typical and maximal kinematic perturbations of the gas,
respectively. Generally, we find that the maximal FWHM is a factor
$2-3$ higher than the average FWHM, with little dependence on the
average or maximal FWHM (upper panel of Fig.~\ref{fig:vsigma}). An
exception is USS~1243$+$036, which has unusually broad line widths of
FWHM$=$2500~km s$^{-1}$ in the central regions
(\S\ref{ssec:induss1243}).

\begin{figure}
\centering
\includegraphics[width=0.45\textwidth]{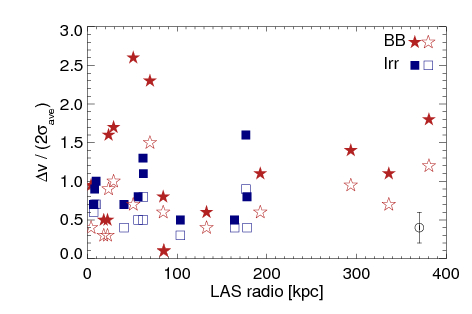}
\caption{
Ratio of ordered to unordered motion,
  $\Delta v$/$\sigma$, as a function of radio size. We show all 25 sources from
  Fig.~\ref{tab:eelrproperties} with measurements of
  $\Delta v$/$\sigma$. Filled and empty symbols show the maximal and average
  line widths, respectively. \label{fig:lasvssigv}}
\end{figure}

We find however higher ratios of $\Delta v/ 2\sigma$ in the sources
with the most extended emission-line gas (lower panel of
Fig.~\ref{fig:vsigma}). The reason for this difference is
two-fold. Firstly, the largest nebulosities have also particularly
large velocity offsets: all galaxies with v/$\sigma$ ratios $>$1.1
have regular velocity fields, and are those which also have $\Delta
v\ge $1000 km s$^{-1}$.  Secondly, they have lower line widths than
the more compact sources. The upper panel of Fig.~\ref{fig:vsigma}
shows that this holds for the average as well as the maximal
FWHM. Spearman's rank test and Kendall's $\tau$, respectively, rule
out with probabilities of $1.1\times 10^{-4}$ and $2.6\times 10^{-4}$,
respectively, that the average v/$\sigma$ are uncorrelated with the
major axis. For the maximum widths, v/$\sigma_{\rm max}$, this can be
ruled out with probabilities of $2.3\times 10^{-3}$ and $1.1\times
10^{-3}$. We did not find a correlation between v/$\sigma$ and radio
size (Fig.~\ref{fig:lasvssigv}).

\section{Feedback}
\label{sec:feedback}
There is little doubt today that the bright, extended emission-line
regions in HzRGs are photoionized by powerful obscured quasars in their
nuclei \citep[e.g.,][]{villar97,humphrey08}. Based on a small number
of HzRGs with imaging spectroscopy (four with extended, two with
compact radio sources), we have previously argued that the energy
injection from the AGN into the gas of the host galaxy is also at the
origin of the irregular gas kinematics with high velocities and line
widths in HzRGs \citep[][]{nesvadba06a,nesvadba08}.
\begin{table*}
\centering
\begin{tabular}{lcccccccc}
\hline 
Source        & [OIII] size & FWHM$_{\rm avg}$ & FWHM$_{\rm max}$    & $\Delta v$   & $v/\sigma_{\rm avg}$ &$v/\sigma_{\rm max}$ & Kin. class \\
       & [kpc$\times$kpc] &  [km s$^{-1}$]& [km s$^{-1}$]    & [km s$^{-1}$]  &          & & \\
\hline
BRL~0128$-$264  & $16.8\pm3.9\times 5.9\pm2.7$   &  1300$\pm$124  &  1910$\pm$182     & 1550$\pm$148   & 1.4$\pm$0.2 & 1.0$\pm$0.2 & BB \\
MRC~0156$-$252  & $68.6\pm3.6\times 22.5\pm3.6$  &  660$\pm$21   &  990$\pm$31      & 1220$\pm$38   & 2.3$\pm$0.1 & 1.5$\pm$0.1 & BB \\
USS~0211$-$122\footnote{For the narrow component}   & $17.0\pm4.0\times6.8\pm1.5$  & 200$\pm$32   &  240$\pm$39    &  124$\pm$20 & 0.6$\pm$0.2 & 0.4$\pm$0.2 & BB \\
MRC~0251$-$273   & $24.1\pm3.3\times14.0\pm2.9$  & 670$\pm$54   & 450$\pm$36   & 780 $\pm$64   & 0.7$\pm$0.4 & 0.6$\pm$0.2 & Irr \\
RC~J0311$+$0507 & $11.4\pm3.6\times10.4\pm3.2$\footnote{[OII]}& 510$\pm$20   & 680$\pm$27 & 200$\pm$8 & 0.5$\pm$0.1 & 0.3$\pm$0.1 & BB \\
MP~J0340$-$6507 & $23.1\pm5.8\times13.6\pm5.4$  & 900$\pm$137   & 1250$\pm$190   & 420$\pm$64   & 0.5$\pm$0.1 & 0.4$\pm$0.1 & Irr \\
PKS~0529$-$549  & $17.6\pm3.0\times 9.6\pm2.7$    & 610$\pm$36   & 820$\pm$48    & 470$\pm$28   & 0.9$\pm$0.1 & 0.7$\pm$0.1 & Irr \\
sTN~J1112$-$2948 & $45.9\pm3.3\times 9.7\pm2.9$    & 437$\pm$23   & 780$\pm$35    & 610$\pm$27   & 1.6$\pm$0.2 & 0.9$\pm$0.2 & BB \\
TXS~1113$-$178  & $9.2\pm2.3\times 8.3\pm2.3$     & 790$\pm$70   & 1130$\pm$100   & 530$\pm$47   & 0.8$\pm$0.3 & 0.6$\pm$0.1 & BB \\
3C~257          & $16.7\pm3.0\times 9.1\pm2.8$    & 1010$\pm$88  & 1410$\pm$123   & 400$\pm$35   & 0.5$\pm$0.1 & 0.3$\pm$0.1 & Irr \\
USS~1243$+$036  & $37.3\pm2.8\times 6.7\pm1.8$    & 880$\pm$22   & 3180$\pm$57   & 1950$\pm$35  & 2.6$\pm$0.2 & 0.7$\pm$0.1 & BB\\
TN~J1338$-$1942 & 9.4$\pm$3.3$\times$6.9$\pm$2.7    & 994$\pm$47   & 1510$\pm$71  & 850$\pm$40   & 1.0$\pm$0.3 & 0.7$\pm$0.1 & Irr \\
USS~1410$-$001  & 40.6$\pm$3.1$\times$13.8$\pm$2.9  & 780$\pm$52   & 1340$\pm$89  & 710$\pm$47   & 1.1$\pm$0.1 & 0.6$\pm$0.1 & BB \\
MRC~1558$-$003  & 40.6$\pm$3.9$\times$21.5$\pm$3.9  & 763$\pm$82   & 1050$\pm$112  & 730$\pm$78   & 1.1$\pm$0.1 & 0.8$\pm$0.1 & Irr\\
USS~1707$+$105  & 24.6$\pm$3.1$\times$4.9$\pm$2.0   & 920$\pm$87   & 1900$\pm$179  & 680$\pm$64   & 0.8$\pm$0.1 & 0.4$\pm$0.2 & Irr \\
3C~362          & 30.5$\pm$4.9$\times$13.6$\pm$4.7  & 1030$\pm$122  & 1625$\pm$192  & 720$\pm$85   & 0.8$\pm$0.1 & 0.5$\pm$0.2 & Irr \\
MP~1758$-$6738  & 31.0$\pm$3.2$\times$14.8$\pm$2.9  & 700$\pm$75   & 1060$\pm$114  & 660$\pm$71  & 1.1$\pm$0.2 & 0.7$\pm$0.1 & BB \\
TN~J2007$-$1316 & 10.1$\pm$3.0$\times$5.9$\pm$1.4   & 970$\pm$140   & 1200$\pm$173  & 100$\pm$73\footnote{Small angular size} & 0.1$\pm$0.1 & 0.1$\pm$0.1 & BB \\
MRC~2025$-$218  & 27.0$\pm$4.2$\times$8.9$\pm$3.4   & 1060$\pm$97  & 1707$\pm$157  & 620$\pm$57   & 0.7$\pm$0.2 & 0.4$\pm$0.1 & Irr &\\
MRC~2104$-$242  & 60.5$\pm$2.8$\times$18.6$\pm$2.3   & 670$\pm$27   & 1237$\pm$52  & 930$\pm$39   & 1.6$\pm$0.2 & 0.9$\pm$0.2 & Irr \\
4C~23.56        & 49.3$\pm$3.4$\times$21.8$\pm$3.1   & 640$\pm$21   & 1060$\pm$36  & 1010$\pm$40  & 1.8$\pm$0.2 & 1.2$\pm$0.1 & BB \\
NVSS~J2135$-$3337& 17.8$\pm$1.8$\times$10.5$\pm$1.8  & 510$\pm$25   & 1140$\pm$43 & 410$\pm$59    & 1.0$\pm$0.2 & 0.4$\pm$0.2 & BB \\
MG~2144$+$1928  & 37.6$\pm$4.4$\times$10.7$\pm$2.7   & 700$\pm$73   & 1930$\pm$164  & 790$\pm$28   & 1.3$\pm$0.1 & 0.5$\pm$0.2 & Irr \\
TN~J2254$+$1857 &  21.1$\pm$3.2$\times$12.1$\pm$2.8  & 850$\pm$63   & 1483$\pm$109  & 800$\pm$59   & 0.5$\pm$0.1 & 0.3$\pm$0.2 & BB\\
MG~2308$+$0336  & 30.5$\pm$3.0$\times$12.5$\pm$2.7   & 850$\pm$53   & 1440$\pm$90  & 1200$\pm$75  & 1.7$\pm$0.2 & 1.0$\pm$0.2 & BB \\
\hline
\end{tabular}
\caption{Properties of extended emission-line regions. The [OIII] size
  is the full projected length of the emission-line region along the
  major and minor axis, respectively, and deconvolved with the size of
  the seeing disk. We only list galaxies of type ``BB'' (bipolar
  velocity fields akin to back-to-back outflows) or ``Irr'' (irregular
  kinematics).}
\label{tab:eelrproperties}
\end{table*}

The results of our present study enable us to expand our previous
analyses in several ways, and to investigate whether the arguments we
put forward in these studies also hold for HzRGs generally. Our
SINFONI maps show a much larger diversity in gas morphologies and
kinematics than in the previously analyzed sources. The three galaxies of
\citet{nesvadba08} showed large bubbles with regular velocity fields,
whereas only about half (17/32) in our present sample with resolved
kinematics are dominated by such monotonic large-scale velocity
fields, and have ratios of bulk to turbulent velocity that are higher 
than in the more compact galaxies. 

Fifteen galaxies have irregular gas kinematics, which are not very
reminiscent of back-to-back outflows. Finding so many sources with
large velocity jumps over small areas near our resolution limit of few
kpc requires the presence of an energy injection mechanism that is
powering the gas kinematics locally and at kpc distance from the AGN
itself.

We see little evidence of multiple stellar components associated with
regions that are kinematically distinct from each other, as we might
expect if these jumps were caused by pre-coalescent major
mergers. Sudden localized jumps in velocity and line width are however
also fully consistent with the jet scenario. Off-nucleus gas
acceleration associated with radio jets has previously been observed
in a number of low-redshift radio galaxies, e.g., IC~5063
\citep[][]{morganti05}, 3C~293 \citep[][]{emonts05}, and 3C~326~N
\citep[][]{nesvadba11c}, suggesting that the global complexity we see
in our sources could be a general signature of interactions between
radio jet and the global interstellar gas of their host galaxy, where
the individual properties of each source might lead to a somewhat
different phenomenology in each individual galaxy.

Both types of morphologies have also been found in hydrodynamic models
of radio jet cocoons expanding through ambient gas.
\citet[][]{sutherland07}, \citet{wagner11}, and \citet{wagner12}
modeled the jet expansion through turbulent, inhomogeneous media,
finding that the jet experiences a first phase of very efficient
momentum and energy deposition as it expands along low-density
channels through the ambient gas disk. Only after breaking out of this
disk and into the more uniform, lower-density intergalactic or
intracluster medium at the end of this flood and channel phase do
well collimated radio jets form. Irregular gas morphologies with
multiple small bubbles, and sudden jumps in velocity or line widths
can be expected during this phase, and would at least qualitatively
correspond to the irregular morphologies that we see in many
sources. This includes not only irregular velocity fields, but also
irregular distributions of line widths, because it is not a priori
clear if these jumps are due to enhanced random motion or turbulence
in small regions of our galaxies, or bulk (out-)flows. Prime examples
for this phase might in particular be 3C~257 and MRC~0251$-$273, where
the jets are not only embedded within the bright emission-line region,
but also misaligned with the major axis of the gas.

After breaking out of the disk (or when disrupting it), the cocoon
inflated by the radio jet may entrain and accelerate clouds of ambient
gas, which are lifted off the disk, fragment under the influence of
Kelvin-Helmholtz instabilities due to the velocity offset between the
cloud and hot wind medium, and form extended filaments of warm ionized
gas over several tens of kpc from the central galaxies
\citep[][]{cooper08, scannapieco15}. An example where sudden jumps in
velocity and line width are directly associated with radio hot spots,
suggesting that gas is being dragged out of the galaxy as the radio
jet is breaking out of the ISM, is MG~2308$+$0336
(\S\ref{ssec:indmg2308}). These simulations also suggest that the
global appearance of this gas preserves the signatures of the initial
gas distribution, i.e., small clumps and filaments of material from a
massive gas disk may be distributed over the entire volume of the hot
wind bubble as suggested by several of our sources (e.g.,
MRC~0156$-$252 or USS~1243$+$036). Alternatively, it may accumulate
along the edges of the bubble, e.g., if backflow from the working
surface of the jet determines the gas kinematics.  This would broadly
correspond to the hourglass-shaped morphology of sources like
MRC~0156$-$252 \citep[][]{gaibler09}, or may produce a radio
morphology that is very asymmetric \citep[][]{gaibler11}.

In the following subsections we will more closely examine the
kinematic properties of the gas and the radio jets, to infer whether
we can find more quantitative evidence for this scenario from the
global properties of the gas, and regardless of the detailed
properties of each individual galaxy. We start by demonstrating that
disk rotation does not match the kinematic properties of the gas in
our sources overall. A complete analysis would also require to compare
with the star formation and bolometic AGN properties of our
sources. For about 20 sources, which are also part of the Herg\'e
sample of HzRGs with Herschel/SPIRE and PACS far-infrared photometry,
we are able to do such a comparison, finding that the radio jet is the
most important source of the energy and momentum in the gas. We
discuss this further in a companion paper (Nesvadba et al. 2016, in
prep.).

\subsection{Kinematic signatures of rotation?}
\label{ssec:rotation}

The ratios of bulk to random motion, $v/\sigma$, provide
information about whether the gas kinematics in our sources might be
dominated by rotation. The light profiles of many HzRGs can be fitted
with de Vaucouleur's profiles \citep[][]{vanbreugel98,pentericci01,
  targett11}, so that the comparison with early-type galaxies seems to
be warranted. \citet{martig09} argued that the gas kinematics in
gas-rich early-type galaxies should reflect the stellar kinematics of
the host galaxy, which is why we compare the gas kinematics in our
galaxies with that of the stars in nearby 'fast rotating' early-type
galaxies from the ATLAS$^{3D}$ survey \citep[][]{emsellem11}.

\begin{figure}
\includegraphics[width=0.5\textwidth]{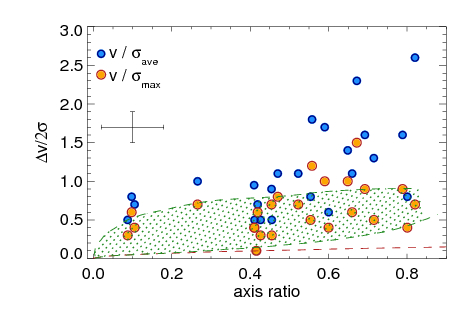}
\caption{Ratio of bulk velocity to dispersion as a function of the
  ellipticity, i.e., $1-D_{\rm min}/D_{\rm max}$, where $D_{\rm min}$ and
  $D_{\rm max}$ are the size of the emission line gas along the minor and
  major axis, respectively. We show all 25 galaxies with emission line
  regions that are resolved along the major and minor axis, and have
  well defined ratios of v/$\sigma$. The green dotted area shows the region
  occupied by the fast rotators in the ATLAS$^{3D}$ survey
  \citep{emsellem11}. The red line delimits the range in parameter
  space for slow rotators.  See \citet{emsellem11} for
  details of the physical properties of these two regions.%
}\label{fig:emsellem}
\end{figure}

In Fig.~\ref{fig:emsellem} we show the ratio of bulk velocity to
velocity dispersion as a function of the ellipticity $\epsilon$ of the
emission-line region, i.e., $\epsilon = 1-D_{\rm min}/D_{\rm max}$, where
$D_{\rm min}$ and $D_{\rm max}$ are the size of the emission line gas along
the minor and major axis, respectively. The figure shows all 25
galaxies with emission-line regions that are spatially resolved along
the major and minor axis. The dotted region shows the upper and lower
envelope of fast rotators by \citet{emsellem11} in the ATLAS$^{3D}$
survey, the red line the upper limit of the region occupied by slow
rotators.  Galaxies falling into this dotted region are within a
characteristic range of inclination, intrinsic ellipticity, velocity
dispersion and rotational velocity which are typical for fast
rotators.  Most of our galaxies fall above the region of this diagram
spanned by the low-redshift galaxies.  The bulk velocities of our
sources are therefore higher than what would be expected from
gravitational motion within an early-type galaxy, in spite of the
unusually broad line widths.  This holds generally for a given
combination of line width, ellipticity, and inclination, although we
cannot reliably determine each of these parameters individually from our
data. Finding ratios of bulk velocities to line widths that are
greater than expected for gravitational motion is also at odds with a
scenario of disk rotation with additional line broadening caused by
local interactions between jet and gas. We note that this
  conclusion depends on two assumptions: Firstly, that our sources
  have similar structural properties to those of \citet{vanbreugel98,
    pentericci01}, and \citet{targett11}, with which we have 15
  sources in common. Secondly: That the structural properties of our
  HzRGs are not different by more than a factor 2-4 from those of
  low-redshift galaxies. We consider this a reasonable assumption,
  because, e.g., observations of the fundamental plane out to z=1-2
  seem to imply much smaller offsets, once passive luminosity
  evolution has been taken into account \citep[e.g.,][]{vandesande14,
    zahid15}

We do obtain a better match when using the maximal line widths
instead.  These widths, however, are only found in small areas of the
emission-line gas, and are significantly higher than those expected
for galaxies with masses of few $10^{11}$ M$_{\odot}$ of
$\sigma=200-300$ km s$^{-1}$ and observed in equally massive galaxies
at z$\sim$1.5 \citep[][]{buitrago13}.

\subsection{Time scales}
\label{ssec:timescales}
\citet{alexander87} and \citet{best95} estimated typical jet advance
speeds of powerful FR~II radio galaxies from the 3~CR of $v_{\rm jet}
= 0.01-0.1 c$, where $c$ is the speed of light. Assuming that the jets
in our sources have expanded at uniform speed throughout their
lifetime, we can use the radio sizes to infer typical age ranges for
our sources. This suggests typical ages of few $10^{6-7}$ yrs and
$10^{5-6}$ yrs for $v_{\rm jet}=0.01\ c$ and $v_{\rm jet}=0.1\ c$,
respectively. Given that the breakout of the radio jet from extended
reservoirs of relatively dense gas is likely to last already about
$10^6$ yrs \citep[][]{wagner12}, it is likely that in particular the
lowest ages are underestimates. In the following we will adopt a
typical age range of $10^{6-7}$ yrs. If the jet is the origin of the
gas kinematics, then jet expansion speeds much lower than $0.01\ c$
(and thus ages much longer than few $10^{6-7} yrs$) are ruled out by
the high gas velocities of few 1000 km s$^{-1}$ we find in some of our
sources.

\subsection{Kinetic jet power}
\label{ssec:kinjetenergy}
Only a small fraction of the kinetic energy of the relativistic
particle content of radio jets is emitted in form of synchrotron
radiation in the centimeter radio regime \citep[e.g.,][and references
  therein]{heckman14}. The ratio of integrated radio luminosity to jet
kinetic power is about 0.1\%, however the detailed outcome for each
source depends on the local magnetic field and the density of the
surrounding medium.  This makes it relatively challenging to estimate
the kinetic power of a radio jet from the observed radio
luminosity. Several empirical methods have been proposed in the
literature to measure this relationship between observed jet power in
the centimeter regime and the intrinsic kinetic power of the
synchrotron electrons. For the purpose of this work, we use the
estimates of \citet{cavagnolo10} and \citet{willott99}.

\citet{cavagnolo10} use the mechanical energy required to inflate
cavities in the X-ray halos of massive low-redshift galaxy clusters to
infer the kinetic power of the radio source. They give a calibration
to estimate the kinetic power, $dE/dt_{\rm kin,jet,C10}$ based on the
measured monochromatic radio power at 1.4~GHz in the rest-frame,

\begin{equation}
dE/dt_{\rm kin,jet,C10}=0.75\times (L_{\rm 1400}-23.8539)+1.91,
\label{eqn:kinjetc10}
\end{equation} 
where $dE/dt_{\rm kj,C10}$ is given in units of $10^{42}$ erg s$^{-1}$,
and $L_{\rm 1400}$, the observed luminosity of the radio jet at 1400~MHz,
in W Hz$^{-1}$.

\citet{willott99} 
calibrate their observations against the rest-frame 151~MHz radio
luminosity, $L_{\rm 151}$, by setting

\begin{equation}
dE/dt_{\rm kj,W00} = 3\times 10^{38} f_c^{3/2} L^{6/7}_{\rm 151},
\label{eqn:kinjetw00}
\end{equation}
where $f_c$ is a proportionality factor, which is most likely around
$f_c=10$ \citep[][]{cattaneo09}.

For both estimates we need to estimate the radio power at the required
rest-frame frequency. We do this by interpolating along the best
linear fit of multifrequency observations of our sources as given in
NED. The 1.4~GHz measurements correspond to 300-400~MHz of observed
frequency for most of our sources, and are approximately matched by
the frequency range of surveys like the Texas survey at 365~MHz or the
Molonglo survey at 408~MHz \citep[][]{large81}. The 151~MHz flux
required for the \citet{willott99} estimate would correspond to
observed radio fluxes at 32-50~MHz, which is lower than the
lowest-frequency measurements that are readily available for large
samples in the south \citep[80~MHz;][]{slee95}.

\begin{figure}
\centering
\includegraphics[width=0.48\textwidth]{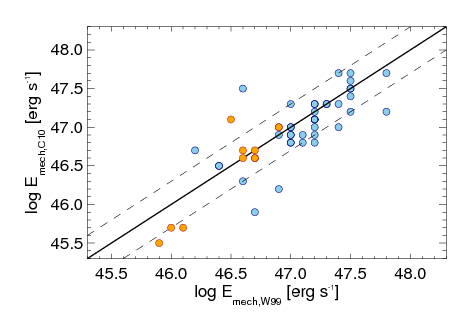}
\caption{Mechanical energy estimates obtained with 
  the formalism of \citet{willott99} and \citet{cavagnolo10} are shown
  along the abscissa and ordinate, respectively. The solid line shows
  the line of equality between both estimates, dashed lines offsets by
  0.3~dex (a factor of~2). The light blue and orange symbols show the
  sources of the present sample and that of \citet{collet14b},
  respectively.\label{fig:emechjet}}
\end{figure}

\begin{figure}
\centering
\includegraphics[width=0.48\textwidth]{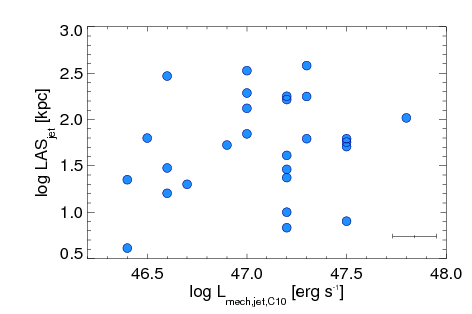}
\caption{Largest angular size of the radio source
  as a function of the jet mechanical energy estimated with the
  formalism of \citet{cavagnolo10}. We show all 49 sources with
  SINFONI data. \label{fig:jetpowersize}}
\end{figure}

In Table~\ref{tab:radiopower} we list the radio luminosities at both
rest-frame frequencies, and the mechanical energy output rates derived with the
\citet{willott99} and \citet{cavagnolo10} approach,
respectively. Fig.~\ref{fig:emechjet} shows the scatter plot of both
estimates. It is reassuring that both lead to similar estimates within
a factor~2 for most sources (0.3~dex, dashed lines in
Fig.~\ref{fig:emechjet}). Sources with larger discrepancies have
relatively shallow spectral indices compared to the generally steep
radio spectral indices of high-redshift galaxies of $\alpha\le-0.8$ to
$-1.0$. Since we have no direct observational constraints about a
possible steepening of these indices towards lower radio frequencies,
we favor the \citet{cavagnolo10} estimates for these
sources. Fig.~\ref{fig:jetpowersize} demonstrates that our sources
sample the overall range in jet kinetic energy fairly evenly at all
radio sizes.

A common worry in estimating kinetic jet powers is that 
  we have to extrapolate from the observed
  GHz regime well into the observed or even rest-frame MHz regime. For
  example, recent LOFAR observations of two nearby FRII radio galaxies
  show a steepening of the spectral index at frequencies between about
  10 and 1000 MHz, which translates into underestimates of about a
  factor 5 in kinetic jet power \citep[][]{harwood16}. On the bright
  side, these measurements give us also the opportunity to investigate
  the systematic uncertainties of our estimates of kinetic jet power
  directly.

We use the energy density estimates of \citet{harwood16} of $1.2\times
10^{-12}$ J m$^3$ in 3C~452, and of $0.28\times 10^{-12}$ J~m$^3$ and
$0.32\times 10^{-12}$ J~m$^3$ in the northern and southern lobe of
3C~223, respectively, and their measured radio sizes,
289.2\arcsec$\times$89\arcsec, and 157.5\arcsec$\times$21.5\arcsec,
and 152\arcsec$\times$25.8\arcsec\ for 3C~452 and 3C~223, respectively,
to estimate a total energy content of $2.3\times 10^{60}$ erg and
$1.8\times 10^{60}$ erg in 3C~452 and 3C~223, respectively. We then use
their projected sizes to estimate a jet age for a fiducial expansion
velocity of $0.1\ c$, finding 14~Myrs and 12~Myrs for 3C~452 and 3C~223,
respectively. This corresponds to kinetic energy injection rates of
5.0 and $4.6\times 10^{45}$ erg s$^{-1}$, respectively. We then use
integrated radio flux measurements from NED between 1.4~GHz and 10~GHz
for the two galaxies to estimate a rest-frame 1.4~GHz jet power in the
same way and for a similar frequency range to our targets, which
have measurements down to observed frequencies of about 70~MHz. We
find 26.3~W Hz$^{-1}$ and 26.1 W Hz$^{-1}$ for 3C~452 and 3C~223,
respectively, and spectral indices of $-1.3$ and $-0.8$, respectively,
not very different from our less powerful sources. With the
\citet{cavagnolo10} approach, this corresponds to 4 and $5\times
10^{45}$ erg s$^{-1}$ for 3C~452 and 3C~223, respectively, comparable to
the rates implied by the results of \citet{harwood16}. This shows
that, although our method is still very approximate, systematic
uncertainties in our estimates of jet power are of comparable
amplitude as other uncertainties.

\subsection{Velocity range and estimates from feedback models}
\label{ssec:wagnercomparison}

\begin{figure*}
\centering
\includegraphics[width=0.48\textwidth]{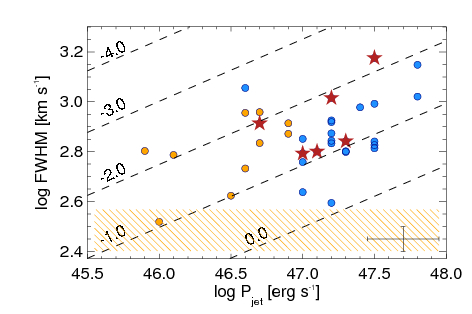}
\includegraphics[width=0.48\textwidth]{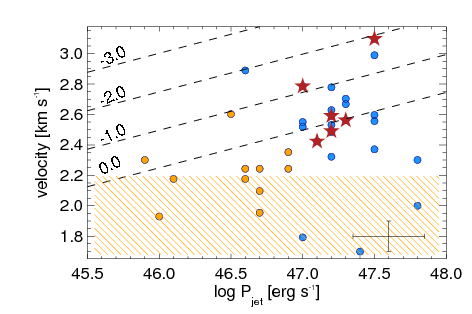}
\caption{Gas velocity as a function of the jet power following
  \citet{wagner12}. The left panel shows FWHM line widths, the right
  panel approximates a putative radial outflow velocity by showing one
  half of the measured velocity offsets. Blue and orange symbols show
  the radio galaxies of the present analysis, and those of
  \citet{collet14b}, respectively. Red stars show galaxies with
  measured Eddington ratios from \citet{nesvadba11a}. The black dashed
  lines show jet powers corresponding to Eddington ratios between
  $10^{-4}$ and unity. The yellow hatched area shows the range spanned
  by the mass selected sample of galaxies without prominent radio
  source of \citet{buitrago13}. \label{fig:wagnerplot} }
\end{figure*}

The expansion velocity of the cocoon is one of the prime quantities
predicted by hydrodynamic models of jets expanding through the ambient
gas, and it is therefore very interesting to compare the range of
velocities we observe with expectations from models.  A similar
velocity range of warm ionized gas clouds can also be expected from
the models of \citet{gaibler09}, who find 
an expansion velocity $\sim 5000$ km s$^{-1}$ of the cocoon. We follow
\citet{cooper08} in parametrizing the acceleration that a dense cloud
experiences through the drag by a hot wind medium as, $f_{\rm th}=3/9 C_D
(\rho_w / \rho_c) v_w^2/R_C$, where $f_{\rm th}$ is the acceleration,
$C_D$ the drag coefficient, $n_c=$500 cm$^{-3}$ the gas density in the
filaments, $R_C$ the size of a cloud, $\rho_w$ and $\rho_c$ the
density in the wind and cloud, respectively, and $v_w$ the wind
velocity. We find accelerations of about $1\times 10^{-12}$ km
s$^{-2}$, and velocities of 300$-$700 km s$^{-1}$ for cloud sizes
between 5 and 10~pc. 

In Fig.~\ref{fig:wagnerplot} we compare the velocity range observed in
our targets with the predictions of \citet{wagner12} for jet
interactions with clumpy, turbulent gas. We plot two samples, the one
presented here, and the one of \citet{collet14b} of radio galaxies
with somewhat lower radio power. The FWHMs of our sources fall into
the velocity range predicted by the model for galaxies with similar
radio power, and Eddington ratios of $10^{-1}$ to $10^{-2}$ (left
panel of Fig.~\ref{fig:wagnerplot}), whereas the velocity offsets
correspond to lower velocities and higher Eddington ratios, and fall
off more quickly with increasing radio power than the line widths. We
caution that these Eddington ratios are between the kinetic jet power
in this case (and not the radiative AGN luminosity, as is more
frequently the case) and the Eddington luminosity of the central
supermassive black hole. For six galaxies (marked as red stars in
Fig.~\ref{fig:wagnerplot}) we have Eddington luminosities measured in
\citet[][their Table~2]{nesvadba11a} and \citet{collet14b}. Using the
jet kinetic energies listed in Table~\ref{tab:radiopower} and line
FWHMs, we find Eddington ratios between 0.4 and 0.1 for all targets,
factors of a few higher than expected from the models. For bulk
velocities, we find Eddington ratios around 1 or even higher. This
suggests that discrepancies between observations and model
expectations do persist, although the good correspondence in the
predicted velocity range is very encouraging, even more so as
observational effects like blurring of localized outflows with
surrounding material and projection effects might in part be the
cause. Another potential source of such discrepancies could however
also be the longer-term evolution of the gas. As \citet{wagner12}
point out, studying the evolution of clumpy, turbulent, and multiphase
gas over time scales longer than few $10^5$ yrs at high resolution is
still computationally very demanding, and most of our nebulae have
dynamical times that are about 10$\times$ longer
(Table~\ref{tab:kinenergyjetgas}).

\subsection{Gas kinetic energy and momentum and transfer efficiencies from the jet to the gas}
\label{ssec:gaskinematics}

Characterizing the kinematics of the gas in the complex emission-line
regions of HzRGs, with their wide ranges of line width and velocity
jumps, with a single (or a small set of) numbers is very challenging.
In addition to the complex intrinsic kinematics, blurring by the
seeing disk and projection effects are likely to play a role. In spite
of these difficulties, we have seen in \S\ref{sssec:linewidths}
that it is possible to identify global trends between radio power and
gas kinematics, even when characterizing the complex kinematic
properties of these galaxies only with a single number.

This encourages us to derive global estimates of the kinetic energy
and momentum in our sources using simple analytical expressions
previously established within the spherical blastwave scenario of
supersonic shocks in the context of supernova explosions. Models of
radio jet cocoons have shown that the global energetics of the
extended gas in HzRGs are not too different from those obtained with a
spherical blastwave model, where the hot cocoon entrains ambient gas
as it expands through the ISM of the galaxy at highly supersonic speed
\citep[e.g.,][]{krause03, wagner12}.  The blastwave approximation
seems to be justified in these models irrespective of the detailed gas
properties or the evolutionary state of the jet. This is perhaps most
surprising for very irregular, clumpy environments, where the jet does
not simply expand near-adiabatially through the ambient gas
\citep[][]{wagner11,wagner12}. 

We adopt the same simple approach as in our previous analyses
\citep[in particular][]{nesvadba06b} to estimate the kinetic energy
injection rate into the gas with the following equations appropriate
for blastwaves \citep[e.g.,][]{dyson80}. Firstly, we set
\begin{equation}
dE/dt = (\Delta v/ 435)^5 \ n_0^{-1} \times 10^{44}\ {\rm erg\ s^{-1}},
\label{eqn:egasshell}
\end{equation}
where $\Delta v$ is the expansion velocity of the bubble, and
$n_0$ the density of the ambient gas (i.e., outside the expanding
bubble). 

Secondly, we can derive an energy injection rate by considering that
the hot gas bubble expands adiabatically, and assuming that the
highest observed gas velocities approximate the expansion velocity of
the bubble:

\begin{equation}
dE/dt = 1.5\times 10^{46}\times r^2\times \Delta v^3 \times n_0\ {\rm erg\ s^{-1}},
\label{eqn:egasbubble}
\end{equation}
where $r$ is the radius of the bubble in units of 10~kpc. In both
equations, $\Delta v$ is given in units of 1000~km s$^{-1}$. 

\begin{figure}
\includegraphics[width=0.48\textwidth]{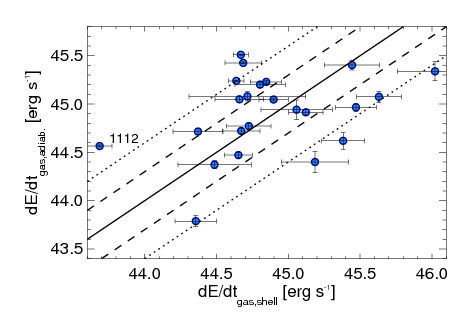}
\caption{
Comparison of the
  results of our two kinetic energy estimates in
  \S\ref{ssec:gaskinematics} for the 23 galaxies with extended
  emission line regions, which have been observed in
  [OIII]$\lambda$5007. The dashed and dotted lines show offsets by
  factors 2 and 4, respectively, from the one-to-one relationship
  (solid line).  \label{fig:energycomparison}}
\end{figure}

In Fig.~\ref{fig:energycomparison} we compare the kinetic energy
distribution of our sources estimated by the two approaches (and using
observed quantities further discussed below), finding a good
correspondence between the two. The slope of the best linear fit,
0.78, is slightly below unity, but mostly driven by the source with
the highest and lowest energy estimate, which have offsets from the
one-to-one relationship that are consistent with the overall scatter
within the sample, and we therefore consider this result consistent
with a one-to-one relationship. Finding a good correspondence between
both estimates implies that the gas velocities and sizes are
consistent with each assumption, and therefore further demonstrates
that the blastwave scenario is a reasonable approximation.

In addition to the energy, the radio jet also injects 
momentum into the gas. We estimate the momentum injection
rates associated with each estimate, $dP/dt$, by setting $dP/dt =
1/v\ dE/dt$.

The next question to address is, which quantity we should use as
velocity estimate. Velocity offsets are perhaps the most obvious
quantity, but also interpretationally difficult: Firstly, in galaxies
with relatively small total velocity offsets ($\le$400-500~km
s$^{-1}$) and gas sizes less than about 10$-$20~kpc, the kinematics
may reflect the radial motion of an outflow, perhaps in combination
with disk rotation, whereas our results in \S\ref{ssec:rotation}
disfavor rotation alone. In back-to-back outflows, gas in either
bubble is driven out at 1/2 the total velocity offset. In winds lifted
off a disk, the observed velocity offset corresponds to the relative
velocity between disk and wind. In both cases, intrinsic velocities
may be factors $2-3$ larger than those observed because of beam-smearing
and projection effects. We have previously discussed these
interpretational complications in more detail in \citet{nesvadba08}
and \citet{collet14b}, and use 1/2 of the total velocity offset in our
calculations here, which gives the more conservative energy
estimates. With all these caveats, and using
Equations~\ref{eqn:egasshell} and \ref{eqn:egasbubble}, we find
kinetic energy injection rates from bulk motion that are between few
$10^{43}$ and $1\times 10^{47}$ erg s$^{-1}$. Corresponding momentum
injection rates are between $10^{34}$ and $10^{39}$ dyn.

Fig.~\ref{fig:fwhmemech} showed that amongst our different measures of
FWHM (average and maximal velocity dispersion and integrated line
widths, see also Table~\ref{tab:eelrproperties}), the average velocity
dispersions are most tightly related to the jet power. We therefore
consider the average widths the most reliable tracers
of gas motion related to the radio jet, on which we base our estimates of
kinetic energy from random motion.  With the two equations presented
above, we estimate kinetic energy injection rates between
$0.6\times10^{44}$ erg s$^{-1}$ and $1\times10^{46}$ erg s$^{-1}$ in
our sample, most galaxies have between 5 and $10\times 10^{44}$ erg
s$^{-1}$. Results for individual galaxies are listed in
Table~\ref{tab:kinenergyjetgas}.

\begin{figure*}
\includegraphics[width=0.48\textwidth]{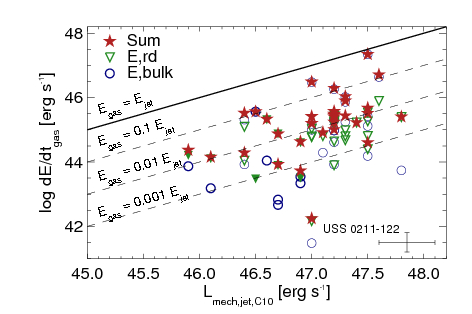}
\includegraphics[width=0.48\textwidth]{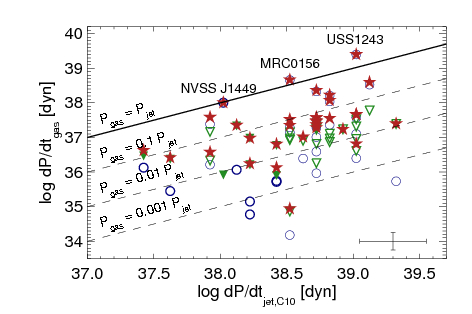}
\caption{{\it(left)} Kinetic energy injection rates from resolved bulk motion
  (dark blue open circles) and unordered motion (upside-down green triangles),
  and the sum of both (red stars), as a function of radio power. 
 The
  solid line shows the kinetic power of the radio jet, and dashed lines
  show ratios of $10^{-1}$, $10^{-2}$, and $10^{-3}$. {\it
    (right)} The same plot showing the momentum injection rates of the
  radio source and the gas. 
}\label{fig:egasvsejet}
\end{figure*}

Another interesting quantity to characterize feedback from radio jets
onto the surrounding gas is the efficiency with which the jet kinetic
energy is being deposited into the gas, i.e., the ratio
$E_{\rm kin,gas}/E_{\rm kin,jet}$. In Fig.~\ref{fig:egasvsejet} we show how
the kinetic energy injection rates estimated from resolved bulk
motion, from the line widths, and the sum of both, respectively, scale
with kinetic jet power. Efficiencies are between $10^{-3}$ and almost
unity in all cases, with most galaxies falling around $10^{-2}$. No
galaxy has an efficiency greater unity, implying that we do not
require a contribution from another mechanism driving the gas based on
these observations and analysis. The scatter is very large, as can be
expected given the considerable uncertanties in the estimates of this
relationship, and may therefore not preclude a much tighter intrinsic
relationship between jet and gas energy. Considering the combined
sample of the sources from the present study and those of
\citet{collet14b}, uncorrelated distributions are ruled out at
significances of few times $10^{-4}$ and better for energy injection
rates derived from bulk kinetic motion, line widths, and the sum of
both. The hydrodynamic models of \citet{wagner12} predict efficiencies
of up to 40\%, about an order of magnitude greater than what we
observe. It is well possible that parts of these discrepancies come
from heating and entrainment of gas in other phases of cold gas
\citep[e.g.,][]{nesvadba10, emonts15}. 

We did not find a relationship between efficiency and jet size. Gas
kinetic energies estimated with the two methods are uncorrelated with
radio size at significances between 0.16 and $>$0.5 with the Spearman
rank test and Kendall's $\tau$.  This result is interesting, but it
would be very naive to conclude that the jet deposits its mechanical
energy at uniform rates throughout its expansion history. Quite to the
contrary, it is reasonable to expect that the efficiency of this
energy deposition drops markedly once the jet has broken out of the
ambient gas. An alternative interpretation is that the gas dissipates
the kinetic energy only relatively slowly compared to the lifetime of
the radio jet. We will further discuss this in
\S\ref{sec:galaxyevolution}.

\begin{figure}
\includegraphics[width=0.48\textwidth]{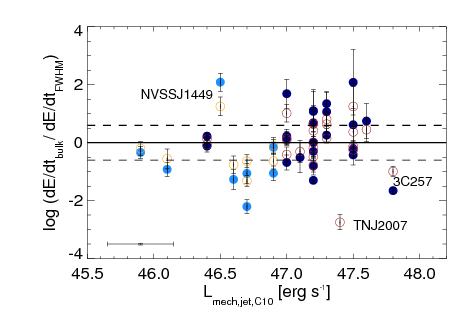}
\caption{Ratio of kinetic energy estimated from the velocity offsets
  and line widths, respectively, as a function of kinetic jet
  power. Filled dark blue and empty red circles show ratios derived
  with Eqn~\ref{eqn:egasshell} and \ref{eqn:egasbubble} for 23
  galaxies with extended emission-line regions in the present sample,
  respectively. We excluded RC~0311$+$0507 for which we only have
  [OII]$\lambda\lambda$3727,3729 with a composite line profile, unlike
  the other sources in our sample. Filled light blue dots and yellow
  open circles show the same estimates for nine galaxies with extended
  emission-line regions discussed in \citet{collet14b}, which have
  significantly lower radio power. The solid line shows the one-to-one
  relationship, dashed lines factors 4 above or below, which
  corresponds to twice the typcial uncertainty expected from
  inclination and resolution effects. 
}\label{fig:ekinratiovsjetpower}
\end{figure}

\section{Role for galaxy evolution}
\label{sec:galaxyevolution}

\subsection{Outflows and turbulence}
\label{ssec:feedbackevolution}

The 'standard' scenario of AGN feedback adopted by most observers and
cosmological models focuses on the gas removal through winds driven by
the AGN through its radiation or the radio jets. In these models, star
formation in galaxies is truncated by starvation as the gas reservoirs
that would otherwise fuel subsequent star formation are being depleted
by the wind. In contrast, several observations of radio galaxies in
the more nearby Universe have shown in recent years that the kinetic
energy injection of radio jets feeds a complex interplay between
turbulence within the gas of the disk and outflows that are lifting
gas above the disk, and which may or may not escape
\citep[e.g.,][]{nesvadba10,nesvadba11c,alatalo11}.  These galaxies are
characterized by small ratios of v/$\sigma$, akin to those in our
galaxies, large gas masses (mostly in diffuse, filamentary, warm
molecular gas at low-z), and shock-like line ratios in large gas
fractions.  The energy carried by the wind appears roughly as large in
these galaxies as that which is driving the turbulence.  A central
characteristic of these galaxies are their strongly enhanced
luminosities of warm (T$>100$ K) H$_2$ emission lines in the
mid-infrared, exceeding expected luminosities from UV heating by
factors of at least a few \citep[][]{ogle10,nesvadba10}. The only HzRG
with an observation of warm mid-IR H$_2$ emission in the literature is
currently MRC~1138$-$262 \citep[][]{ogle12}, the first galaxy we also
observed as part of our SINFONI survey. Turbulence and winds could be
closely related in these galaxies. \citet{scannapieco13} point out
that turbulence in the clumpy ISM of starburst galaxies could strongly
heat small parcels of gas, thereby enhancing the efficieny of mass
loading of the hot wind medium with ambient gas, and playing an
important role in launching collimated winds.

The most direct observable of turbulence in high-redshift galaxies
before more mid-IR spectroscopy of HzRGs will become available with
the James-Webb Space Telescope are their broad line widths. It is
therefore very interesting that the emission lines in HzRGs, including
MRC~1138$-$262, reach very large, and almost constant, values of about
$500-800$ km s$^{-1}$ over several orders of magnitude in radio power
\citep[see also][]{collet14b}. The ratios of bulk to random motion,
$v/2\sigma$, are as low as $v/2\sigma=0.3-0.4$ in several of our
sources (Table~\ref{tab:eelrproperties}). This implies that these are
not self-gravitating disks, although at least some of the gas may
still be bound within the gravitational potential of the galaxy. The
large line widths imply that at least parts of this gas
along the line wings also exceed the escape velocity of their host
galaxy \citep[about $500-700$ km s$^{-1}$, e.g.,][]{nesvadba06a}.

To further investigate the role of turbulence and bulk velocity, we
can investigate how each depends on other properties of our
sources. In \S\ref{sssec:voversigma} we showed that the ratios of
$v/2\sigma$ do scale with increasing size of the emission-line regions
(or elongation, Fig.~\ref{fig:emsellem}), but not with the size of the
radio source (Fig.~\ref{fig:lasvssigv}). It is also very interesting
to compare how the kinetic energy in bulk and random motion scale with
radio power. In Fig.~\ref{fig:ekinratiovsjetpower} we plot the ratio
of the kinetic energy injection rates into the gas estimated from the
Gaussian line widths ($\sigma=$FWHM/2.355) and those derived from
$\Delta v/2$ as a function of radio power. At all values along the
abscissa we note a subset of sources which have much lower energy
injection rates from ordered than unordered motion, although both
forms of energy injection increase roughly at the same rate with radio
power. Galaxies with large ratios of ordered to unordered motion seem
to be generally associated with radio sources with kinetic power
greater than about $10^{47}$ erg s$^{-1}$. This corresponds to gas
kinetic energy injection rates greater than about $10^{44-45}$ erg
s$^{-1}$. The only exception is NVSS~J144932$-$385657, which is the
largest source in the sample of \citet{collet14b}, which has a radio
power of $2.5\times 10^{46}$ erg s$^{-1}$ with the estimate of
\citet[][but $1.3\times 10^{47}$ with the estimate of
  \citealt{willott99} ]{cavagnolo10}. This discontinuity is a new
effect which could only be discovered observationally with a large
sample such as ours, and which has to our knowledge not been
anticipated in hydrodynamic simulations or observations of other types
of AGN, and deserves a more detailed discussion.

\begin{table*}
\centering
\begin{tabular}{lcccccc}
\hline 
Source & age          & $\log$ E$_{\rm gas,FWHM,1}$ & $\log$ E$_{\rm gas,FWHM,2}$ & $\log$ E$_{\rm gas,\Delta v, 1}$ & E$_{\rm gas,\Delta v,2}$\\
       & [$10^7$ yrs] &                          &                  &                              & \\
\hline
BRL~0128-264  & 1.5   &  44.6$\pm$0.2 &  45.2$\pm$0.2  &  45.3$\pm$0.2 &  45.7$\pm$0.2  \\
MRC~0156-252  & 12.0  &  44.9$\pm$0.2 &  45.6$\pm$0.2  &  46.6$\pm$0.2 &  46.6$\pm$0.2 \\
MRC~0251-273  & 4.1   &  44.0$\pm$0.2 &  44.7$\pm$0.2  &  44.7$\pm$0.2 &  45.1$\pm$0.2  \\
MP~J0340-6507 & 3.0   &  44.4$\pm$0.3 &  45.1$\pm$0.2  &  43.1$\pm$0.2 &  44.3$\pm$0.2 \\
PKS~0529-549  & 3.3   &  43.6$\pm$0.2 &  44.3$\pm$0.2  &  43.4$\pm$0.2 &  44.2$\pm$0.2 \\
TXS~1113-178  & 1.3   &  43.4$\pm$0.2 &  44.1$\pm$0.2  &  42.9$\pm$0.2 &  43.8$\pm$0.2 \\
3C~257        & 2.0   &  44.2$\pm$0.2 &  44.9$\pm$0.2  &  42.6$\pm$0.2 &  43.9$\pm$0.2 \\
USS~1243+036  & 4.9   &  44.7$\pm$0.5 &  45.4$\pm$0.2  &  46.8$\pm$0.2 &  46.7$\pm$0.2 \\
USS~1410-001  & 6.0   &  44.7$\pm$0.2 &  45.4$\pm$0.2  &  44.8$\pm$0.2 &  45.4$\pm$0.2 \\
MRC~1558-003  & 6.1   &  44.6$\pm$0.2 &  45.3$\pm$0.2  &  44.9$\pm$0.2 &  45.5$\pm$0.2  \\
USS~1707+105  & 3.1   &  44.4$\pm$0.2 &  45.1$\pm$0.2  &  44.1$\pm$0.2 &  45.0$\pm$0.2 \\
3C~362        & 3.4   &  44.8$\pm$0.3 &  45.$\pm$0.2   &  44.4$\pm$0.2 &  45.2$\pm$0.2  \\
MP~1758-6738  & 5.1   &  44.3$\pm$0.2 &  45.0$\pm$0.2  &  44.5$\pm$0.2 &  45.1$\pm$0.2 \\
MRC~2025-218  & 1.2   &  44.0$\pm$0.2 &  44.6$\pm$0.2  &  43.1$\pm$0.2 &  44.2$\pm$0.2 \\
MG~2037-0011  & $<$1.0&  43.8$\pm$0.3 &  44.5$\pm$0.2  &  41.8$\pm$0.3 &  43.3$\pm$0.2 \\
MRC~2104-242  & 10.4  &  44.8$\pm$0.2 &  45.5$\pm$0.2  &  45.9$\pm$0.2 &  46.1$\pm$0.2 \\
4C~23.56      & 8.9   &  44.6$\pm$0.2 &  46.3$\pm$0.2  &  45.9$\pm$0.2 &  46.1 $\pm$0.2 \\
MG~2144+1928  & 6.2   &  44.5$\pm$0.2 &  44.1$\pm$0.2  &  45.1$\pm$0.2 &  45.5 $\pm$0.2 \\
TN~J2254+1857 & 2.9   &  44.2$\pm$0.2 &  44.9$\pm$0.2  &  44.4$\pm$0.2 &  45.0 $\pm$0.2 \\
MG~2308+0336  & 4.1   &  44.5$\pm$0.2 &  45.2$\pm$0.2  &  45.6$\pm$0.2 &  45.9 $\pm$0.2 \\
\hline
NVSS~J0024    & 2.5   & 44.3$\pm$0.2  & 45.0$\pm$0.2   &  43.1$\pm$0.2  & 44.3$\pm$0.2 \\
NVSS~J0040    & 3.6   & 43.8$\pm$0.2  & 44.5$\pm$0.2   &  42.7$\pm$0.2  & 43.8$\pm$0.2 \\
NVSS~J0129    & 2.4   & 43.9$\pm$0.2  & 44.5$\pm$0.2   &  41.6$\pm$0.2  & 43.2$\pm$0.2 \\
NVSS~J0304    & 3.6   & 43.1$\pm$0.2  & 43.9$\pm$0.2   &  42.1$\pm$0.2  & 43.2$\pm$0.2\\
NVSS~J1449    & 11.8  & 43.6$\pm$0.2  & 44.3$\pm$0.2   &  45.7$\pm$0.2  & 45.5$\pm$0.2 \\
NVSS~J2342    & 4.6   & 43.0$\pm$0.2  & 43.6$\pm$0.2   &  42.8$\pm$0.2  & 43.5$\pm$0.2 \\
CEN~J949      & 2.8   & 43.2$\pm$0.2  & 43.9$\pm$0.2   &  42.3$\pm$0.2  & 43.4$\pm$0.2 \\
CEN~J952      & 4.9   & 43.6$\pm$0.2  & 44.2$\pm$0.2   &  43.2$\pm$0.2  & 44.0$\pm$0.2 \\
CEN~J949      & 8.7   & 43.2$\pm$0.2  & 43.9$\pm$0.2   &  42.0$\pm$0.2  & 43.2$\pm$0.2 \\
\hline\end{tabular}
\caption{
Jet ages and kinetic energy of the gas. See text for details. 
\label{tab:kinenergyjetgas}}
\end{table*}

Fig.~\ref{fig:ekinratiovsjetpower} shows the presence of sources
dominated by kinetic energy in unordered motion at all jet
power. At a jet power above $10^{47}$ erg s$^{-1}$, however, the
number of sources dominated by their bulk outflow energy becomes
significant. It is possible that this discontinuity marks a threshold
at which the momentum injection into the ambient gas becomes so strong
that the disk no longer remains globally stable. In the most powerful
sources, the disk gas would become overall unbound.  At energies below
this threshold, more localized outflows may exist in an overall
turbulent disk, which may closely co-exist with the jet cocoon as
previously discussed by \citet{nesvadba10, nesvadba11c}, and
\citet{collet14b}. This could also explain why outflows in
low-redshift radio galaxies, which do not reach such high jet power,
appear typically to be more localized, very different from the
extended bubbles in the most powerful high-redshift galaxies.

We are not aware of a hydrodynamic simulation that would show such an
effect for powerful AGN hosts, but we can build upon a very recent
analytical approach to study in a more quantitative way how turbulence
and outflows may depend on each other in these HzRGs.  Based upon the
notion by \citet{thompson14} that turbulence might cause
super-Eddington winds in sub-Eddington starbursts, \citet{hayward15}
presented an analytic model of feedback which includes outflows as
well as turbulence. Turbulence creates a multi-phase environment,
where cells of gas with different velocity, density, temperature, and
phase co-exist. A given momentum injection rate might already be
sufficient to remove gas from comparably low-density cells, even
though the momentum injection rate is not sufficient to remove
average-density gas.

\citet{hayward15} developed their algorithm for self-regulated
star formation, however, their basic equations are appropriate for
blastwave scenarios generally, and can therefore easily be adapted to
our present needs. We will in the following use some of their
equations together with our measurements to investigate whether the
increased importance of gas outflows compared to turbulence with
increasing radio power is consistent with such a scenario.

\begin{figure}
\centering
\includegraphics[width=0.5\textwidth]{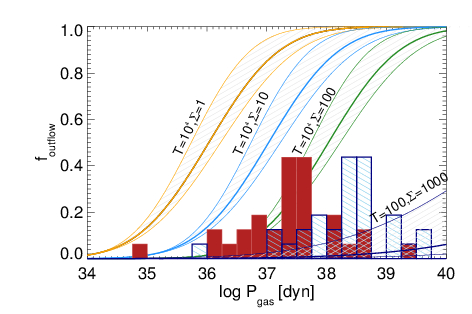}
\caption{
Fraction of gas with densities below the critical value for outflows
as a function of the momentum injection, following
\citet{hayward15}. We show three ranges of mass surface density
(orange, light blue, and green hatched regions, respectively) with
$\Sigma = 1$~M$_{\odot}$ pc$^{-2}$, 10~M$_{\odot}$ pc$^{-2}$, and
100~M$_{\odot}$ pc$^{-2}$, respectively. For each region, the left,
central, and right lines represent turbulent velocities that
correspond to FWHM=500, 800, and 1000 km s$^{-1}$, respectively. All
are given for temperatures of $10^4$~K, corresponding to typical
electron temperatures in the warm ionized gas in HzRGs. The red
histogram shows the distribution of total gas momentum measured in the
galaxies in our sample (Fig.~\ref{fig:egasvsejet}). The dark blue
region and light blue hatched histogram assume that the total gas
content is dominated by molecular and cold neutral gas, with a total
gas mass about 10$\times$ greater than that of warm ionized gas,
T=1000~K and $\Sigma = 1000$~M$_{\odot}$ pc$^{-2}$.}
\label{fig:hayward}
\end{figure}

The most important relationship for our purposes is equation (9) of
\citet{hayward15}, which parametrizes the mass outflow fraction,
$f_{\rm out}$, as
\begin{equation}
f_{\rm out} = \frac{1}{2} \left[{1 - {\rm erf} \left(\frac{-2\ x_{\rm out} +\ \sigma^2_{\rm ln \Sigma_g} }{2\ \sqrt{2}\sigma_{\rm ln \Sigma_{g}}}\right)}\right]
\end{equation}
through the ratio $x_{\rm out}$ between the maximal surface density of gas
that can reach escape velocity, $\Sigma^{max}_{\rm g}$, and the average gas
mass surface density, $\langle \Sigma_g\rangle$. 
$\Sigma^{max}_{\rm g}$ is set by the momentum injection rate density into the gas, 
$\dot{\Sigma}_p$, which we take to be in the range observed from the
momentum of the warm ionized gas in our galaxies, $10^{35-38}$ dyn,
divided by the surface area of the emission line nebulae
(Table~\ref{tab:eelrproperties}), the orbital frequency, $\Omega$, and
circular velocity, $v_c$ of the galaxy. We adopt $v_c=350$ km s$^{-1}$
\citep[][]{nesvadba11b}, corresponding to the velocity dispersion
expected for a galaxy with stellar mass of few $\times 10^{11}$
M$_{\odot}$ \citep[as measured for HzRGs
  by][]{seymour07,debreuck10}. We calculate the orbital frequency at a
radius of 10~kpc, corresponding to the fiducial size of the
emission-line nebulae we adopt in these calculations.  $\sigma_{\rm ln \Sigma_g}$
 is the width of the density distribution of the gas mass
surface density set by the turbulence, $\sigma_{\rm ln \Sigma_g}^2\approx
1+R\ M^2/4$. Here, $M$ is the Mach number of the turbulent gas, given by
the turbulent velocity dispersion, $\sigma_T$, and the sound speed,
$c_s$, $M=\sigma_T/c_s$. $R$ depends on the Mach number and
power-law index of the turbulence, $\alpha$ \citep[][]{hayward15}.
See \citet{hayward15} for the detailed derivation of these relationships.

In Fig.~\ref{fig:hayward} we show how the mass outflow fraction
depends on the momentum injection rate into the gas, for ranges of
characteristic values of velocity dispersion (FWHM$=$500 km s$^{-1}$,
800 km s$^{-1}$, and 1000 km s$^{-1}$), and sound speed, $c_s$ (11.4
km s$^{-1}$ and 3.7 km s$^{-1}$ for $T=10^4$ K and $T=10^3$ K,
respectively). $\langle\Sigma_g\rangle$ is the average gas mass
surface density over the total column, which we assume to be $D=10$
kpc, corresponding to the minor axis sizes of the emission-line
nebulae. We measured the gas mass surface densities of warm ionized
gas in our galaxies, typically few to few times $10$ M$_{\odot}$ pc$^{-2}$,
from the ionized gas mass estimates in \S\ref{sssec:gasmasses} (see
also Table~\ref{tab:gasmasses}) and the projected sizes of
[OIII]$\lambda$5007 line emission given in
Table~\ref{tab:eelrproperties}. 

Fig.~\ref{fig:hayward} shows that for galaxies with characteristics
that are similar to those in our HzRGs, the fractions of outflowing
gas increase rapidly with increasing momentum injection rate in the
range that we observe in our galaxies. The black, light blue, and red
areas represent gas densities of 1, 10, and 100~M$_{\odot}$ pc$^{-2}$,
and are each given for three turbulent velocities (which correspond to
observed FWHMs of 500 km s$^{-1}$, 800 km s$^{-1}$, and 1000~km
s$^{-1}$, respectively). We assume $T=10^4$~K for all surface mass
densities. The solid red histogram in Fig.~\ref{fig:hayward} shows the
observed momentum of the warm ionized gas in our galaxies,
corresponding to about $1-10$\% of the momentum provided by the radio
jet, for a jet expansion speed of $0.1c$ (see also
Fig.~\ref{fig:egasvsejet}).

The dark blue histogram in Fig.~\ref{fig:hayward} accounts for the
additional gas phases that we are likely missing. In turbulent
multiphase gas, we would expect that this gas has broadly similar
kinematic properties to the warm ionized gas, as observed in nearby
radio galaxies \citep[][]{nesvadba11c}, and consistent with the broad
line widths of [CI] and even CO observed in HzRGs of about 500-1000 km
s$^{-1}$. From the ratios of ionized to molecular gas mass in
Table~\ref{tab:gasmasses} and other initial estimates of gas masses in
other phases in high-z radio galaxies \citep[e.g., warm molecular gas
  measured from the H$_2$ 0-0 S(3) line or cold neutral gas measured
  from atomic carbon, Gullberg et al. 2015, A\&A submitted,
][]{ogle12}, we expect to find that the total gas mass surface density
will probably be higher by about an order of magnitude, if taking
these gas phases into account. Hence, gas mass surface densities would
be in the range $10^{2}-10^{3}$ M$_{\odot}$ pc$^{-2}$, and the
momentum about an order of magnitude greater than that of the warm
ionized gas alone.

For lower gas temperatures the required momentum injection rates to
drive outflows with high mass-loading factors increase rapidly, as
seen from the dark-blue rimmed area, which uses $T=10^3$~K, a gas mass
surface density of $\Sigma\sim 10^3$ M$_{\odot}$ pc$^{-2}$, and
otherwise similar parameters to the previous cases. This could imply
that molecular outflows in HzRGs are indeed less efficient compared to
outflows of warm ionized gas, unless the gas is being heated to
temperatures above about 100~K (e.g., through the AGN radiation,
cosmic rays, or shocks).

We note that these outflow rates are given relative to the total
density distribution of the gas, and concern only gas that can be
accelerated during one turbulent crossing time, $t\sim
l_{\rm Eddy}/\sigma_T$. Assuming that turbulence is injected on kpc
scales, this corresponds to about $1\times 10^6$ yrs, leaving time for
repeated cycles, if the AGN lifetime is long enough. Even outflow
rates of few 10\% of the total gas mass could therefore effectively
drain the ISM of our HzRGs during an AGN life time. 

Of course, all these estimates are highly simplified, and detailed
hydrodynamic simulations would be necessary to obtain firm
quantitative estimates of how turbulence may boost the efficiency of
AGN-driven winds. Nonetheless, our simple calculations show that such
a scenario could plausibly explain the increasing importance of
galaxy-wide outflows at the highest jet energies. We emphasize that in
the scenario adopted here, the molecular and warm ionized gas do not
represent two strictly isolated entities, but are part of a
multiphase medium, where the local density and temperature
fluctutations are large enough that gas might change its phase. This
might almost be a prerequisite in our case, and help explain the large
numbers of recombining photons we see from the warm ionized gas. With
a recombination cross section of few $10^{-14}$ s$^{-1}$
\citep[][]{osterbrock89}, and electron densities between $10^{2}$ and
$10^{3}$, the recombination time of H$\alpha$, $\tau_{\rm
  rec}=1/n\ \alpha$, is very short compared to the lifetime of the
radio jet, about $10^{3-4}$ yrs. Here, $n$ is the electron density,
and $\alpha$ the recombination cross section. This would imply
implausibly large reservoirs of gas that is cooling down, of-order
$10^{2-3}$ times greater than the observed HII masses, or about
$10^{12-13}$ M$_{\odot}$ in the most extreme cases, exceeding the
dynamical masses we measure in some of our galaxies by at least an order
of magnitude \citep[e.g.,][]{nesvadba07b,collet14b}. The alternative
explanation would be that each hydrogen atom undergoes multiple cycles
of heating and cooling during the jet lifetime, cycling (at least)
through the neutral and ionized phase, as would be naturally expected
in a turbulent multiphase medium
\citep[e.g.,][]{guillard09,nesvadba10,scannapieco15}. As an outcome of
this, fairly cold, fairly dense molecular gas as probed by mid-level
rotational CO line emission may indeed not be a major tracer of
outflows in HzRGs, because the more tenuous gas is preferentially
accelerated. This is also the gas that cannot self-shield, and
therefore is most susceptible to photoionzation by AGN photons
\citep[see our discussion of large ionized gas masses in
  \S\ref{sssec:gasmasses} and ][]{hopkins10}.

These results add further to the growing evidence that turbulence
plays a major role for our understanding of AGN feedback. Galaxies
with kinetic jet power $>10^{47}$ erg s$^{-1}$ are exceedingly rare,
however, galaxies with few $10^{46}$ erg s$^{-1}$ of radio power,
which we discuss in more detail in \citet[][]{collet14b}, are abundant
enough to represent a general phase in the formation and early
evolution of massive galaxies at high redshift.

\subsection{ISM metal enrichment}
\label{ssec:metallicity}

Gas-phase metal abundances in the interstellar gas are important
probes of the integrated past growth history of galaxies. The
relationship between the current ISM or stellar metal content and
total stellar mass formed in galaxies, the mass-metallicity
relationship therefore depends sensitively not only on the total mass
of stars formed, but contains also the fingerprints of the growth
history of these galaxies. Gas infall through accretion or outflows
from AGN and star formation should lower the total metal budget in
galaxies. We will now discuss how the metallicities implied by the
line ratios discussed in \S\ref{ssec:diagnostics} may further our
understanding of the evolutionary state of HzRGs.

\begin{figure}
\centering
\includegraphics[width=0.5\textwidth]{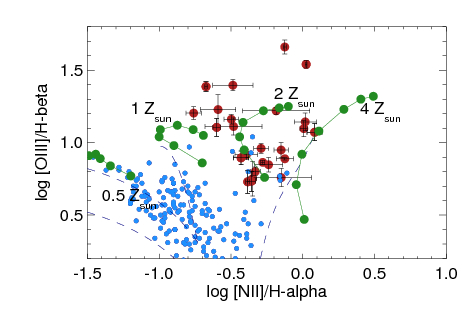}
\caption{
[NII]/H$\alpha$ vs. [OIII]/H$\beta$
  diagnostic diagram with our sources as already shown in
  Fig.~\ref{fig:metallicity}, but now also with the tracks for
  different metallicities overplotted that \citet{groves06} modeled
  for galaxies with narrow-line regions of different metallicity
  (green dots and lines) and a range of ionization parameter
  \citep[see~Tab.~2 in ][for details]{groves06}. The range of our
  sources corresponds well to metallicities between 1 and 4$\times$
  the solar value. Dashed dark blue lines indicate the star-forming
  and AGN sequence from \citet{kewley13a}. Light blue dots show the
  sample of \citet{steidel14} of UV/optically selected star-forming
  galaxies at similar redshifts, and with detections of all four
  lines.}
\label{fig:groves}
\end{figure}

In \S\ref{ssec:diagnostics} we showed the diagnostic ``BPT'' diagrams
constructed from [NII]$\lambda$6583, H$\alpha$, [OIII]$\lambda$5007,
and H$\beta$, and that our galaxies fall within the range expected for
galaxies at z$\sim$2.5, which is significantly offset from the AGN
branch of galaxies in the low-redshift Universe. \citet[][]{groves06},
\citet{kewley13a}, and \citet{kewley13b} argued, that this offset is
mainly an effect of low metallicities in high-redshift gas photoionzed
by AGN compared to local AGN narrow-line regions, which often have
metallicities of few times the solar value of 12$+$[O/H]$=$8.96
\citep[][]{allende01}. They argue that this offset comes in particular
from a lowering in the [NII]/H$\alpha$ ratio, which is dominated by
the lower metallicities at high redshift. Compared with the diagrams
of \citet{groves06} we find that our galaxies fall mainly within a
metallicity range between 1-4$\times$ the solar value
(Fig.~\ref{fig:groves}). 

Metallicities of about solar have previously
been derived for HzRGs in a similar redshift range by \citet{vernet01}
and \citet{humphrey08}, based on rest-frame UV lines, in particular
[NV] and HeII. These lines have high ionization potentials, requiring a
hard, intense radiation field, which suggests they are most
sensitive to the very inner regions of the galaxy near the AGN. It is
therefore interesting that also the optical lines, probing less
extreme environments more representative of the global ISM of our
galaxies, suggest solar to supersolar metallicity, including galaxies
which have very extended emission-line regions that reach far beyond
the effective radii of massive galaxies of few kpc. \citet{villar02}
also suggested based on a rest-UV longslit spectrum of the z=2.6 HzRG
TXS~0828+193 that metals of at least 1/3 Z$_{\odot}$ are likely to be
present within the high surface-brightness gas out to few 10s of kpc
from the galaxy's nucleus. This suggests that the gas in the extended
emission-line regions has already been strongly processed by stars, as
expected in the outflow scenario. 

Several arguments have been put forward in recent years that HzRGs
must be in the last stages of their assembly, including their high
stellar and dynamical masses \citep[][]{seymour07, debreuck10,
 nesvadba07b}, the proximity of their black hole and bulge masses to
the relationship found in local galaxies \citep[][]{nesvadba11a}, and
their black hole and bulge growth rates \citep[][]{drouart14}. If
these galaxies have indeed not significantly increased their stellar
mass between z$\sim$2 and today, then this implies that the metal
enrichment in their interstellar gas must already be at least as high
as that found in the photospheres of stars in massive galaxies
today. 

\citet{gallazzi05} used the SDSS DR2 to estimate stellar
metallicities of 175128 nearby galaxies, finding a typical metallicity
range of one to two times the solar value for galaxies with stellar
masses of $M_{\rm stellar}=10^{11.5}$ M$_{\odot}$ and above. Finding
gas-phase metallicities that are as high or even higher in our sources
supports the view that the active phase of star formation in HzRGs
must be very near its end. Finding somewhat higher gas-phase than
stellar metallicities is in fact expected, because stars continue to
expel metals as they evolve, even if no new stars are
formed. \citet{gallazzi05} find an offset of about 0.5~dex between the
stellar and gas-phase metallicities in low-redshift galaxies at a
given stellar mass. If we take our results at face value, then we find
a gas-phase metallicity enhancement of a factor of 2 at z$=$2 compared
to a factor 3 at z$=$0. This underlines again the absence of
subsequent star formation in massive galaxies between z$=$2 and z$=$0.

\section{Summary}
\label{sec:summary}

We presented an analysis of 33 powerful radio galaxies at z$\sim$2
with previously unpublished SINFONI imaging spectroscopy of the bright
rest-frame optical emission lines from warm ionized gas. Including
already published sources with similar data sets, our total sample
size is 49, corresponding to about 1/3 of the total number of such
galaxies currently known in this redshift range. We studied the gas
surface brightness in [OIII]$\lambda\lambda$4959,5007 and H$\alpha$,
diagnostic line ratios, extinction, and the spatially resolved and
integrated kinematic properties of the gas. Our main results are as follows:

\begin{itemize}
\item
HzRGs are commonly surrounded by extended emission line regions with
complex, irregular gas kinematics. FWHM line widths are typically
between 500 and 1000 km s$^{-}$, velocity fields consist either of two
bubbles with a bipolar velocity field or are irregular, potentially
with multiple bubbles. This qualitatively corresponds to the flood
and channel phase predicted by the hydrodynamic models of
\citet{sutherland07,wagner11,wagner12}.
\item
The ratio of bulk velocity to velocity dispersion in our sample does
not scale with ellipticity in the way expected for rotationally or
dispersion-dominated early-type galaxies. This disfavors gravitational
motion as the main cause of the emission-line kinematics in our
sources. The same is suggested by the higher line widths compared to
mass-selected galaxies in the same mass range with SINFONI data, and
without obvious (radio-loud or radio-quiet) AGN.
\item
We find a number of empirical trends and correlations between the gas
and jet properties. These include (1) good alignment between the gas and
radio jet axis to within less than about $20^\circ$ for most
sources. (2) The size of the emission line regions, which is always
less than the size of the radio jets, except in a few sources where
the jet size is smaller than the typical size of high-redshift
galaxies. (3) Aspect ratios of the emission-line regions and jet size
scale broadly as expected from hydrodynamic models of light jets with
high density contrast. 
(4) H$\alpha$ and [OIII] morphologies and
line profiles are very similar. (5) Kinetic energy of the gas and the
radio source are strongly correlated, where the gas kinetic energy
corresponds to about $10^{-2}$ to $10^{-3}$ of that of the radio jet.
\item
At all jet kinetic energies we find sources where the kinetic energy
derived from the line widths (unordered motion) exceeds that from
global, ordered, bulk flow (ordered motion). While it is not clear
whether the unordered motion corresponds to turbulence in the strict
physical sense, or is dominated by unresolved, local outflows, we do
see a discontinuity at jet kinetic energies of around $1\times
10^{47}$ erg s$^{-1}$, where the number of galaxies with higher
kinetic energy in bulk flow than unordered motion starts to
dominate. We suggest that this could be a transition between galaxies
where the ISM is being stirred up by the radio jet, but where
interactions are only local, and galaxies, where the energy injection
from the radio jet is so strong that the global ISM in the host is
being disrupted.
\item
Using the recent redshift-dependent diagnostic diagrams of
\citet{kewley13a}, we find that the ratios of bright optical emission
lines, [NII]$\lambda$6583,H$\alpha$,[OIII]$\lambda$5007, and H$\beta$,
are consistent with expectations of cosmological models regarding the
global metallicity evolution of galaxies. Gas-phase metallicities in
our sources are about $1-2\times$ Z$_{\odot}$, similar to the
luminosity-weighted metallicities of the stellar populations in
massive galaxies at low redshift. 
This is expected if the outflowing gas in our sample galaxies
represented the same ISM from which these stars formed.
\item
Overall, we find a good global correspondence between the observed velocities and
velocities predicted by recent hydrodynamic models. The axis ratios of
the emission-line regions correspond roughly to the density constrast
between jet plasma and ambient medium expected for light jets at high
redshift.
\end{itemize}

Generally speaking, our observations are in agreement with
expectations from hydrodynamic models, however, we also find that the
efficiencies of the kinetic energy transfer from the jets to the gas,
and the velocities do differ. The discontinuity in the ratio between
the kinetic energy from global bulk flow and localized unordered
(perhaps turbulent) motion has so far not been observed, and deserves
further scrutiny. Our current results suggest that it is the outcome
of the very high energy injection rates from the most powerful radio
jets, which might be sufficient to disrupt the global interstellar
medium of the most massive high-redshift galaxies. This means also
that the interplay between bulk flow and turbulence is an important,
so far rather neglected, aspect of AGN feedback at high
redshift. Taken together, all of this suggests that AGN feedback is a
rich, complex phenomenon, that we are only starting to understand.

\section*{Acknowledgments}
We would like to thank the staff at Paranal observatory for having
carried out the observations, and the ESO OPC for the extraordinary
allocation of observing time. We thank the referee for insightful
comments, which helped improve the paper. We also thank I.~Tanaka for
interesting discussions and sharing some of his results on 3C~23.56,
and Zelenkova and Parijskij for sharing their radio map of
RC~0311$+$0507 with us. NPHN wishes to thank C. Carilli for
interesting discussions and for sharing his VLA data of parts of our
sample. She would also like to thank G. Bicknell, A. Wagner, M. Krause
and V. Gaibler for enlightening discussions about radio jets, and
L.~Kewley for equally interesting discussions about metallicities.  CC
wishes to acknowledge support from the Ecole Doctorale Astronomie \&
Astrophysique de l'Ile de France.
\bibliographystyle{aa}
\bibliography{hzrg}

\onecolumn
\begin{landscape}
\begin{table*}
\centering
\begin{tabular}{lccccccccccc}
\hline 
Source           & L([OIII])         & L(H$\alpha$)    & L(H$\beta$)            & $A_V$ &L(H$\alpha$,c)       & L(H$\beta$,c)  & M$_{\rm WIM}$   & M$_{\rm CO}$  & $M_{\rm WIM}/M_{\rm CO}$ & Reference   \\
             & [$10^9$ L$_{\odot}$] & [$10^9$ L$_{\odot}$] & [$10^9$ L$_{\odot}$]     & [mag] & [$10^9$ L$_{\odot}$]  & [$10^9$ L$_{\odot}$] & [$10^9$ M$_{\odot}$] &  [$10^{10}$ M$_{\odot}$] &  [M$_{\odot}$] & \\
\hline
MRC~0114$-$211   & 7.3$\pm$0.1    & 15.2$\pm$0.2   & 0.9$\pm$ 0.1 & 4.65$\pm$0.22 & 197.1$\pm$21.7 & 68.4$\pm$8   & 4.9$\pm$0.5 & 3.6$\pm$0.7 & 0.14$\pm$0.03   & E14   \\
BRL~0128$-$264   & 38$\pm$1.2     & 13.4$\pm$0.8   & 5.4 $\pm$2.2 & 0.0    & 13.4$\pm$0.8  & 5.4$\pm$2.2    & 0.3$\pm$0.02 & \dots       & \dots  & \dots \\
MRC~0156$-$252   & 161 $\pm$4     & 36.6$\pm$1.1   & 9.7$\pm$5.2  & 0.73$\pm$0.06 & 54.8$\pm$12.6  & 19.0$\pm$4.4 & 1.4$\pm$0.3  & 7.4$\pm$1.5 & 0.02$\pm$0.006   & E14   \\
USS~0211$-$122   & 67$\pm$0.4     & 21.9$\pm$0.2   & 6.2$\pm$0.8  & 0.55$\pm$0.13 & 29.7$\pm$0.3  & 10.3$\pm$0.6 & 0.8$\pm$0.05     & 3.7         & 0.02   & E14   \\
MRC~0251$-$273    & 91$\pm$2.1     & \dots          & 6.0$\pm$2.3  & \dots  & \dots & \dots  & $>0.4$& \dots     & \dots  & \dots \\
MP~J0340$-$6507  & 98$\pm$1.1     & 35.3$\pm$1.0   & 15.5$\pm$1.2 & 0.0    & \dots & \dots  & 0.9$\pm$0.02    & \dots     &  2.2   & E14   \\
PKS~0529$-$549   & 101.3$\pm$0.4  & 56.5$\pm$1.7   & 13.8 $\pm$0.2& 0.94$\pm$0.013 & 95.3$\pm$1.4  & 33.1   &  2.4$\pm$2.4  & \dots       & \dots  & \dots \\
5C~7.267         & 1.6$\pm$0.5    & 3.0$\pm$0.4    & \ldots       & \ldots & \ldots& \ldots & $>$0.1& \dots & \dots     &  \dots \\
TN~J1112$-$2948  & 35.5$\pm$0.9   & \dots          & 3.2$\pm$0.9  & \dots  & \dots & \dots  & $<0.3$&  \dots     & \dots  & \dots \\
TXS~1113$-$178   & 24.6$\pm$0.3   & 8.23$\pm$0.4   & 2.4$\pm$0.3  & 0.5$\pm$0.06   & 10.8$\pm$0.6  & 3.8$\pm$0.2    & 0.3$\pm$0.02     & \dots       & \dots  & \dots \\
3C~257           & 111$\pm$1.3    & 44.8$\pm$0.1   & 12.5$\pm$1.4 & 0.6$\pm$0.03   & 62.0$\pm$8.3  & 21.5$\pm$2.9   & 1.5$\pm$0.2   & \dots       & \dots  & \dots \\
USS~1243$+$036   & 87.4$\pm$1.5   & \dots          & 8.53$\pm$2.1 & \dots  & \dots & \dots  & $<0.6$& \dots       & \dots  & \dots \\
MG~1251$+$1104   & 118$\pm$7.6    & 62.8$\pm$9.7   & \dots        & \dots  & \dots & \dots  & 1.6$\pm$0.2    & \dots       & \dots  & \dots \\
MRC~1324$-$262   & 40.2$\pm$2.2   & 15.6$\pm$1.8   & 5.2 $\pm$2.5 & 0.1$\pm$0.2 & 16.6$\pm$8  & 5.8$\pm$2.8 & 0.4$\pm$0.2    & \dots       & \dots  & \dots \\
USS~1410$-$001   & 163$\pm$1.4    & 52.0$\pm$0.9   & 10.2$\pm$1.2 & 1.5$\pm$0.05   & 121.5$\pm$12 & 42.2$\pm$4.2   & 3.$\pm$0.3    & \dots       & \dots  & \dots \\
MRC~1558$-$003   & 104$\pm$1.6    & 46.6$\pm$1.6   & 11.4$\pm$1.3 & 0.9$\pm$0.05   & 78.4$\pm$9  & 27.2$\pm$3 & 2.$\pm$0.2    & \dots       & \dots  & \dots \\
USS~1707$+$105   & 26.4$\pm$0.6   & 7.8$\pm$0.6    & 1.6 $\pm$0.4 & 1.4$\pm$0.16   & 17.1$\pm$2  & 5.9$\pm$0.7 & 0.4$\pm$0.05    & \dots       & \dots  &       \\
3C~362           & 80.4$\pm$0.9   & 34.2$\pm$0.9   & 3.2 $\pm$1.0 & 3.5$\pm$0.3    & 239.8$\pm$150 & 83.2$\pm$53   & 6.0$\pm$4.0    & \dots       & \dots  & \dots \\
MP~1758$-$6738   & 28.6$\pm$0.6   & 9.5$\pm$1.1    & \dots        & \dots  & \dots & \dots  & 0.2$\pm$0.02   & \dots       & \dots  & \dots \\
TN~J2007$-$1316  & 105$\pm$1.8    & \dots          & 20.0 $\pm$1.8& \dots  & \dots & \dots  & 1.5$\pm$0.2   &  \dots       & \dots  & \dots \\
MRC~2025$-$218   & 116 $\pm$1.5   & 25.1$\pm$3.2   & \dots        & \dots  & \dots & \dots  & 0.6$\pm$0.08   & 3.8$\pm$1.0 & 0.015$\pm$0.002  & E14   \\
MG~2037$-$0011   & 1.9 $\pm$0.2   & 2.0 $\pm$0.2   & \dots        & \dots  & \dots & \dots  & 0.05$\pm$0.005 &  \dots       & \dots  & \dots \\
MRC~2104$-$242   & 48.5$\pm$0.5   & 28.3$\pm$0.6   & 3.3$\pm$0.4  & 2.9$\pm$0.05   & 140.5$\pm$16.9 & 48.8$\pm$6 & 3.5$\pm$0.06      & $<$2        & $\ge$0.2 & E11 \\
4C~23.56         & 133 $\pm$4.7   & 36.7$\pm$2.7   & 10.3$\pm$4.7 & 0.6$\pm$0.2    & 50.4$\pm$22.7  & 17.5$\pm$7.9   & 1.3$\pm$0.4   & \dots       & \dots  & \dots \\
NVSS~J2135$-$3337& \dots          & 4.2$\pm$0.2    & \dots        & \dots  & \dots & \dots  & 0.1$\pm$0.005   & \dots       & \dots  & \dots \\
MG~2144$+$1928   & 96.2 $\pm$3.1  & \dots          & 5.4$\pm$0.8  & \dots  & \dots & \dots  & 0.4$\pm$0.06   &  \dots       & \dots  & \dots \\
MRC~2224$-$273   & \dots          & 31.4 $\pm$0.8  & \dots        & \dots  & \dots & \dots  & 0.4$\pm$0.01   &  $<$2        & $\ge$0.2 & E14 \\
TN~J2254$+$1857  & 5.0$\pm$0.4    & 5.8$\pm$0.5    & \dots        & \dots  & \dots & \dots  & 0.15$\pm$0.01  &  \dots       & \dots  & \dots \\
MG~2308$+$0336   & 62.6 $\pm$1.4  & 28.2 $\pm$0.9  & 7.57$\pm$1.2 & 0.7$\pm$0.07   & 41.3$\pm$6.6  & 14.4$\pm$2.3 & 1.0$\pm$0.2   & \dots       & \dots  & \dots \\
\hline
\end{tabular}
\caption{
Line luminosities and warm ionized gas masses for fiducial electron densities of n$=$500 cm$^{-3}$. Error bars include measurement errors only.
\label{tab:gasmasses}}
\end{table*}
\end{landscape}

\appendix
\twocolumn
\section{Individual Objects}
\label{sec:individualobjects}

\subsection{MRC~0114-211} 
\label{ssec:indmrc0114}
MRC0114$-$211 at z$=$1.4 is the lowest-redshift target in our sample. 
 \citet[][see also \citealt{seymour07}]{debreuck10} find a
stellar mass of M$_{\rm stellar}=2.5\times10^{11}$ M$_{\odot}$ and mid-IR
emission dominated by a stellar continuum.  It has a compact steep
radio spectrum in the GHz regime, with two hotspots offset by
$<$0.5\arcsec\ \citep{mantovani94} and with a high rotation measure of
RM$=$646~rad~m$^{-2}$ in the rest-frame, probably a signature of dense
gas surrounding the radio source. \citet{debreuck10} found a faint
radio core, which is resolved only in their 8.46~GHz VLA A-array
imaging. The radio map shown in the right panel of Fig.~\ref{fig:maps}
was smoothed to 0.5\arcsec\ to be more comparable with the SINFONI
data, and therefore shows only a single radio source.

At z$=$1.4, we cover all bright rest-frame optical lines from H$\beta$
to the rest-frame near-infrared lines of
[SIII]$\lambda\lambda$9069,9513. Line emission extends over
1.6\arcsec$\times$1.1\arcsec, not much larger than the seeing disk of
a size of 1.0\arcsec$\times$1.1\arcsec. The deconvolved size along the major
axis, which is roughly aligned east-west, is 1.2\arcsec, corresponding
to 9.7~kpc. We consider this size an upper limit.

All emission lines are adequately fit with single Gaussian
components. We find FWHM=673$\pm$6~km s$^{-1}$ for
[OIII]$\lambda$5007, and use the same width for the other lines as
well. This enables us in particular to decompose H$\alpha$ and
[NII]$\lambda\lambda$6548,6583, which are strongly blended. This
approach does not produce significant fit residuals. The results of
our fits to the integrated spectrum are summarized in
Table~\ref{tab:emlines}.

The line-free continuum image of MRC~0114$-$211 shows two peaks at a
relative projected distance of 1.3\arcsec\ (corresponding to 11~kpc at
z$=$1.41). The fainter peak is centered on the emission-line region,
the brighter peak falls outside.  If not an interloper, then this may
be a nearby, perhaps interacting galaxy.  We do not detect line
emission associated with the bright continuum emitter in any of our
bands. 

\subsection{BLR~0128-264}
\label{ssec:indblr0128} 
\begin{figure*}
\includegraphics[width=0.24\textwidth]{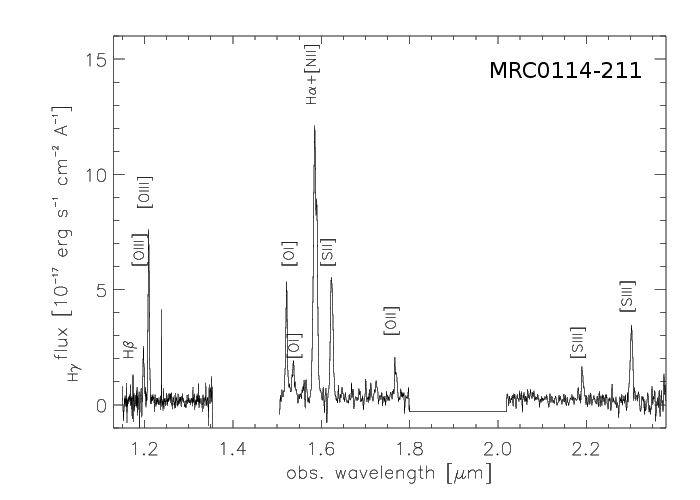}
\includegraphics[width=0.24\textwidth]{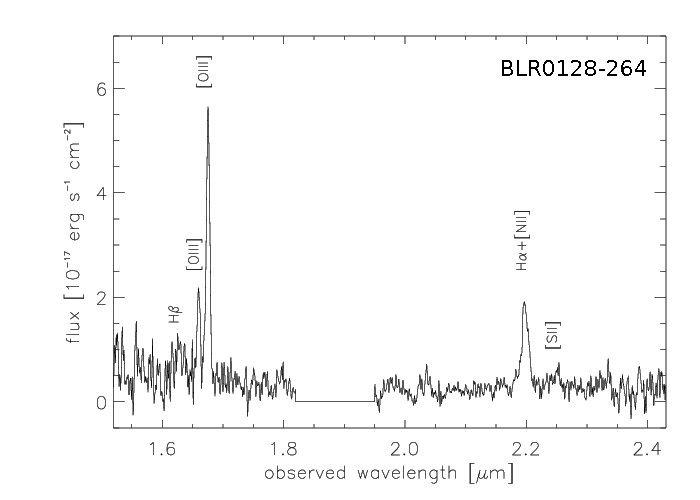}
\includegraphics[width=0.24\textwidth]{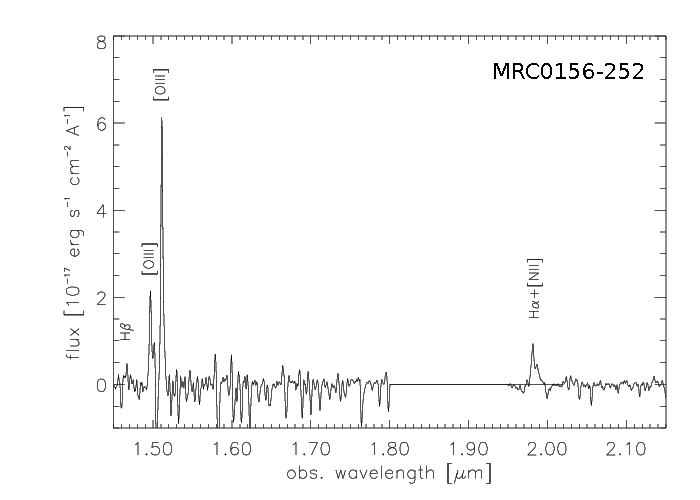}
\includegraphics[width=0.24\textwidth]{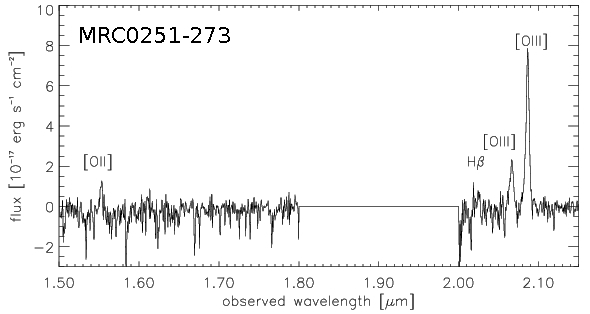}\\
\includegraphics[width=0.24\textwidth]{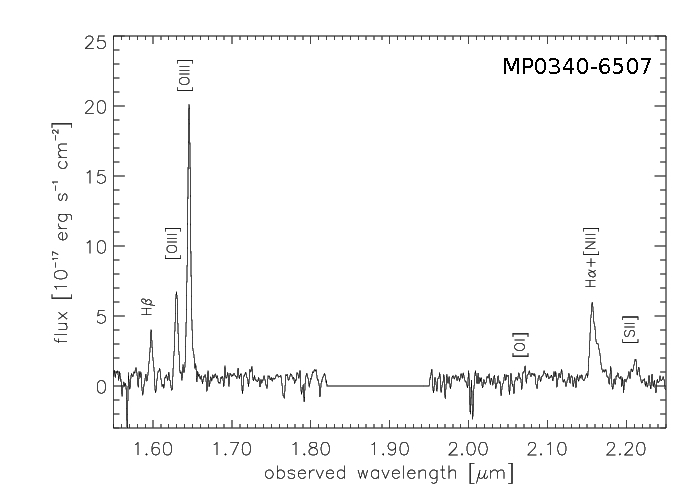}
\includegraphics[width=0.24\textwidth]{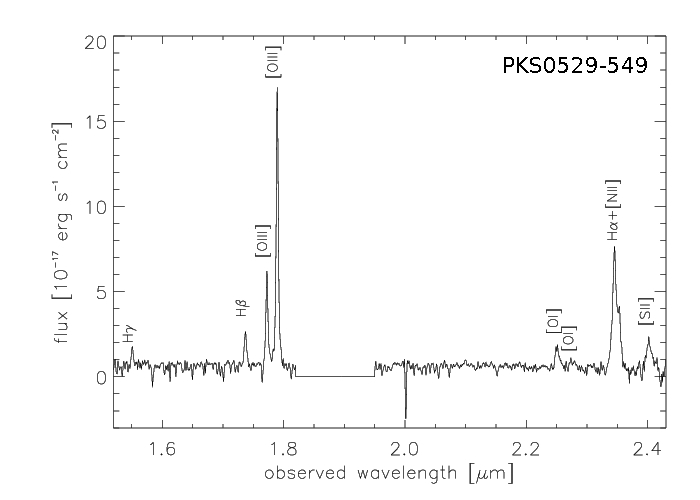}
\includegraphics[width=0.24\textwidth]{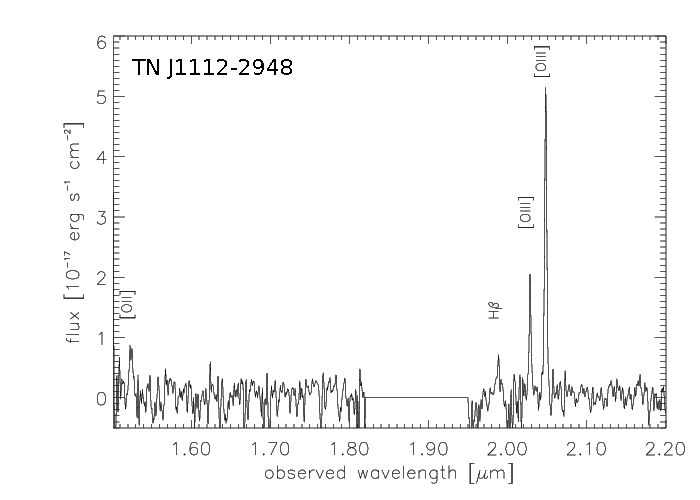}
\includegraphics[width=0.24\textwidth]{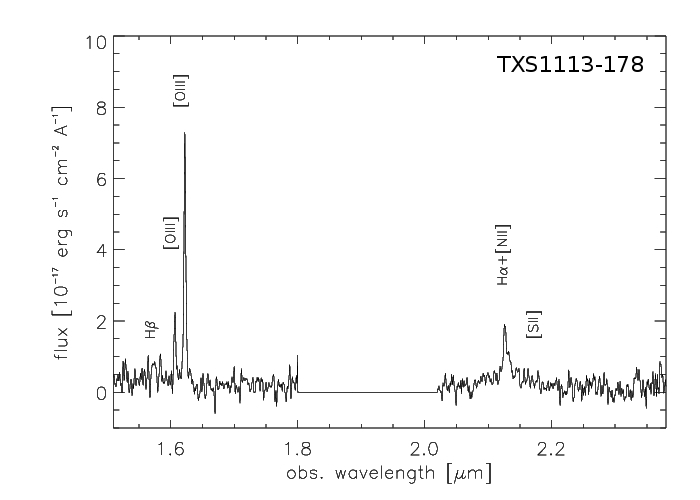}\\
\includegraphics[width=0.24\textwidth]{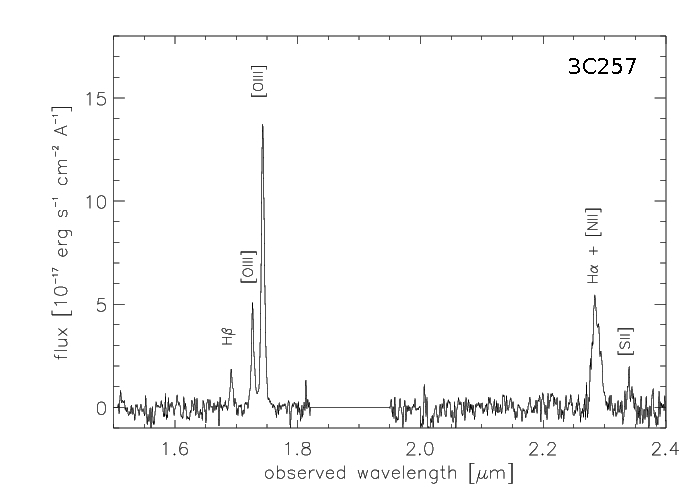}
\includegraphics[width=0.24\textwidth]{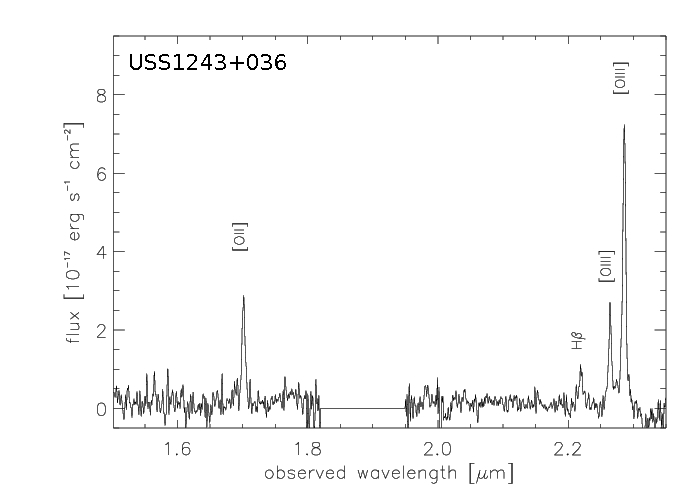}
\includegraphics[width=0.24\textwidth]{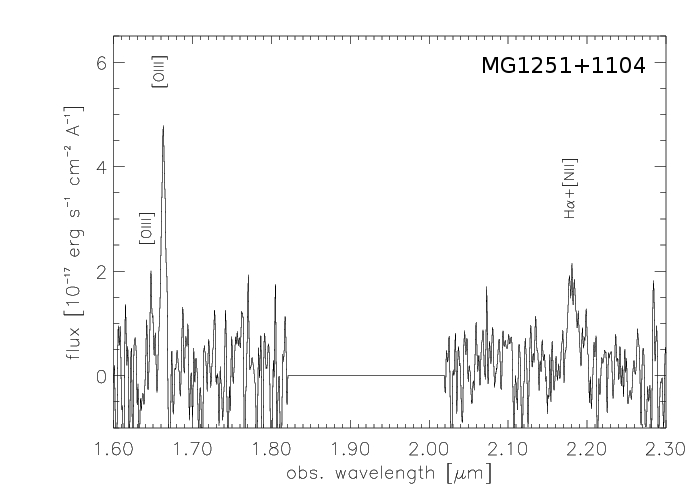}
\includegraphics[width=0.24\textwidth]{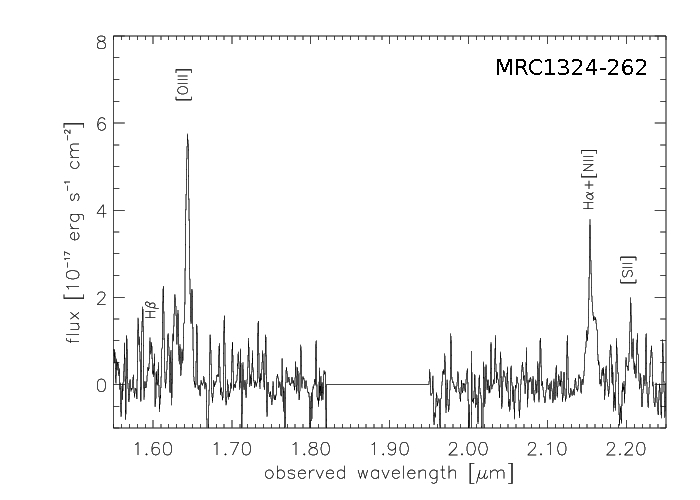}\\
\includegraphics[width=0.24\textwidth]{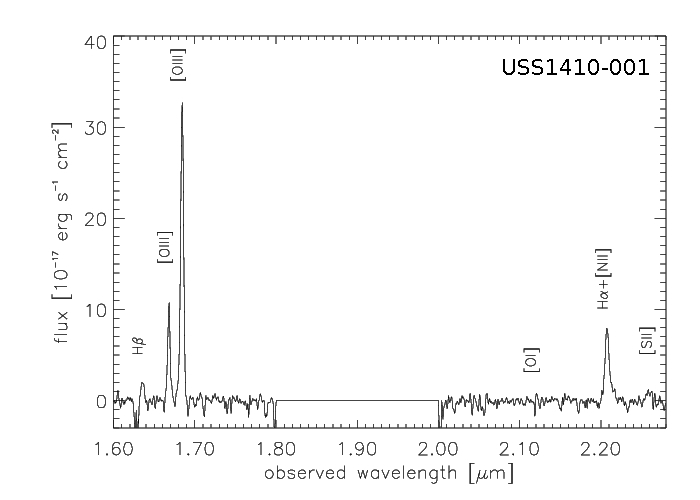}
\includegraphics[width=0.24\textwidth]{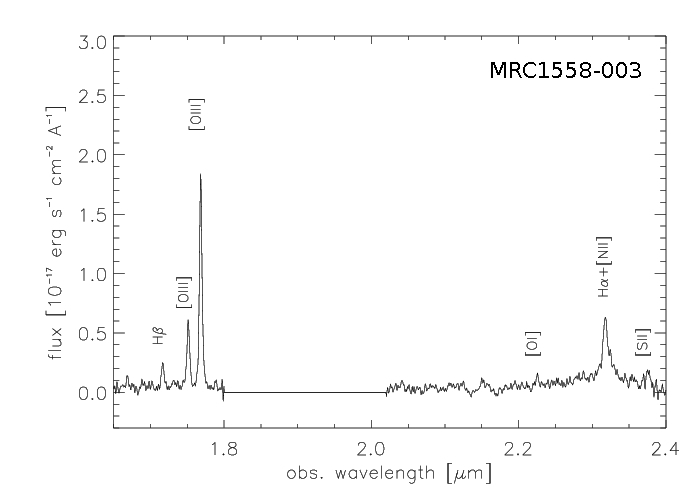}
\includegraphics[width=0.24\textwidth]{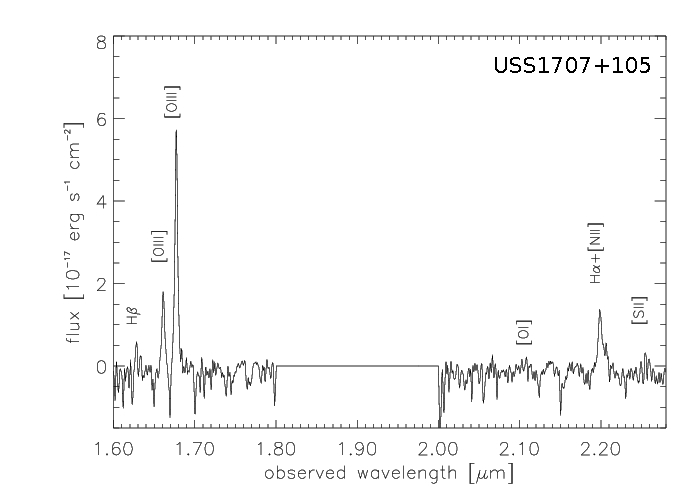}
\includegraphics[width=0.24\textwidth]{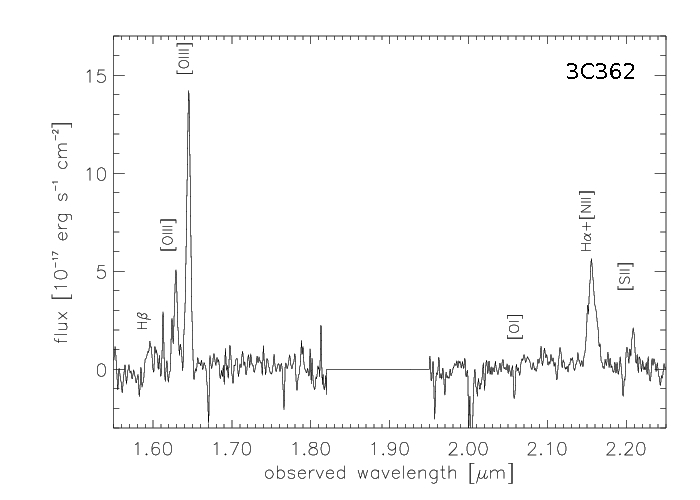}\\
\includegraphics[width=0.24\textwidth]{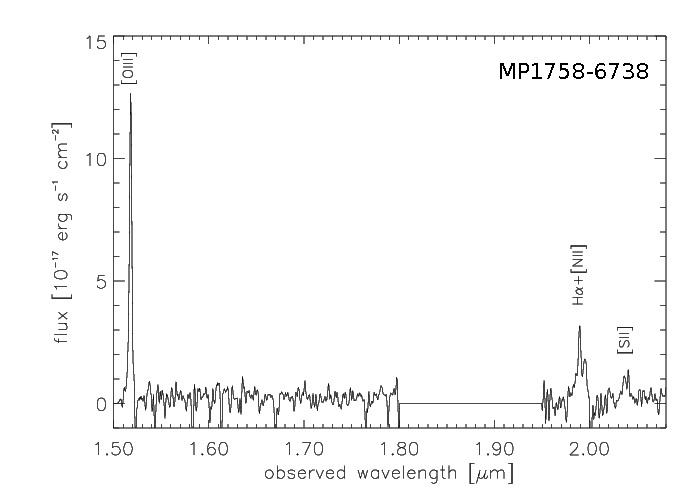}
\includegraphics[width=0.24\textwidth]{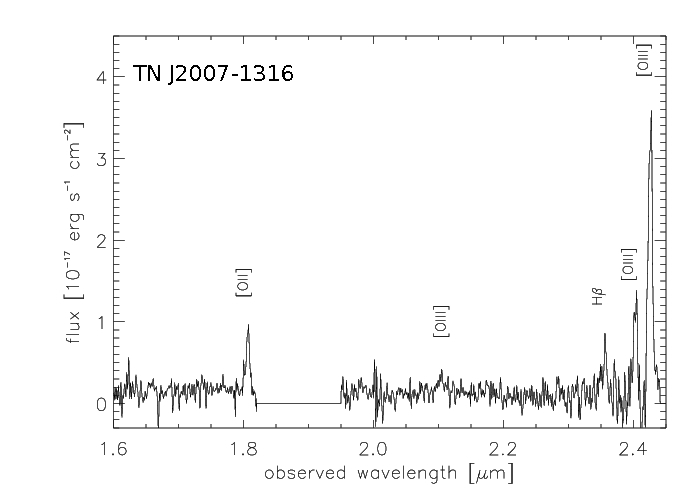}
\includegraphics[width=0.24\textwidth]{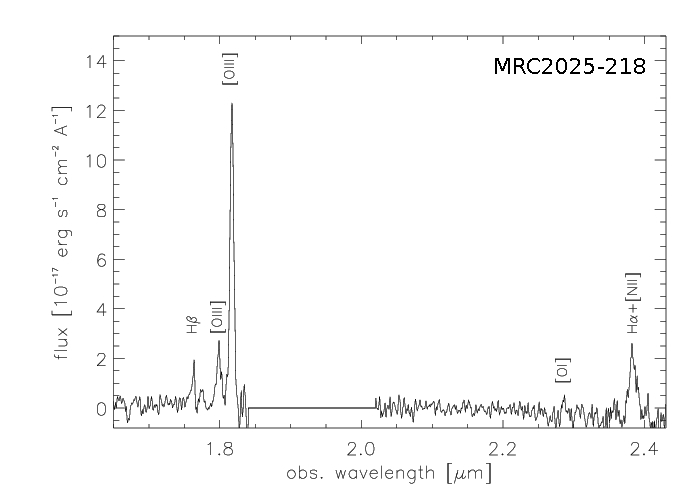}
\includegraphics[width=0.24\textwidth]{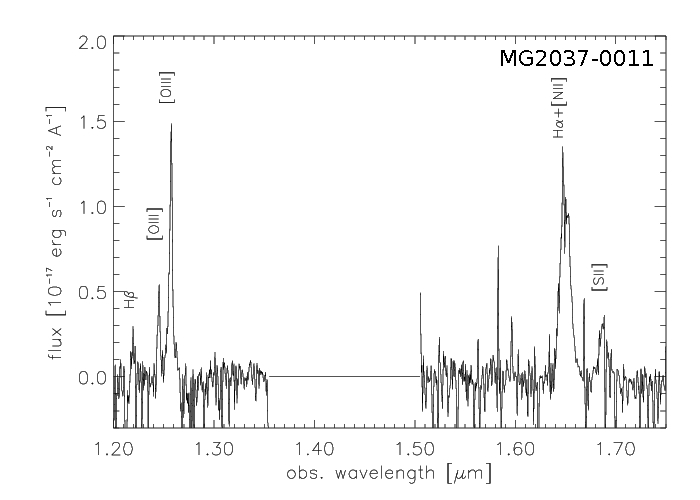}\\
\includegraphics[width=0.24\textwidth]{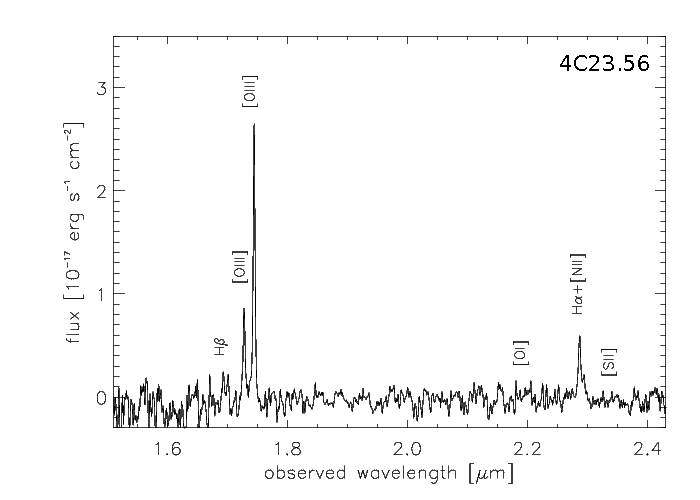}
\includegraphics[width=0.24\textwidth]{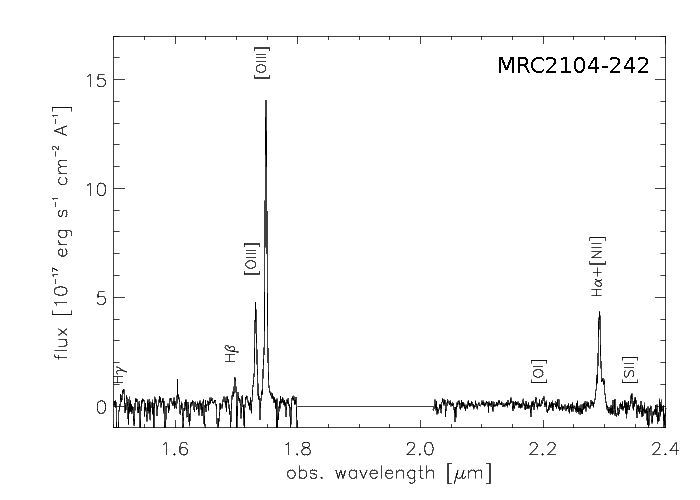}
\includegraphics[width=0.24\textwidth]{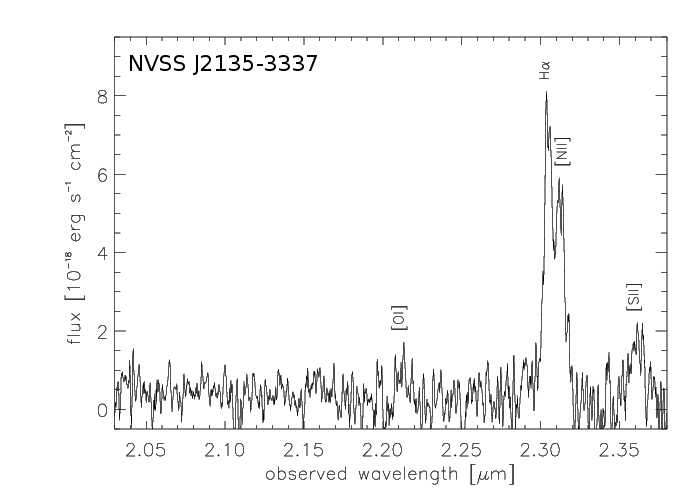}
\includegraphics[width=0.24\textwidth]{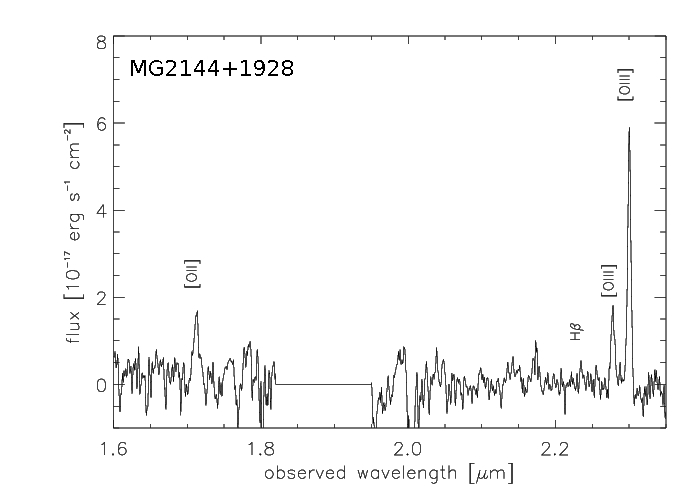}
\includegraphics[width=0.24\textwidth]{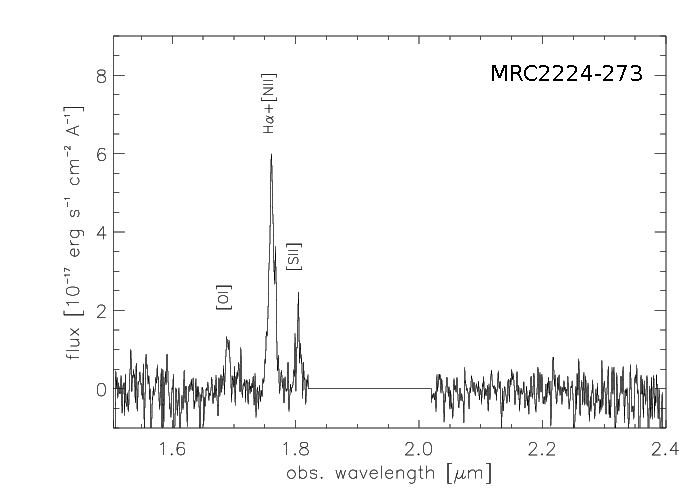}
\includegraphics[width=0.24\textwidth]{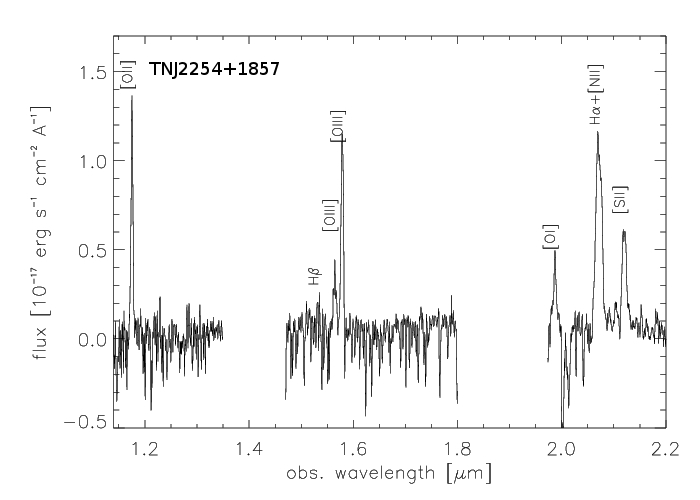}
\includegraphics[width=0.24\textwidth]{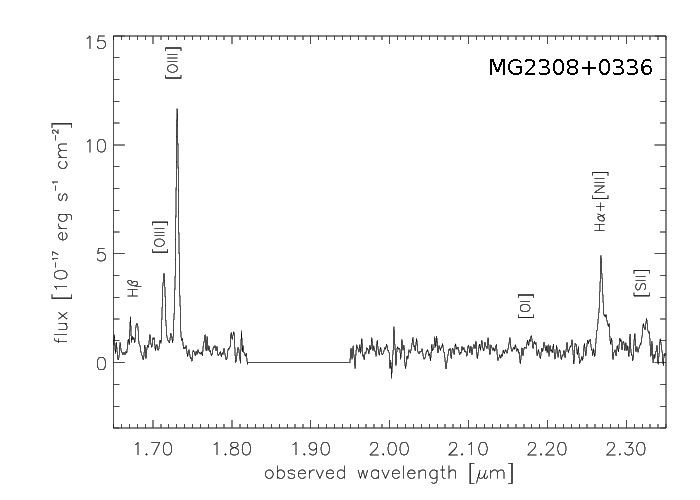}
\caption{
Integrated spectra of our galaxies.}
\label{fig:intspec}
\end{figure*}

BLR~0128$-$264 at z$=$2.35 has very extended radio jets with a
projected size of 35.8\arcsec, corresponding to 294~kpc at z$=$2.35,
and is one of our most powerful radio galaxies, with P$_{\rm 500}=29.1$~W
Hz$^{-1}$ at 500~MHz in the rest-frame. With SINFONI we detect
[OIII]$\lambda\lambda$4959,5007, H$\beta$ and H$\alpha$, as well as
[SII]$\lambda\lambda$6716,6731, and [NII]$\lambda$6583
(Fig.~\ref{fig:intspec}). The lines are well fit with a single
component of FWHM$=$1136$\pm$20 km s$^{-1}$ (the error is that
measured for [OIII]$\lambda$5007). The fitting 
parameters for all lines are given in Table~\ref{tab:emlines}.

Line emission in BLR~0128-264 is spatially resolved but not very
extended with a size of 3.2\arcsec$\times$1.5\arcsec\ along the major
and minor axis, respectively. Compact continuum emission is detected
roughly in the center of the emission line region. The emission line
morphology is fairly regular, albeit with higher [OIII] surface
brightness towards the north (Fig.~\ref{fig:maps}). The velocity field
of BLR~0128$-$264 has a remarkably large gradient of $\Delta$v$\sim$ 1500~
km~s$^{-1}$. Resolved FWHM line widths are between 700 and 2200~km
s$^{-1}$. The largest widths coincide with the continuum position, and
may be partially broadened by the steep and large velocity
gradient. The right panel of Fig.~\ref{fig:maps} shows that the
direction towards the radio hotspot is roughly aligned with the major
axis of the emission-line region and the velocity gradient, in
particular for the north-western jet.

\subsection{MRC~0156-252}
\label{ssec:indmrc0156} 
MRC~0156$-$252 at z$=$2.02 is one of our galaxies with very extended
emission-line gas. Line emission is bright, but given the somewhat
difficult redshift, we only detect [OIII]$\lambda\lambda$4959,5007,
H$\beta$, H$\alpha$, and [NII]$\lambda6583$ in the integrated spectrum
(Fig.~\ref{fig:intspec}). Line properties are listed in
Table~\ref{tab:emlines}.  We probe [OIII]$\lambda$5007 over an area of
69~kpc$\times$23~kpc along the major and minor axis, respectively. The
emission-line region corresponds to a bright inner region, which
extends over 1.5\arcsec (13~kpc), and two extended emission-line
plumes, with the larger plume to the south. The continuum is compact
and roughly centered on the central part of the emission-line
region. MRC~0156-252 offers a direct line of sight into the AGN, as
seen through nuclear H$\alpha$ broad-line emission
\citep[][]{nesvadba11a}. It is therefore possible that parts of the
emission-line morphology in the central region are influenced by the
wings of the unresolved nuclear point source, although we removed this
component from the data cube before fitting the residual line emission
(\S\ref{ssec:blrremoval}). We suspect this is the case in particular
for the asymmetric [OIII]$\lambda$5007 surface-brightness
distribution. Although forbidden lines like [OIII]$\lambda$5007 do not
probe the broad-line region itself, they may still show a near-nuclear
component from the inner narrow-line region, where we expect a boost
of bright high-ionization lines like [OIII]$\lambda$5007. The extended
lobes are too far away from the nucleus to be possibly affected by
residuals of a nuclear point source.

MRC~0156$-$252 has a regular velocity gradient throughout the central
regions and the extended lobes with a total velocity offset of 1600~km
s$^{-1}$, one of the largests in our sample. The line widths are
broadest near the center, with up to FWHM$\sim$900 km s$^{-1}$, and
are between 600 and 800~km s$^{-1}$ in the two
lobes. \citet{pentericci01} found that the brightest Ly$\alpha$
emission in MRC~0156$-$252 is associated with the north-western radio
lobe, and suggested that the jet may be deflected off a dense gas
cloud at that position. \citet{emonts14} detected CO(1-0) emission at
the same position. 

Other examples, like MRC~1138$-$262, where the brightest Ly$\alpha$
emission is also found associated with a gas cloud that is offset from
the radio galaxy itself \citep[][]{kurk04}, also show large velocity
offsets relative to the ambient gas
\citep[][]{nesvadba06a,kuiper11}. This is not the case here, where the
gas near the north-western lobe appears very extended, with a smooth
velocity gradient of 800 km s$^{-1}$, which is similar to that in the
south-west of the galaxy. This is not what would be expected from a
gas cloud within the surrounding dark-matter environment of
MRC~0156$-$252, but rather ambient gas that has been swept up by the
radio jet cocoon.

The radio morphology shown in the right panels of Fig.~\ref{fig:maps}
shows that the northern rim of the emission-line region aligns very
well with the extended radio structure. The south-western radio
hotspot is unresolved, fainter, and falls outside the emission-line
gas. The line widths are high in the south-western emission-line
region, about 650 km s$^{-1}$, although less so than in the
north-eastern region, where they reach $\ge$800 km s$^{-1}$.

\subsection{USS~0211-122} 
\label{ssec:induss0211}

\begin{figure}
\centering
\includegraphics[width=0.45\textwidth]{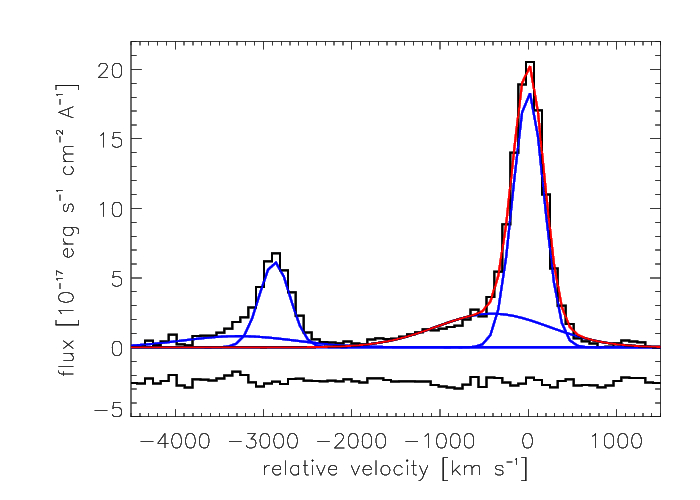}
\caption{Line profile of [OIII]$\lambda\lambda$4959,5007 in
  USS~0211$-$122 and best fit. The lower spectrum shows the fit
  residuals. 
}
\label{fig:0211profilefit}
\end{figure}

USS~0211$-$122 at z$=$2.34 has a moderately extended radio source with a
size of 16\arcsec\ (133~kpc at z$=$2.34) and a moderate radio power
(for our sample) of 28.5~W~Hz$^{-1}$ at 500~MHz in the rest-frame. Its
line emission is however unusual compared to the other sources in our
sample. All identified emission lines
([OIII]$\lambda\lambda$4959,5007, H$\beta$, H$\alpha$,
[NII]$\lambda$6583, [SII]$\lambda\lambda$6716,6731) are very narrow
with FWHM$=$275~km~s$^{-1}$. We do see broad blue wings, as shown in
Fig.~\ref{fig:0211profilefit} for the integrated spectrum. When
fitting the data cube with two Gaussian components, we find that the
broader component is not spatially resolved.  The wing is well fit with
a Gaussian profile with FWHM$=$1400~km s$^{-1}$ and a blueshift of
 $-418$~km s$^{-1}$ from the narrow line component, consistent with a
wind originating from the central regions of this galaxy. The redshift and
width of the broad component are fixed to the best-fit parameters of
the integrated spectrum. 

The maps of the narrow component show gas extended over
2.3\arcsec$\times$1.1\arcsec\ (corresponding to a deconvolved size of
19~kpc$\times$9 kpc) and elongated along an axis that goes from
south-east to north-west, roughly aligned with the radio jet axis. The
velocity field shows a fairly regular gradient of 120~km s$^{-1}$,
although the lowest velocities are found in the nothern part of hte
western bubble, and velocities in the outermost parts of the western
bubble are comparable to the highest velocities in the eastern
emission-line regions. Line widths are very different on the two sides
of the nucleus. Towards the east, they have FWHM$\sim300$ km s$^{-1}$,
compared to FWHM$=400-500$ km s$^{-1}$ on the western side, with the
largest values reached in the most peripheral gas.

\begin{figure}
\centering
\includegraphics[width=0.45\textwidth]{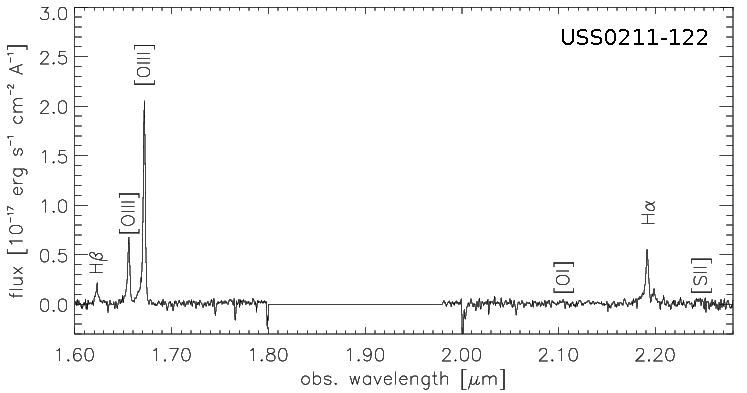}
\caption{Integrated spectrum of USS0211$-$122.
}
\label{fig:0211intspec}
\end{figure}

\subsection{MRC~0251-273}
\label{ssec:indmrc0251} 

MRC~0251$-$273 at z$=$3.17 has a fairly compact radio source with an
upper limit of LAS$<$3.9\arcsec\ \citep{debreuck10} and a FIR
luminosity $<1\times10^{13}$ L$_{\odot}$ measured with SCUBA
\citep{reuland04}. It is undetected with SPIRE with an upper limit of
$L^{IR}_{\rm tot}=7.6\times 10^{12}$ L$_{\odot}$ of infrared luminosity
\citep[][]{drouart14}.

\begin{figure*}
\includegraphics[width=0.48\textwidth]{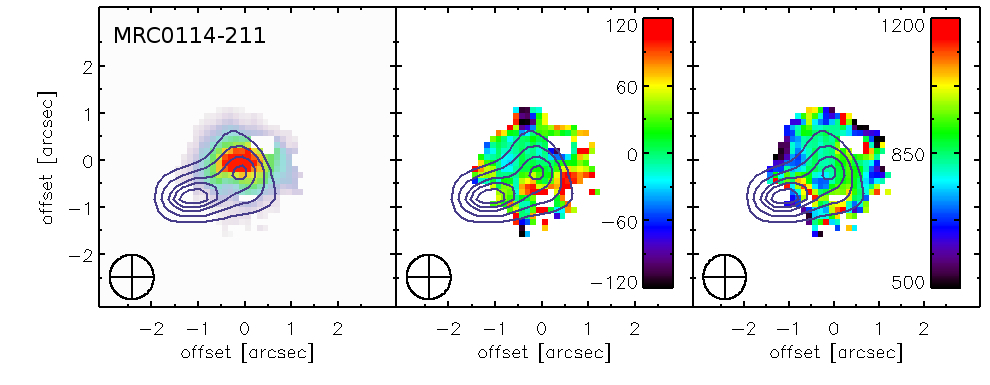}  \includegraphics[width=0.48\textwidth]{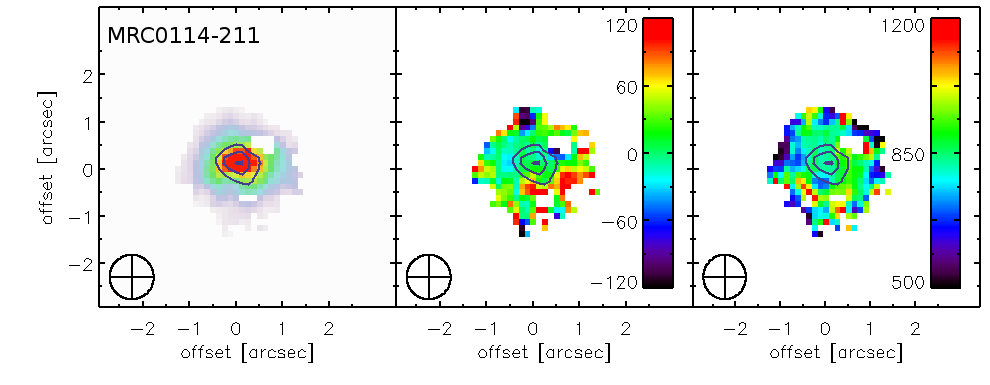}\\
\includegraphics[width=0.48\textwidth]{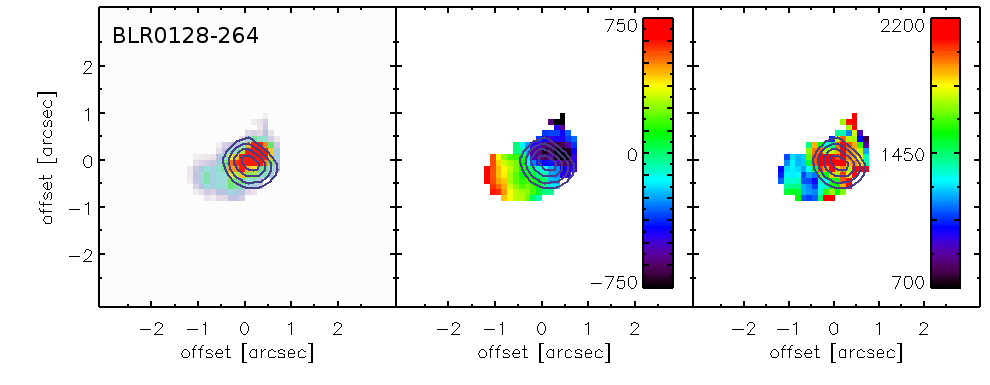}  \includegraphics[width=0.48\textwidth]{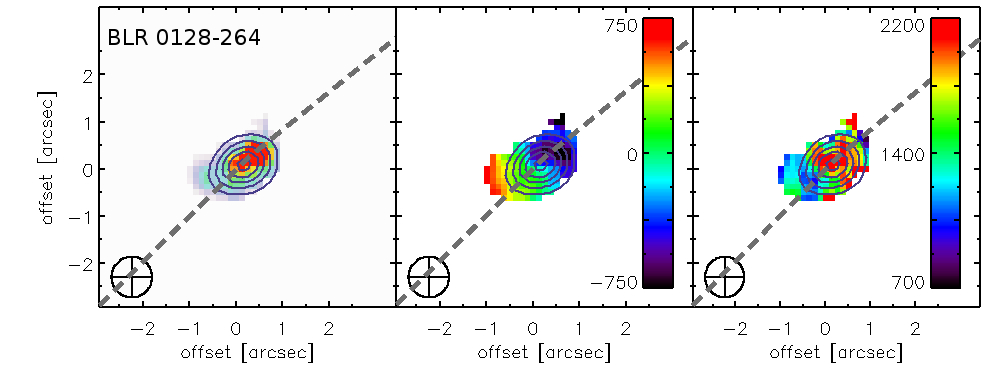}\\
\includegraphics[width=0.48\textwidth]{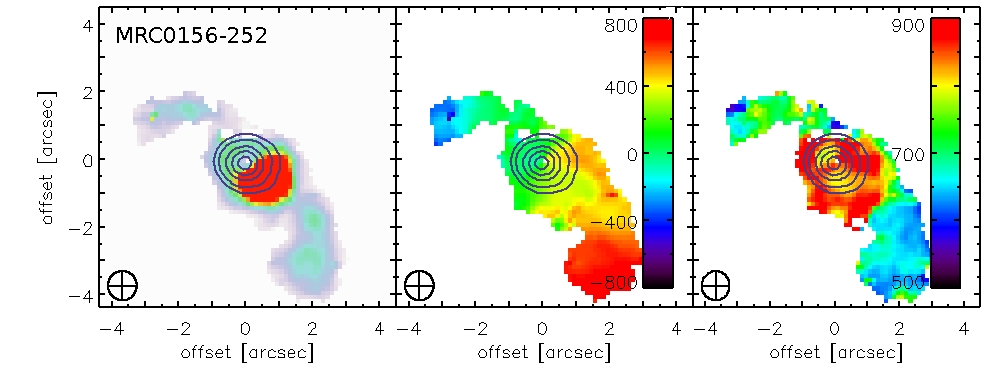}  \includegraphics[width=0.48\textwidth]{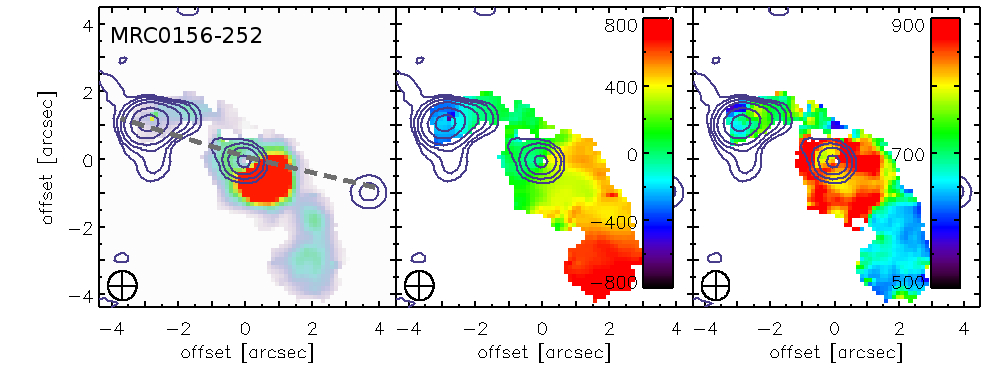}\\
\includegraphics[width=0.48\textwidth]{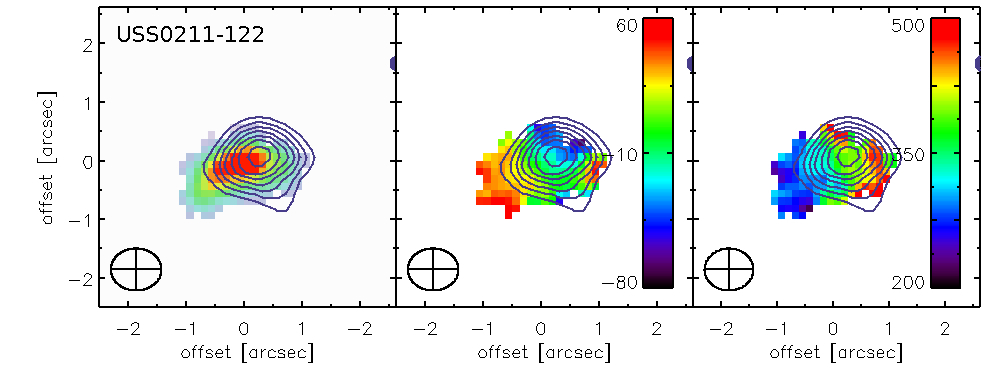}  \includegraphics[width=0.48\textwidth]{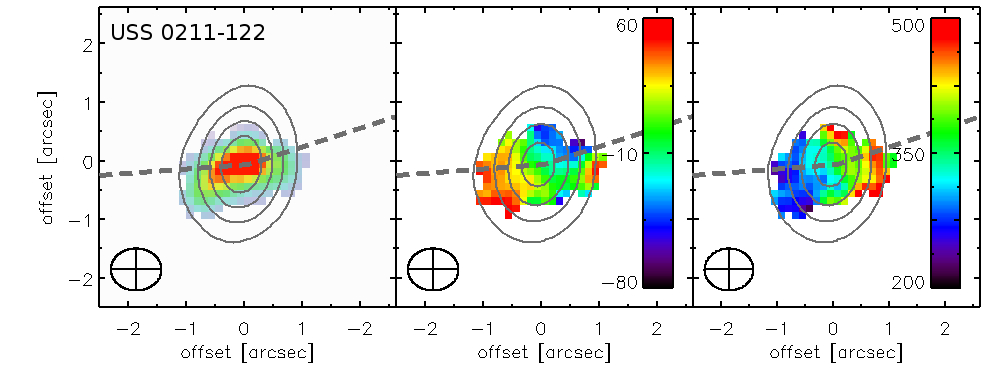}\\
\includegraphics[width=0.48\textwidth]{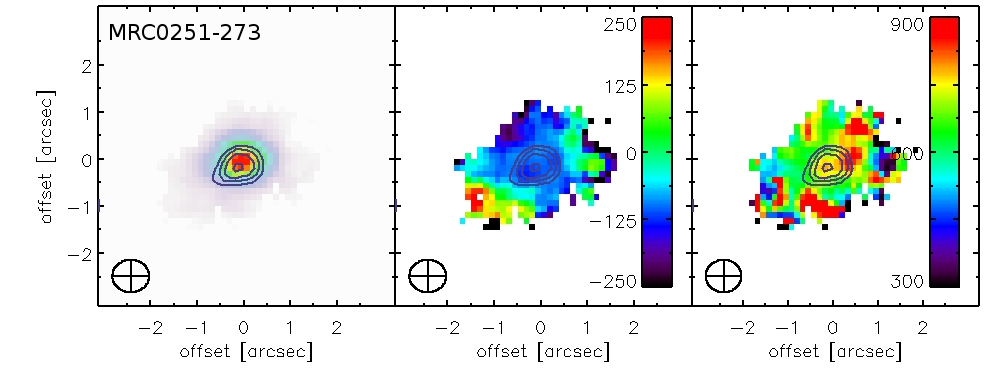}  \includegraphics[width=0.48\textwidth]{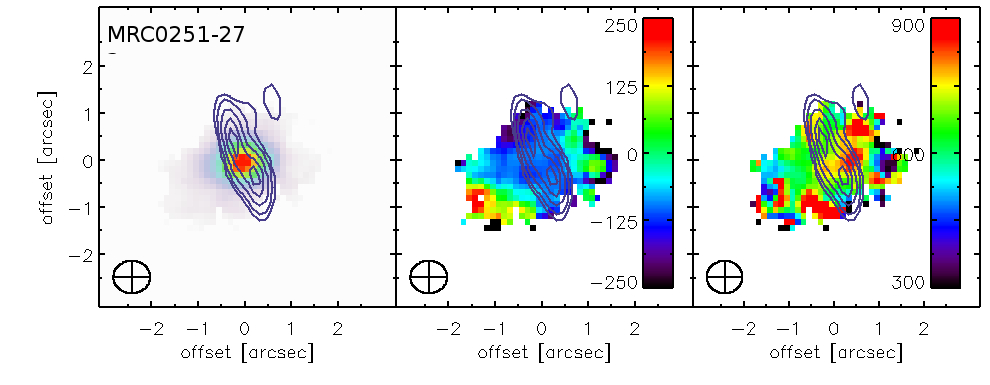}\\
\includegraphics[width=0.48\textwidth]{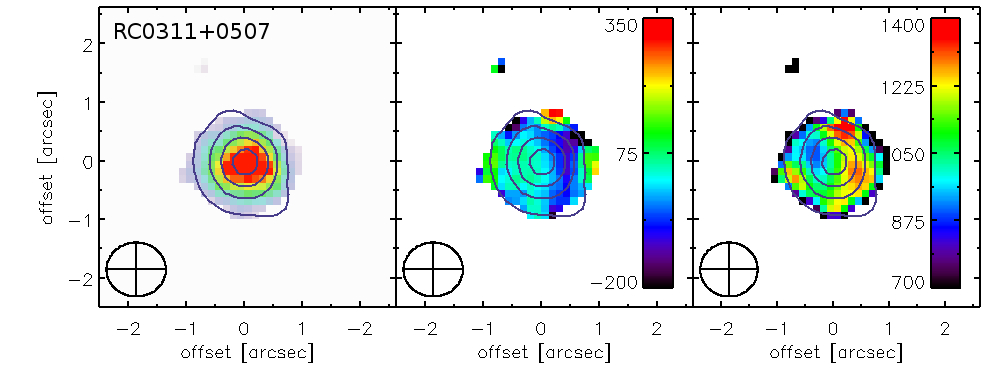}   \includegraphics[width=0.48\textwidth]{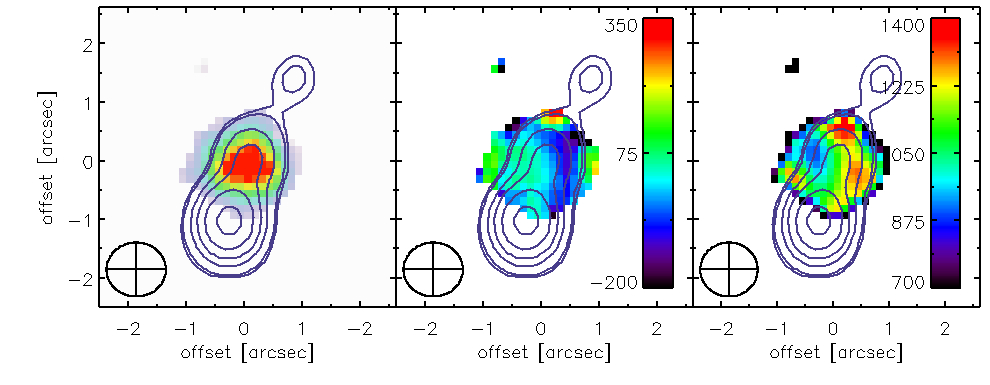}\\
\includegraphics[width=0.48\textwidth]{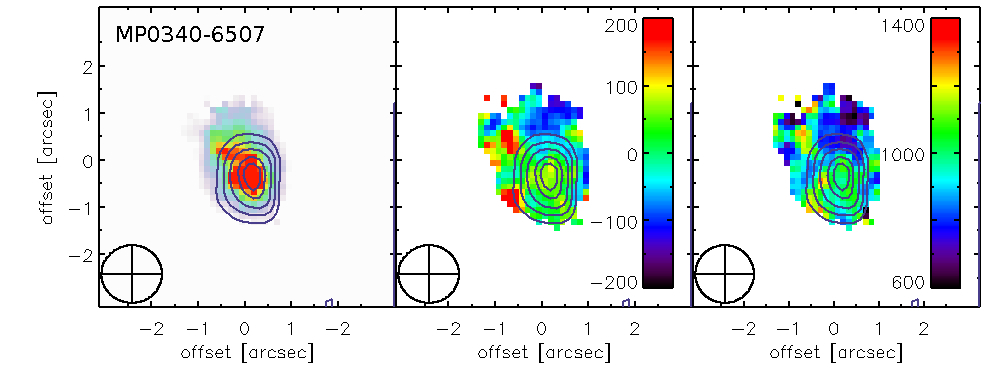} \\
\caption{
Maps of our galaxies in order of RA, with the
  rest-frame optical continuum shown as contours in the left, and the
  GHz radio continuum in the right panel. In each panel, maps show
  (from {\it left to right}) Emission-line morphology, velocities
  relatve to the average redshift of all pixels covered by the
  emission-line region, and FWHM line width. Relative velocities and
  line widths are given in km s$^{-1}$. In galaxies where we did not
  detect the rest-frame optical continuum, we show the emission-line
  morphology instead of the optical continuum (with red contours, the
  continuum is shown as blue contours). Coordinates are given relative
  to the position listed in Table~\ref{tab:sample}. Contour levels are
  arbitrary, their main purpose is to guide the eye.}
\label{fig:maps}
\end{figure*}

\begin{figure*}
\includegraphics[width=0.48\textwidth]{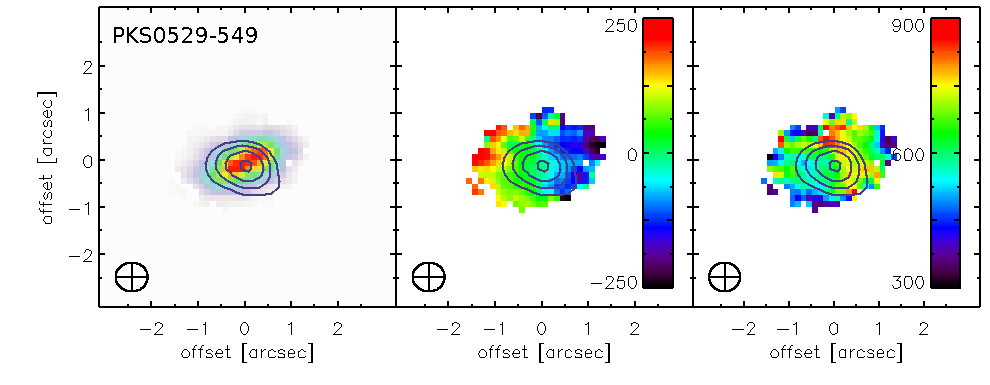}  \includegraphics[width=0.48\textwidth]{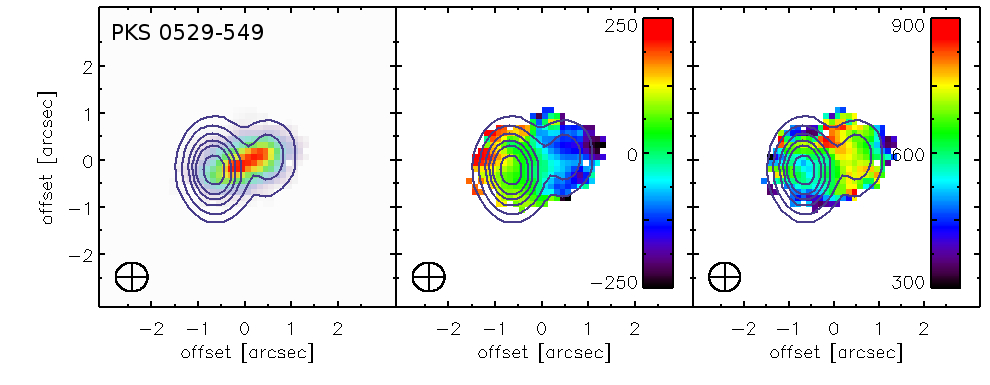}\\
\includegraphics[width=0.48\textwidth]{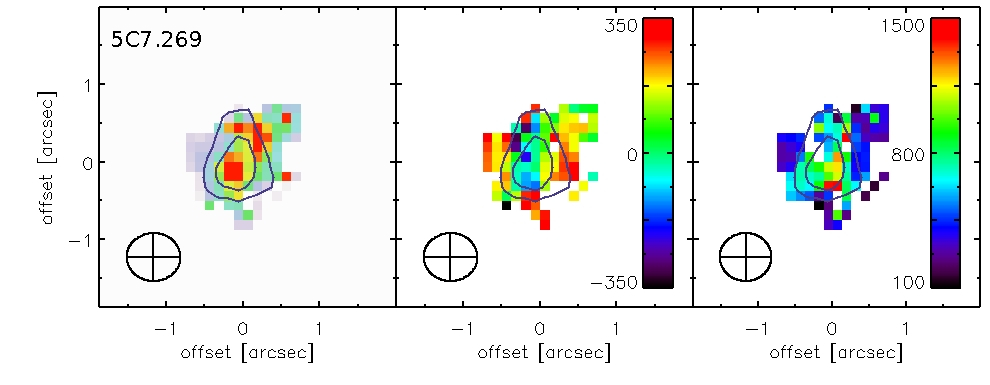} \includegraphics[width=0.48\textwidth]{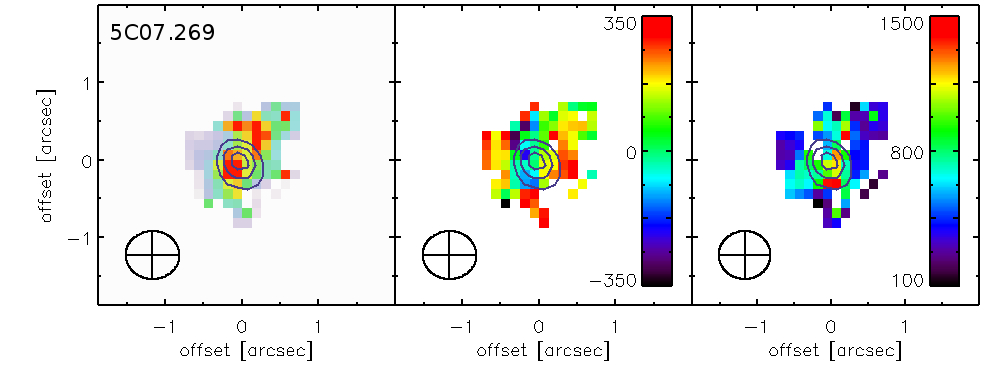}\\
\includegraphics[width=0.48\textwidth]{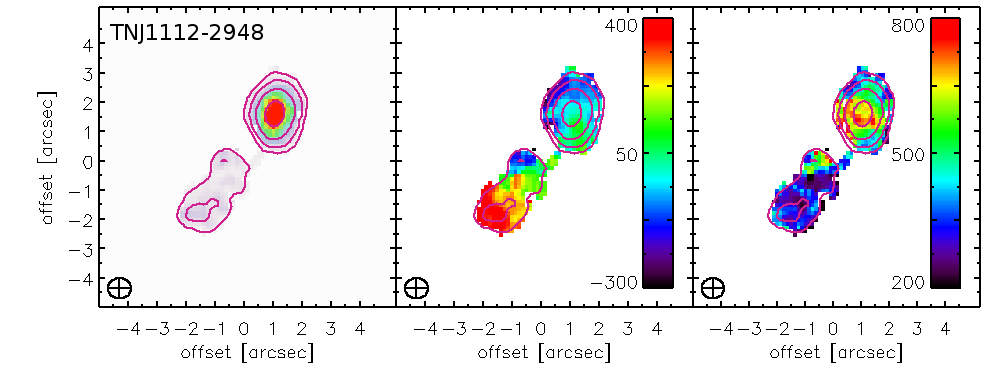}  \includegraphics[width=0.48\textwidth]{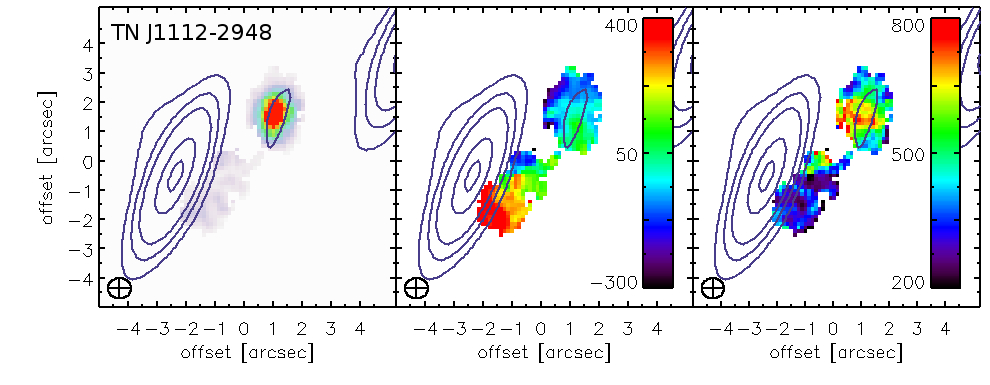}\\
\includegraphics[width=0.48\textwidth]{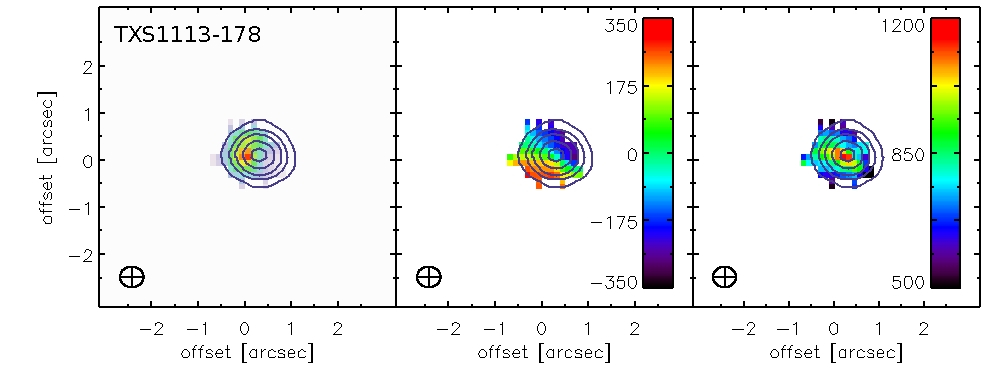}  \includegraphics[width=0.48\textwidth]{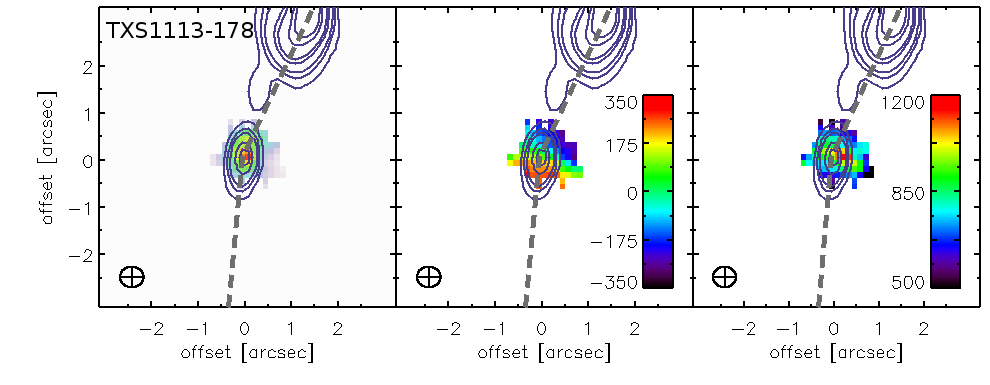}\\
\includegraphics[width=0.48\textwidth]{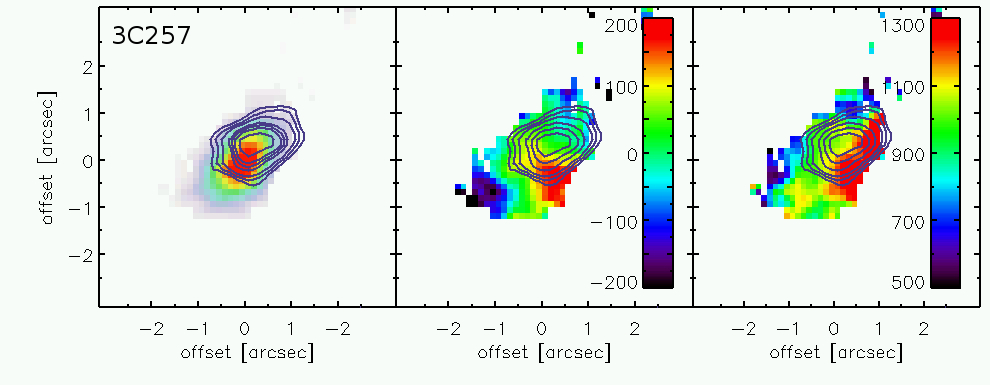}    \includegraphics[width=0.48\textwidth]{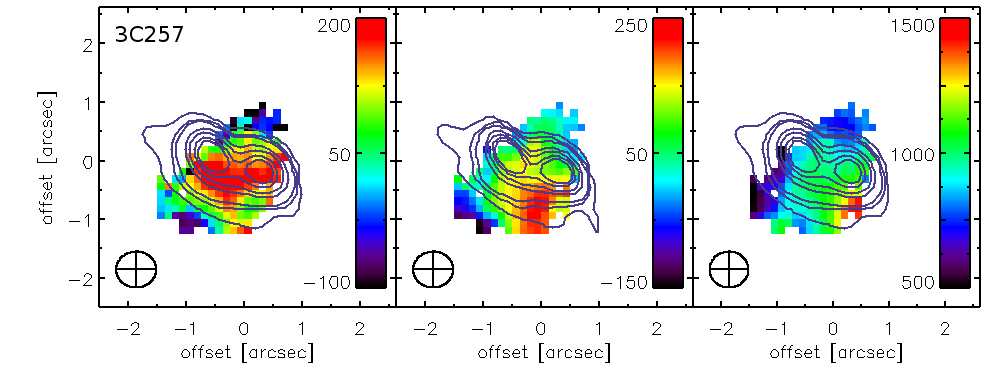}\\
\includegraphics[width=0.48\textwidth]{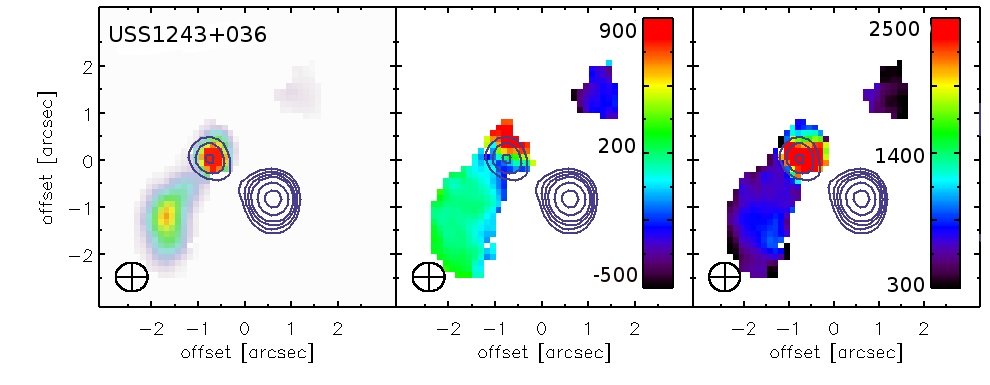} \includegraphics[width=0.48\textwidth]{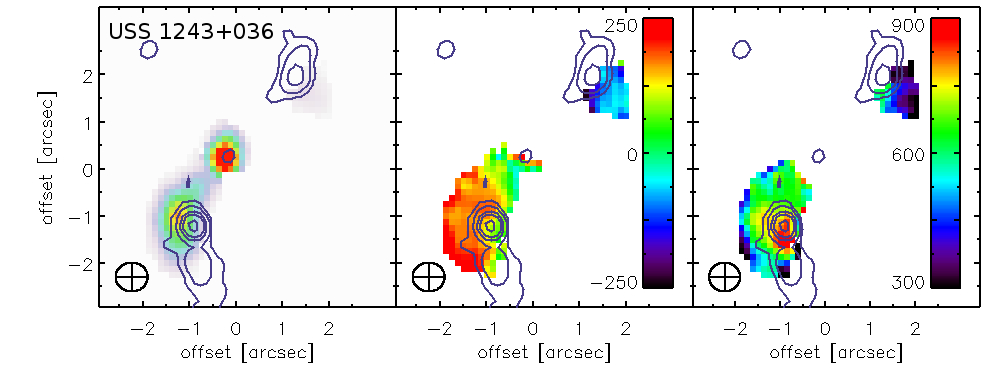}\\
\includegraphics[width=0.48\textwidth]{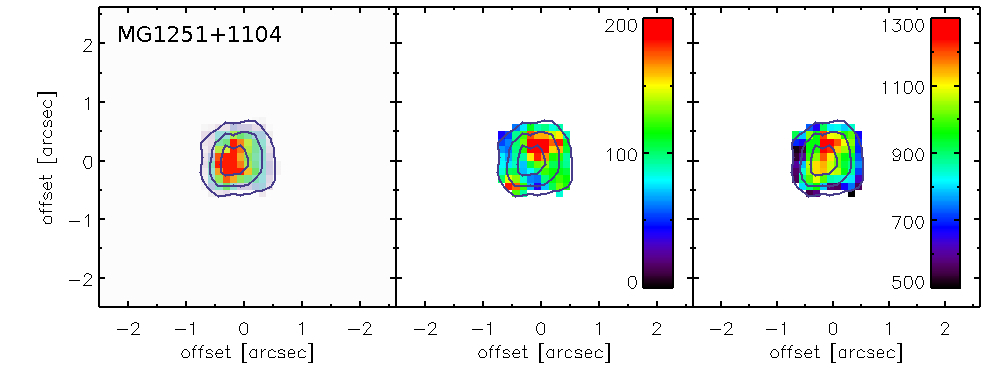}\\
\caption{
Maps of our galaxies in order of RA, with the
  rest-frame optical continuum shown as contours in the left, and the
  GHz radio continuum in the right panel. In each panel, maps show
  (from {\it left to right}) Emission-line morphology, velocities
  relatve to the average redshift of all pixels covered by the
  emission-line region, and FWHM line width. Relative velocities and
  line widths are given in km s$^{-1}$. In galaxies where we did not
  detect the rest-frame optical continuum, we show the emission-line
  morphology instead of the optical continuum (with red contours, the
  continuum is shown as blue contours). Coordinates are given relative
  to the position listed in Table~\ref{tab:sample}. Contour levels are arbitrary,
  their main purpose is to guide the eye. For MG1251$+$1104 we do not have a radio image
  available. It is listed as a point source with an upper limit to the source size of 1.2\arcsec\
  in \citet{debreuck00}.}
\label{fig:maps2}
\end{figure*}

\begin{figure*}
\includegraphics[width=0.48\textwidth]{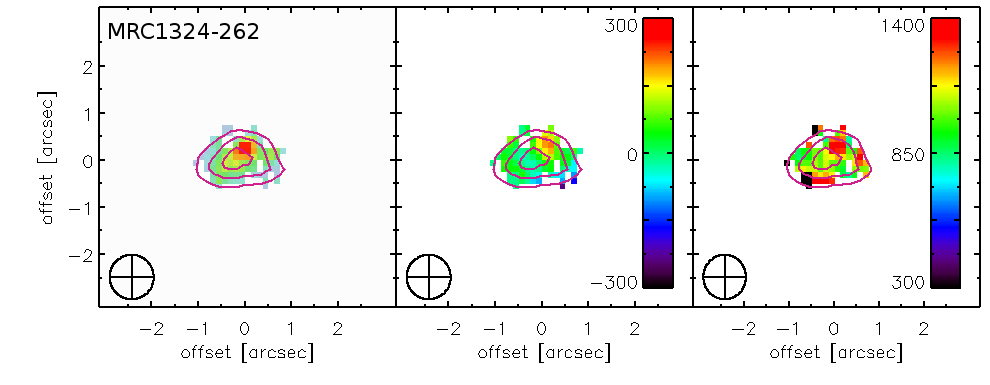}  \includegraphics[width=0.48\textwidth]{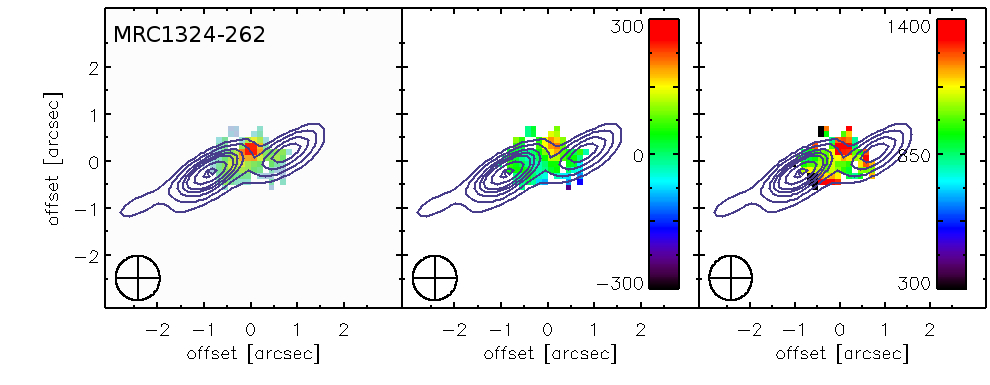}\\
\includegraphics[width=0.48\textwidth]{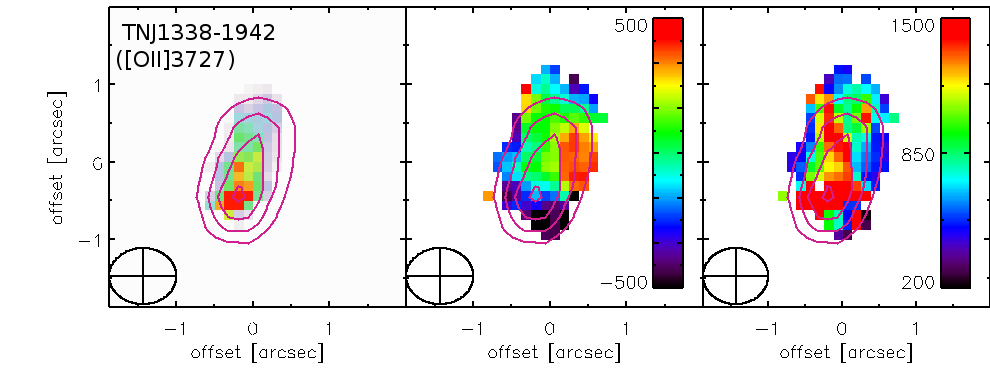} \includegraphics[width=0.48\textwidth]{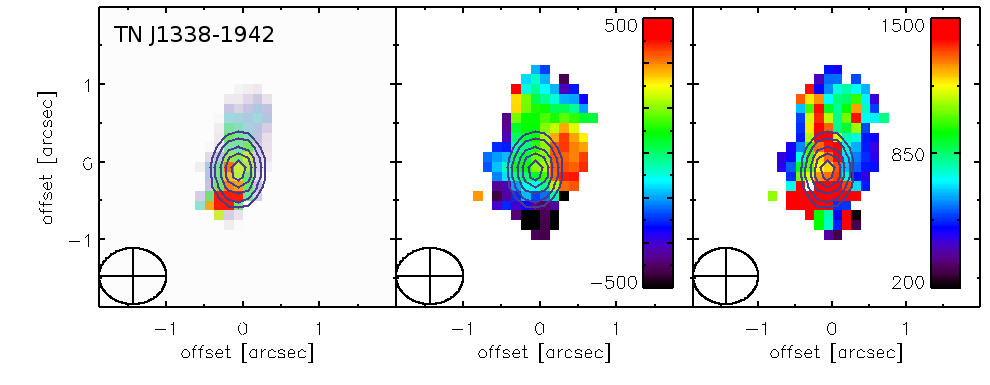}\\
\includegraphics[width=0.48\textwidth]{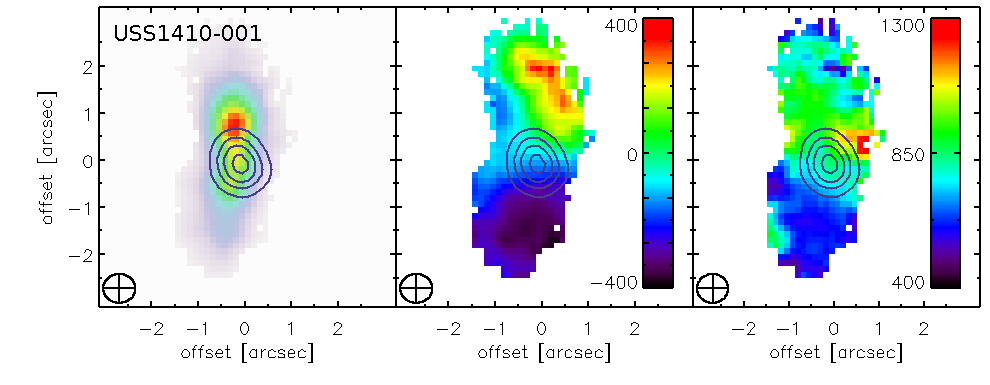} \includegraphics[width=0.48\textwidth]{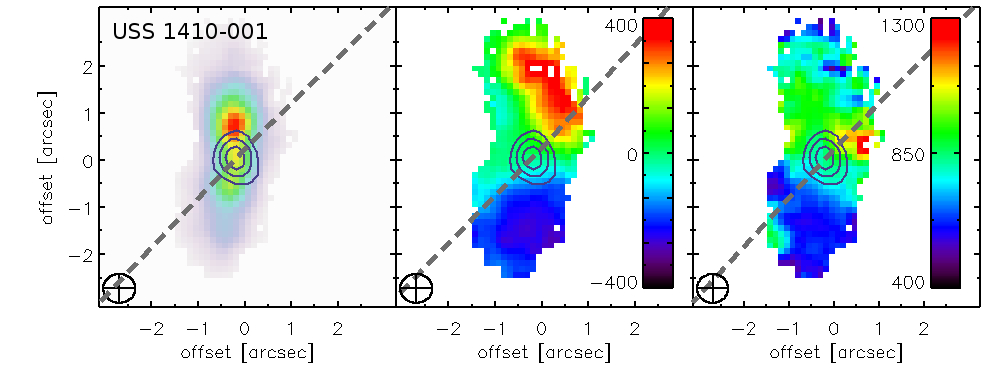}\\
\includegraphics[width=0.48\textwidth]{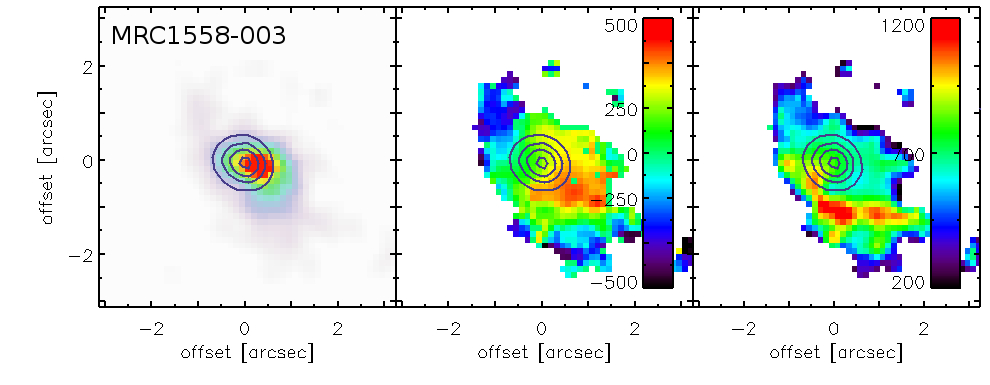} \includegraphics[width=0.48\textwidth]{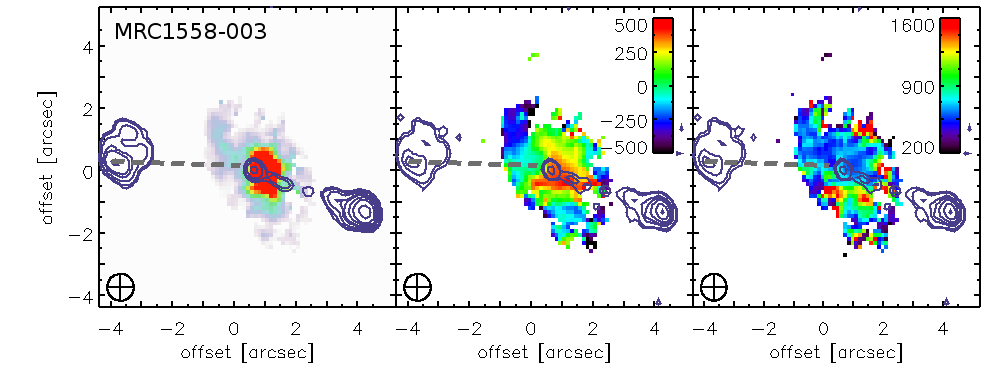}\\
\includegraphics[width=0.48\textwidth]{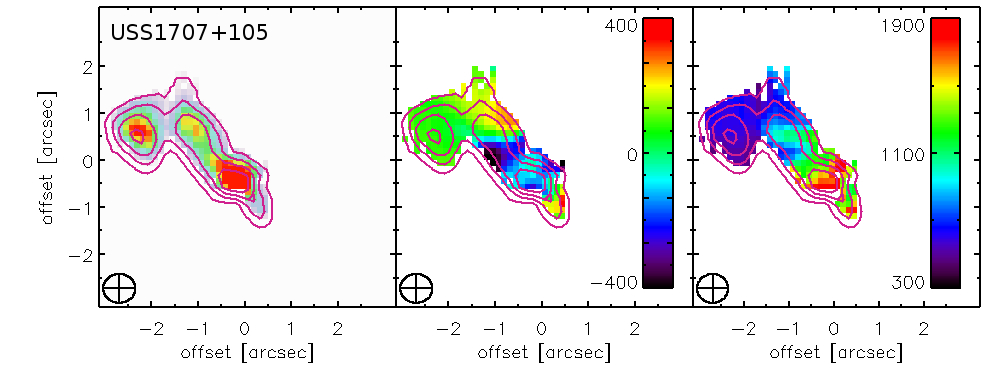} \includegraphics[width=0.48\textwidth]{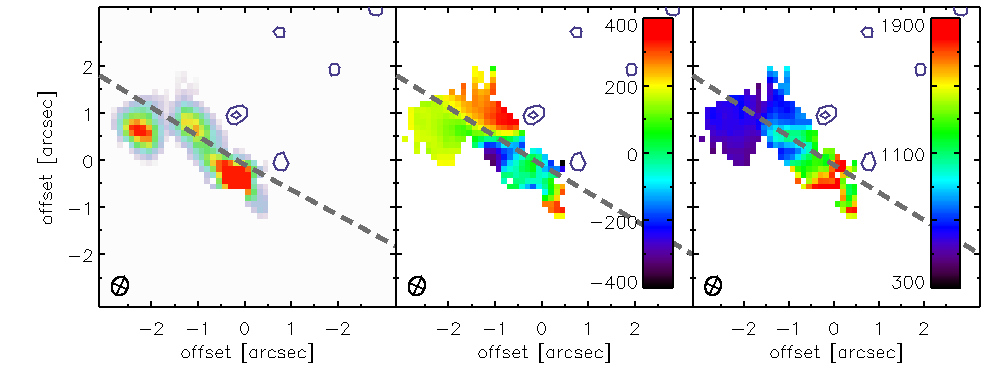}\\
\includegraphics[width=0.48\textwidth]{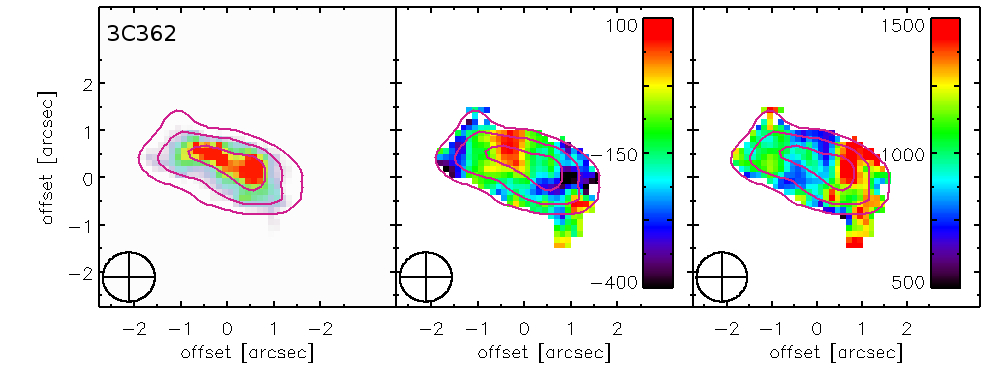}\includegraphics[width=0.48\textwidth]{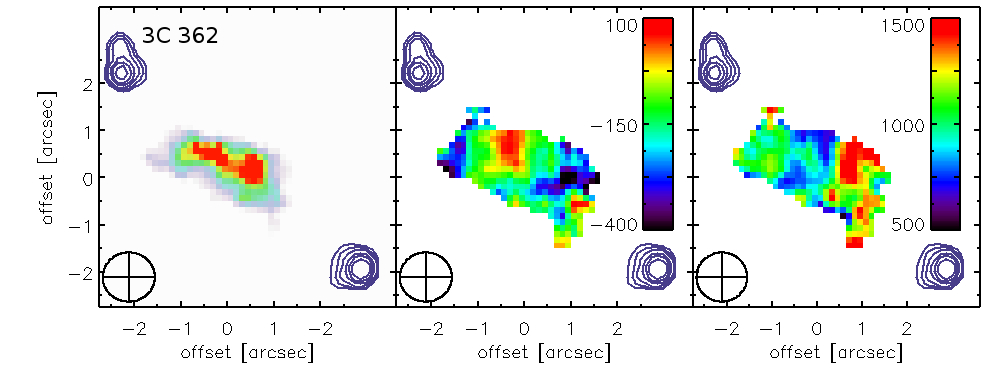}\\
\includegraphics[width=0.48\textwidth]{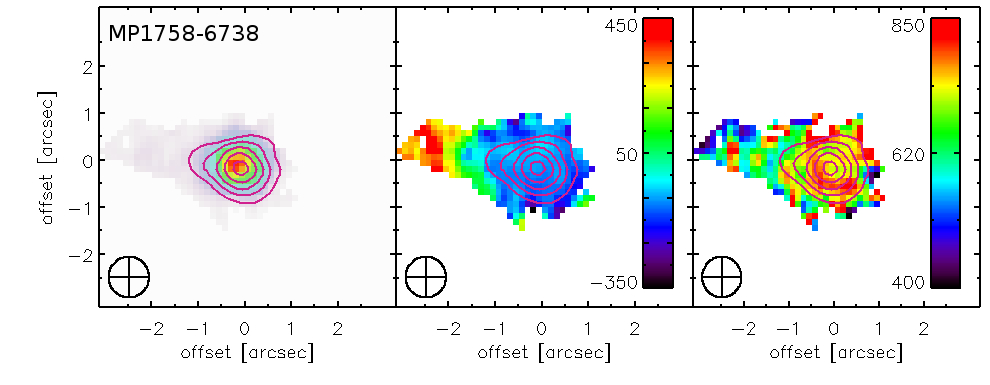}\\
\caption{
Maps of our galaxies in order of RA, with the rest-frame optical
continuum shown as contours in the left, and the GHz radio continuum
in the right panel. In each panel, maps show (from {\it left to
  right}) Emission-line morphology, velocities relatve to the average
redshift of all pixels covered by the emission-line region, and FWHM
line width. Relative velocities and line widths are given in km
s$^{-1}$. In galaxies where we did not detect the rest-frame optical
continuum, we show the emission-line morphology instead of the optical
continuum (with red contours, the continuum is shown as blue
contours). Coordinates are given relative to the position listed in
Table~\ref{tab:sample}. Contour levels are arbitrary, their main
purpose is to guide the eye. The radio map of MP~1758$-$6738 has a
very large beam of 24.4\arcsec\ and is not shown.}
\label{fig:maps3}
\end{figure*}

\begin{figure*}
\includegraphics[width=0.48\textwidth]{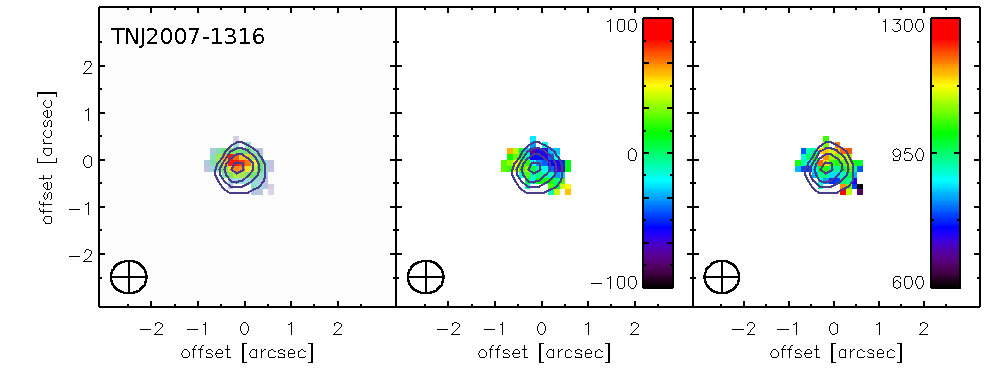} \includegraphics[width=0.48\textwidth]{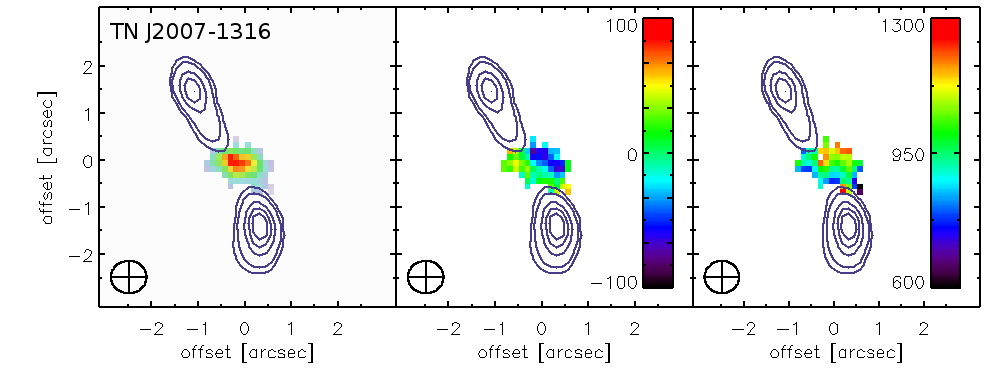}\\
\includegraphics[width=0.48\textwidth]{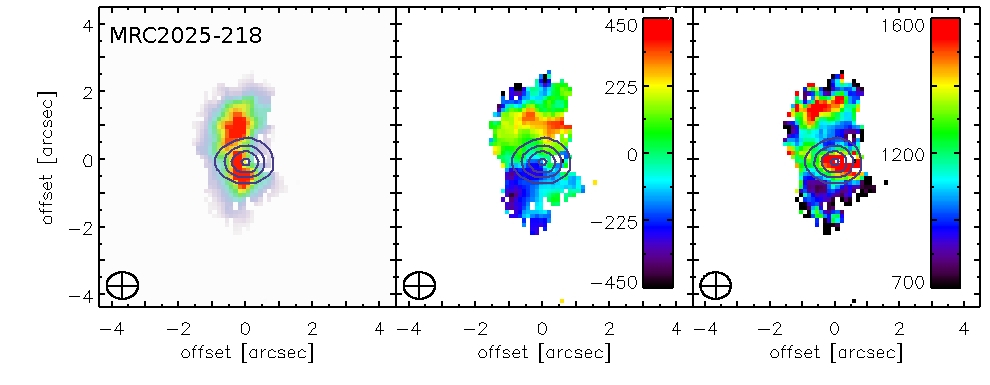} \includegraphics[width=0.48\textwidth]{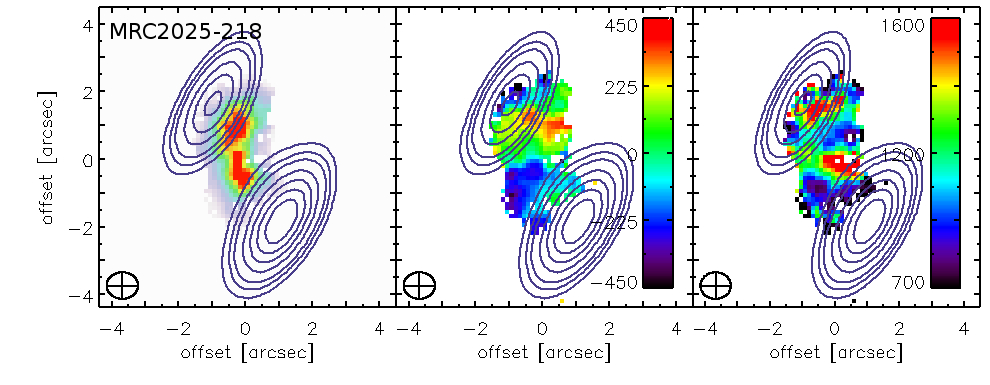}\\
\includegraphics[width=0.48\textwidth]{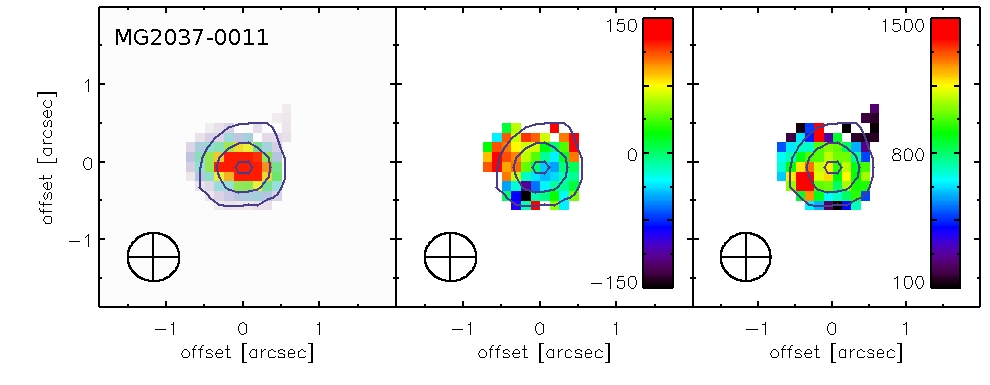}  \includegraphics[width=0.48\textwidth]{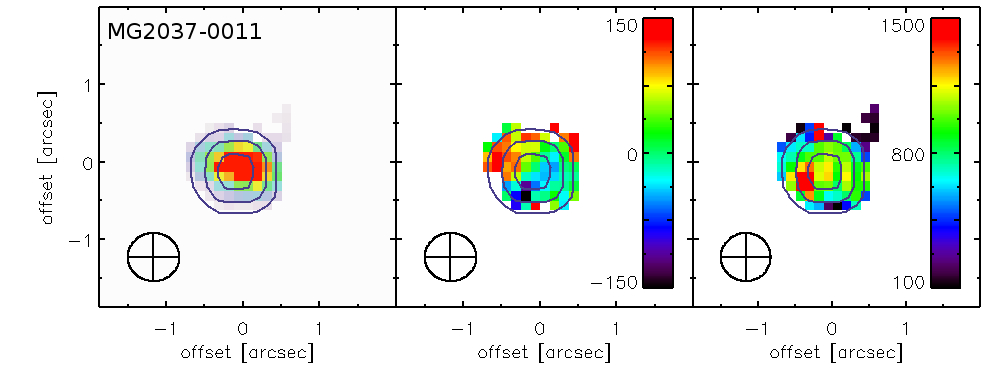}\\
\includegraphics[width=0.48\textwidth]{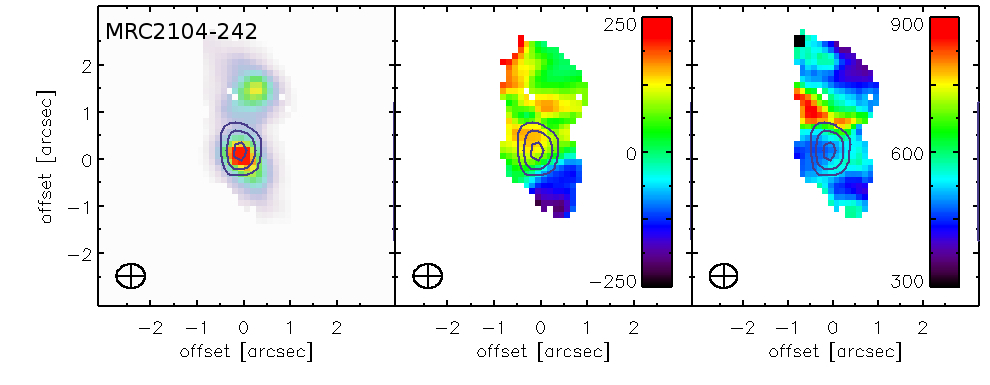}  \includegraphics[width=0.48\textwidth]{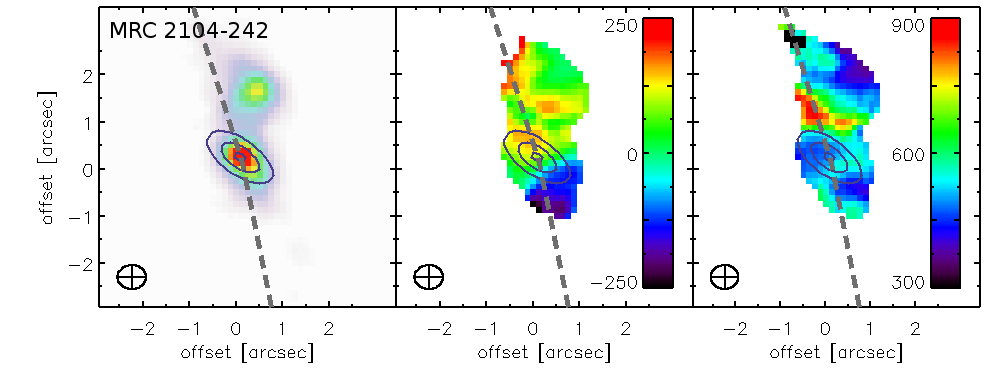}\\
\includegraphics[width=0.48\textwidth]{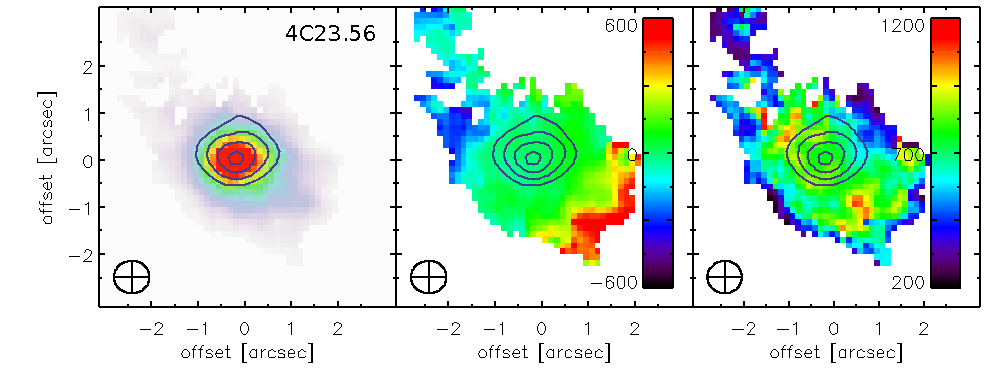}   \includegraphics[width=0.48\textwidth]{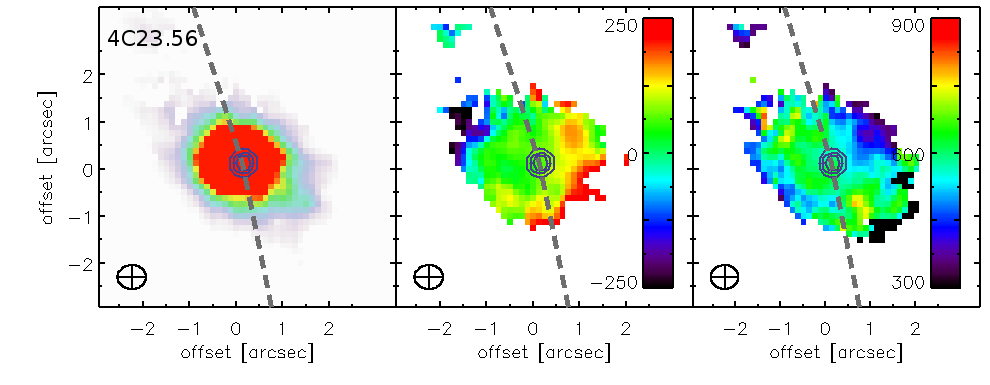}\\
\includegraphics[width=0.48\textwidth]{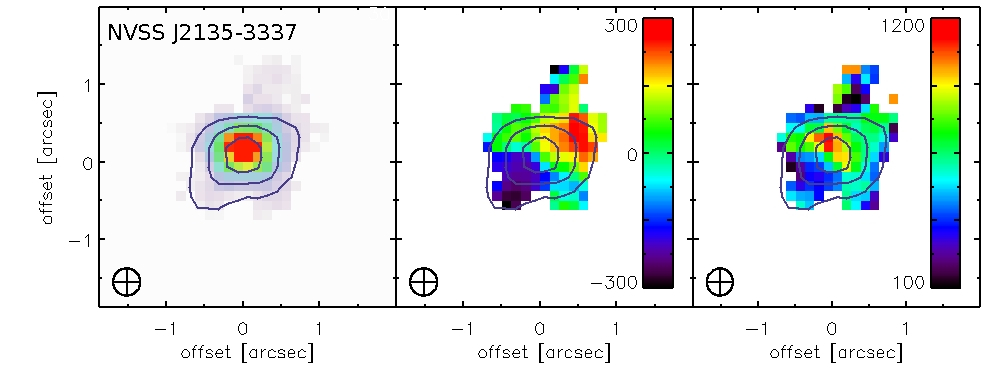} \includegraphics[width=0.48\textwidth]{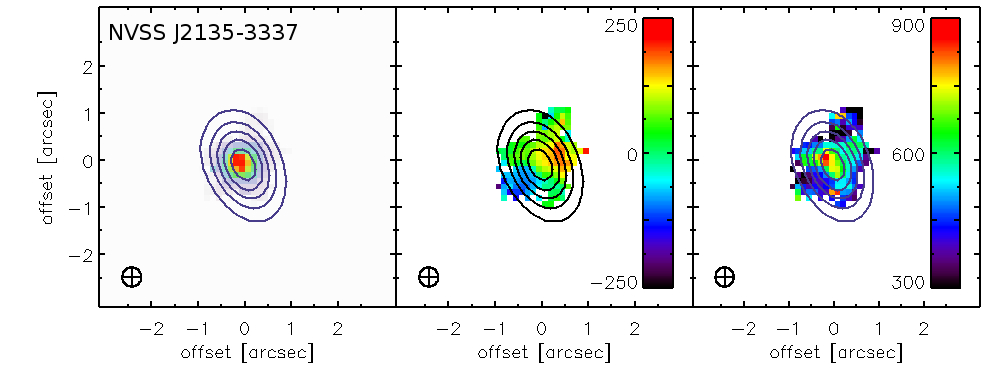}\\
\includegraphics[width=0.48\textwidth]{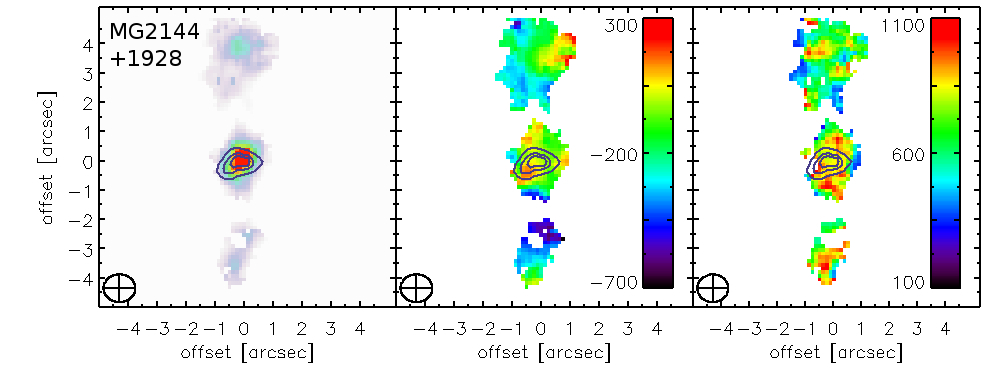} \includegraphics[width=0.48\textwidth]{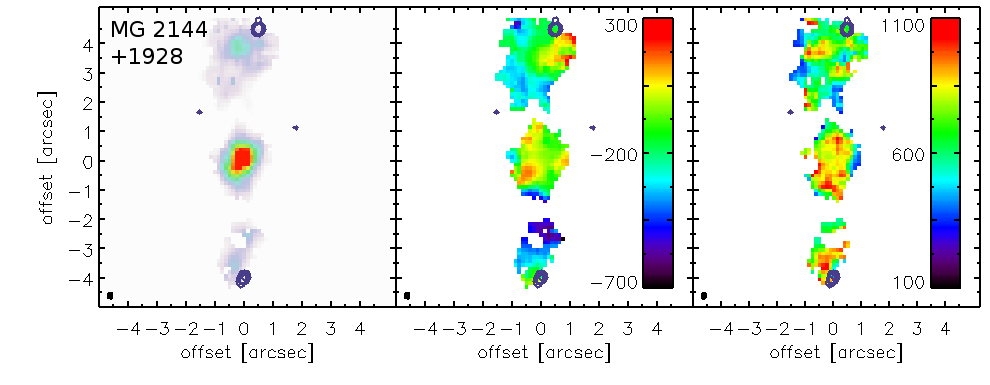}\\
\caption{
Maps of our galaxies in order of RA, with the rest-frame optical
continuum shown as contours in the left, and the GHz radio continuum
in the right panel. In each panel, maps show (from {\it left to
  right}) Emission-line morphology, velocities relatve to the average
redshift of all pixels covered by the emission-line region, and FWHM
line width. Relative velocities and line widths are given in km
s$^{-1}$. In galaxies where we did not detect the rest-frame optical
continuum, we show the emission-line morphology instead of the optical
continuum (with red contours, the continuum is shown as blue
contours). Coordinates are given relative to the position listed in
Table~\ref{tab:sample}. Contour levels are arbitrary, their main
purpose is to guide the eye. We have no good radio map of
MRC~2224$-$273, but the source has been listed as compact by
\citet{debreuck00} with an upper size limit of 0.4\arcsec.}
\label{fig:maps4}
\end{figure*}

\begin{figure*}
\includegraphics[width=0.48\textwidth]{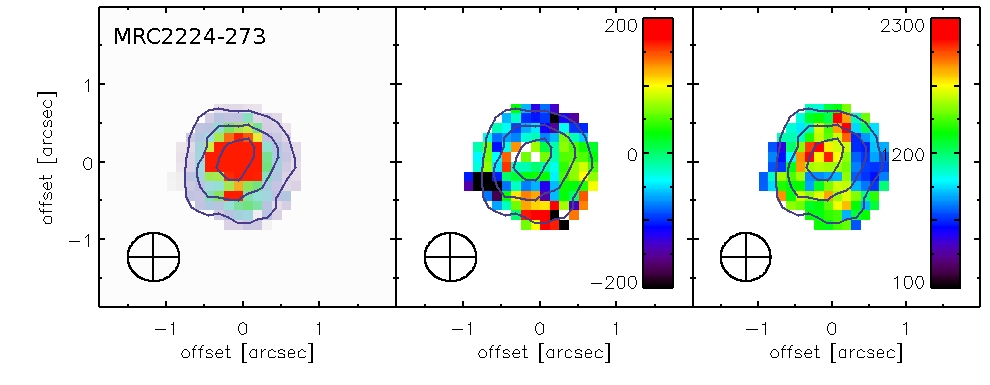}\\
\includegraphics[width=0.48\textwidth]{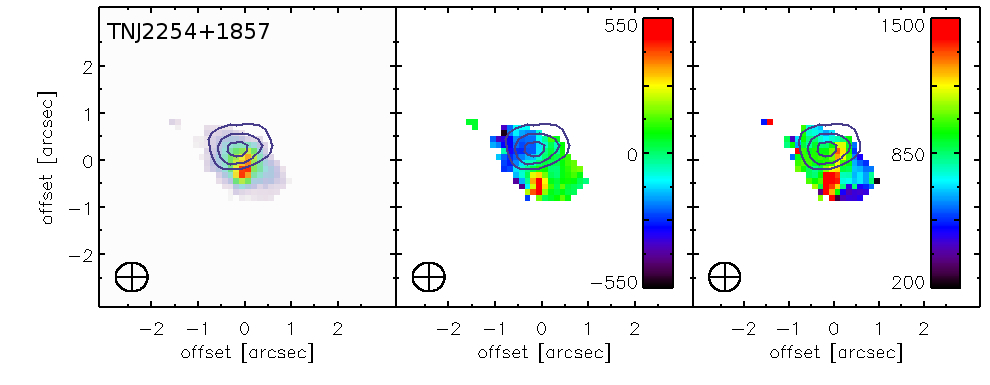}   \includegraphics[width=0.48\textwidth]{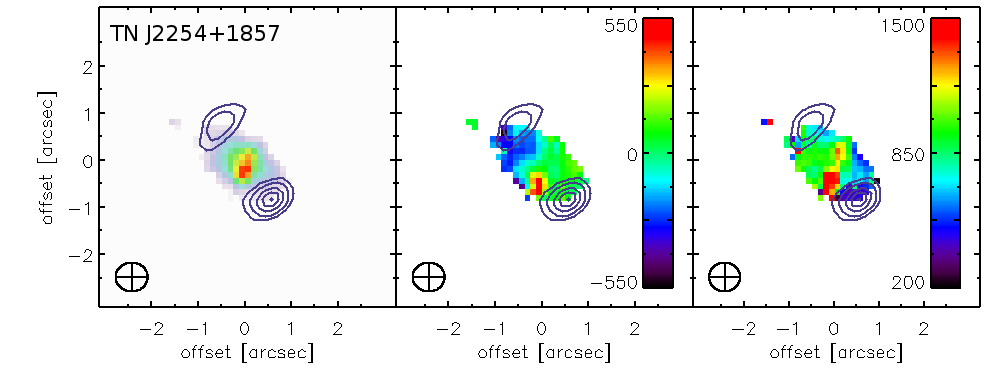}\\
\includegraphics[width=0.48\textwidth]{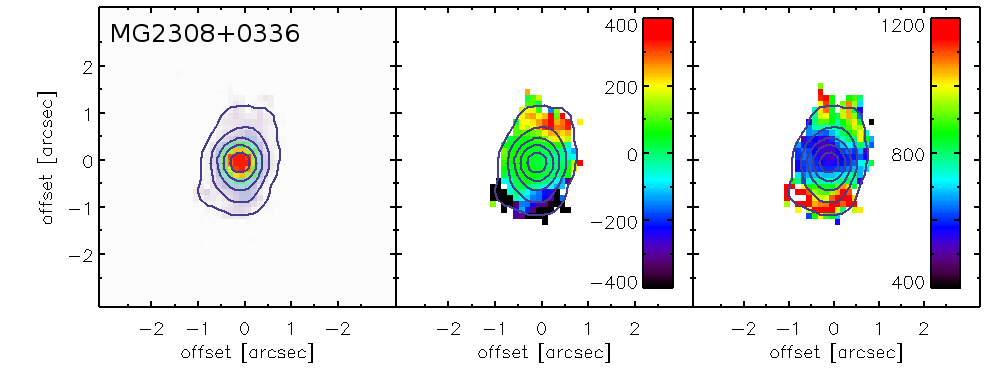}    \includegraphics[width=0.48\textwidth]{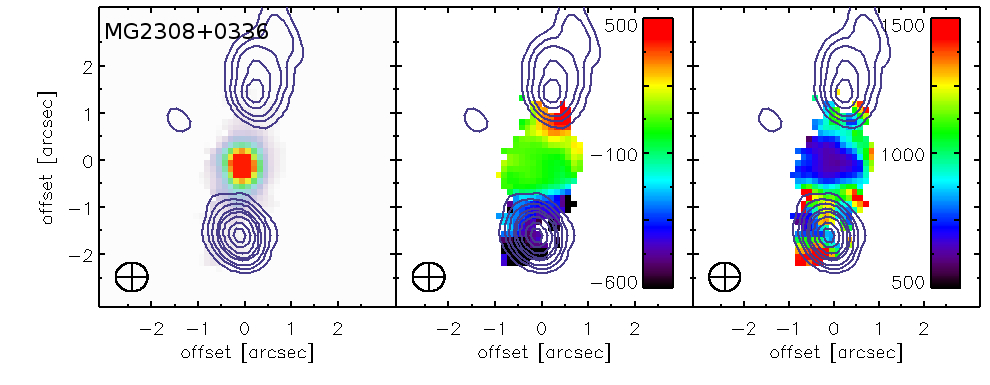}\\
\caption{
Maps of our galaxies in order of RA, with the rest-frame optical
continuum shown as contours in the left, and the GHz radio continuum
in the right panel. In each panel, maps show (from {\it left to
  right}) Emission-line morphology, velocities relatve to the average
redshift of all pixels covered by the emission-line region, and FWHM
line width. Relative velocities and line widths are given in km
s$^{-1}$. In galaxies where we did not detect the rest-frame optical
continuum, we show the emission-line morphology instead of the optical
continuum (with red contours, the continuum is shown as blue
contours). Coordinates are given relative to the position listed in
Table~\ref{tab:sample}. Contour levels are arbitrary, their main
purpose is to guide the eye.}
\label{fig:maps5}
\end{figure*}

It is a bright, marginally spatially resolved line
emitter at a seeing of 0.9\arcsec$\times$0.8\arcsec, associated with a
fairly bright continuum (Fig.~\ref{fig:maps}). Velocities and
line widths are uniform, with FWHM=681$\pm$9 km s$^{-1}$ (the error
refers to [OIII]$\lambda$5007). At the redshift of MRC~0251$-$273 we
detect H$\beta$ and [OIII]$\lambda\lambda$4959,5007 in the K and
[OII]$\lambda\lambda$3727 in the H-band. The integrated spectrum is
shown in Fig.~\ref{fig:intspec}. The two components of the [OII]
doublet are blended due to the fairly large intrinsic width of the
lines.

The line emission extends over an area of
3.5\arcsec$\times$2.5\arcsec\ on the sky, corresponding to
26~kpc$\times$18~kpc at z$=$3.17, and after deconvolving with the size
of the seeing disk. The gas is overall well centered on the unresolved
continuum source. The velocity offsets are relatively small, with a
total offset of 500~km s$^{-1}$. There is no single, montonic
gradient, but a blueshift near the center, with more redshifted gas
(by $250-500$~km s$^{-1}$) on either side. The FWHM line widths are
between 600 and 900~km s$^{-1}$, with faint regions of more narrow
widths at the very periphery of the emission line region
(FWHM$\sim$300~km~s$^{-1}$). The radio map in the right panels
of Fig.~\ref{fig:maps} shows a small double-peaked structure that is
embedded within the emission-line gas and well aligned with the region
of blueshifted gas around the center of the emission-line region. The
areas of broadest line width are mostly found adjacent to the radio
contours.

\subsection{RC~J0311$+$0507}
\label{ssec:indrcs0311} 

\begin{figure}
\centering
\includegraphics[width=0.45\textwidth]{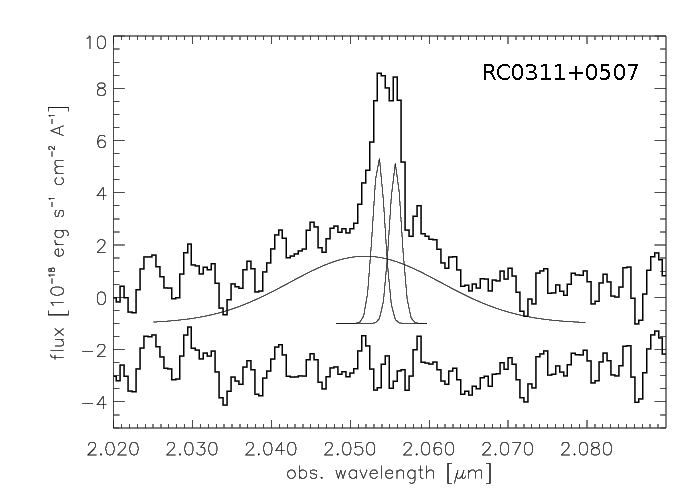}\\
\caption{Line profile of [OII]$\lambda\lambda$3726,3729 in the
  integrated spectrum of RC~J0311$+$0507.}

\label{fig:0311intspec}
\end{figure}

RC~J0311$+$0507 (4C$+$04.11) had previously a spectroscopic redshift
of z$=$4.514 measured only from Ly$\alpha$ \citep{pariiski98}. It has
recently been described in more detail by \citet{parijskij14}, who
find a very asymmetric radio source extending nearly north-south, with
a size of 2.8\arcsec\ (18.5~kpc). We identify
[OII]$\lambda\lambda$3726,3729, and [NeIII]$\lambda\lambda$3869,3968
in the K and H-band (Fig.~\ref{fig:0311intspec}). This is consistent
with the previously published redshift, which was based only on
Ly$\alpha$. [OII]$\lambda\lambda$3726,3729 is found at good S/N in the
integrated spectrum (Fig.~\ref{fig:0311intspec}). RC~J0311$+$0507 has
a complex line profile, with a broad component of FWHM$=$1400 km
s$^{-1}$, superimposed on a component which is narrow enough
(FWHM$=$112 km s$^{-1}$) for the doublet to be spectrally
resolved. Our three-component Gaussian fit (modeling the broad
component with only one line, and each of the narrow components of the
doublet individually).

[OII]$\lambda\lambda$3726,3729 is density sensitive, and the flux ratio
F(3727)/F(3729)=0.96$\pm$0.18 suggests electron densities between 200
and 1000 cm$^{-3}$, within the range found previously from
[SII]$\lambda\lambda$6716,6731 measurements \citep{nesvadba06a,
  nesvadba08, collet14a}. To our knowledge, this is the first estimate
of the electron density in an AGN host galaxy at z$\ge$4.

The line emission in RC~J0311$+$0507 appears fairly compact at the
spatial resolution of our data
(FWHM$=$1.0\arcsec$\times$0.9\arcsec\ along right ascension and
declination, respectively, corresponding to 6$\times$5.5 kpc at
z$=$4.51). We marginally detect continuum emission at the center of
the line emission. The emission-line maps are shown in
Fig.~\ref{fig:maps}. Although nominally, the source appears spatially
unresolved, the velocity map (measured from [OII]) does show a
gradient of $\sim 220$ km s$^{-1}$. The gas is more blueshifted in
roughly the eastern half of the source, where \citet{parijskij14}
detect the radio continuum in high-resolution observations with
MERLIN. The presence of a velocity gradient suggests that the source
is marginally spatially resolved with an intrinsic size that is very
near the size of the seeing disk. 

\citet{collet14a} used Monte-Carlo
simulations to investigate how low spatial resolution affects
measurements of velocity gradients in marginally spatially resolved
SINFONI maps akin to our data. Their analysis suggests that the
intrinsic velocity gradient projected onto the line of sight could be
about a factor~2 greater. A velocity offset of 400-500~km s$^{-1}$
would however not strongly affect the FWHM line widths, which are
above 900-1000 km s$^{-1}$ on average, and up to 1300 km s$^{-1}$ in
the western part of the emission-line region.

\subsection{MP~J0340$-$6507}
\label{ssec:indmpj0340}

MP~J0340$-$6507 at z$=$2.28 shows extended
(3.8\arcsec$\times$1.8\arcsec, corrected for the size of the seeing
disk), line emission along an axis going from north to south. The 
[OIII]$\lambda\lambda$4959,5007 emission-line region is asymmetry, and 
shifted by 0.5\arcsec\ towards
south-west. The continuum is fairly bright and peaks at the same
position as the line emission. We note a small velocity gradient of
$\approx$200 km s$^{-1}$. Line widths are FWHM$=$800$-$1000 km
s$^{-1}$. Unfortunately, we do not have a radio map of this galaxy
with a beam comparable to the resolution of the SINFONI data.

\subsection{PKS~0529-549}
\label{ssec:indpks0529}
PKS~0529$-$549 at z$=$2.57 has extended [OIII]$\lambda\lambda$4959,5007
and H$\alpha$ morphologies and a regular velocity gradient of 460 km
s$^{-1}$ (Fig.~\ref{fig:maps2}). Line widths are of-order
FWHM$\sim$700~km~s$^{-1}$ (Table~\ref{tab:emlines}).

In order to extract a maximal set of constraints from our data cubes
of PKS~0529-549, we constructed two cubes. One cube of
45~min exposure time at good seeing
(0.7\arcsec$\times$0.6\arcsec), which we used to produce the maps of
the gas kinematics of the bright [OIII]$\lambda\lambda$4959,5007
lines. In addition we constructed a cube of the full data set of
5.3~hrs to produce an integrated spectrum (where the spatial
resolution is of minor importance) to measure the fluxes of fainter
lines (Fig.~\ref{fig:intspec}). In total we detect H$\gamma$,
H$\beta$, H$\alpha$, [OIII]$\lambda\lambda$4959,5007,
[OI]$\lambda\lambda$6300,6363, [NII]$\lambda$6583, and
[SII]$\lambda$6724. The lines of the [SII] doublet are blended owing
to the intrisically broad lines. For H$\alpha$ and [NII]$\lambda$6583
the blending is not very strong.

We detect the continuum in the 'deep' cube which is unresolved
at a spatial resolution of 1.2\arcsec$\times$1.2\arcsec\ and well
centered relative to the emission-line region and velocity gradient. 

\citet{broderick07} found a very high rotation measure of $-9600$ rad
m$^{-2}$, the largest found so far in a radio galaxy at z$\sim$2,
suggesting a particularly strong magnetic field, dense environment, or
both, and a radio source with largest angular size of
1.2\arcsec. \citet{humphrey11} found bright 1.1~m emission from cold
dust with AzTEC.

The radio morphology shows a small double source, which has a very
similar size to the emission-line region and is well aligned with
it. The velocity jump seen in the gas appears to occur between the two
radio lobes, consistent with expectations from a back-to-back
outflow. The broadest line widths are found inbetween the two radio
lobes, which are each associated with gas with FWHM$\sim 700-800$ km
s$^{-1}$.

\subsection{5C~07.269}
\label{ssec:ind5c07269}
5C~07.269 at z=2.22 is one of our fainter sources in the radio, with
$\log P_{\rm 500}=$27.8~W~Hz$^{-1}$ at 500~MHz in the rest-frame, and a largest
angular size of 7.6\arcsec\ in the radio \citep[][]{debreuck10}. The
integrated spectrum shows [OIII]$\lambda\lambda$4959,5007 in the H
band, and H$\alpha$, [NII]$\lambda\lambda$6548,6583, and
[SII]$\lambda\lambda$6716,6731 in the K-band. We also detect the faint
continuum near the geometric center of the emission-line region.

The emission-line maps in Fig.~\ref{fig:maps2} show a source that is
extended along an axis that goes from the south-east to the
north-west. The velocity distribution is irregular, with a total
velocity range of $\pm$350 km s$^{-1}$ around an average redshift
z$=$2.22505, but much of that scatter might be due to the low
signal-to-noise ratio of this source. The distribution of line widths
appears more regular, with a maximum of 1100~km s$^{-1}$ near the
center, and more moderate velocities of $200-300$~km s$^{-1}$ in the
periphery. The broadest gas near the center seems to be associated
with the small, unresolved radio core (right panels of
Fig.~\ref{fig:maps2}).

\subsection{MRC~1017-220}
\label{ssec:indmrc1017}

In MRC1017$-$220 at z$=$1.7, H$\alpha$ falls very near the red end of
the H-band at $\lambda=1.8203\mu$m, and [OIII]$\lambda$5007 just
outside of the J-band, so that the emission lines are heavily affected
by telluric effects. This makes it impossible to measure the
emission-line kinematics, line morphologies, or line ratios for this
galaxy. \citet{nesvadba11a} found and discussed a broad H$\alpha$ line
in MRC~1017$-$220, which extends into cleaner parts of the
spectrum. Given these observational complications, we did not include
this source in our overall analysis.

\subsection{TN~J1112-2948}
\label{ssec:indtnj1112}
Line emission in TN~J1112$-$2948 at z$=$3.1 is very extended with a
size of 6.7\arcsec$\times$2.2\arcsec\ (52.3~kpc $\times$17.3~kpc at
z$=$3.1, Fig.~\ref{fig:maps2}). We detect
[OIII]$\lambda\lambda$4959,5007, H$\beta$, and [OII]$\lambda$3727. The
emission-line maps shown in Fig.~\ref{fig:maps2} show high
surface-brightness line emission in a resolved area around the
nucleus, extending over
2.3\arcsec$\times$2.0\arcsec\ (18~kpc$\times$16~kpc at z$=$3.1) with a
small velocity gradient of 180~km s$^{-1}$ and line widths between
FWHM$\sim$600 and 800 km s$^{-1}$. Towards south-east, we see an
extended gas plume extending out to a distance of 38~kpc from the
nucleus, with velocities of up to 400~km s$^{-1}$ and FWHM line widths
of 300$-$500~km s$^{-1}$. The extended line emission in
TN~J1112$-$2948 is asymmetric about the radio core, which is
associated with the northern cloud. We do not detect another plume
towards the north-west, although we obtained a second SINFONI pointing
to cover this area specifically. The southern cloud is directly
adjacent to the southern radio hotspot, although not perfectly
aligned. The overall velocities in TN~J1112$-$2948 are monotonically
increasing from the north-west to the south-east with a total gradient
of 700 km s$^{-1}$.
\citet[][]{reuland04} detected TN~J1112$-$2948 with SCUBA, Herschel
photometry has unfortunately not been obtained for this galaxy.

\subsection{TXS~1113-178}
\label{ssec:indtxs1113}
TXS~1113$-$178 at z$=$2.24 shows a broad H$\alpha$ emission line coming
from the nucleus \citep[][]{nesvadba11a}, which made it necessary to
remove the nuclear point source before analyzing the extended line
emission in this source (\S\ref{ssec:blrremoval}). We do detect
residual line emission associated with the continuum emission from the
nucleus, although it is fairly compact, extending
over 1.3\arcsec$\times$1.1\arcsec. This corresponds to
9.2~kpc$\times$8.3~kpc after deconvolution with the size of the seeing
disk, which was exceptionally small for this data set,
0.6\arcsec$\times$0.5\arcsec. We identify
[OIII]$\lambda\lambda$4959,5007, H$\beta$, H$\alpha$, and
[NII]$\lambda$6583 in the integrated spectrum shown in
Fig.~\ref{fig:intspec}. The line properties are also listed in
Table~\ref{tab:emlines}.

In spite of the low spatial resolution of the data, we do detect a
velocity gradient of 700~km s$^{-1}$ from north-west to south-east,
and potentially a gradient in FWHM line width of 1200~km s$^{-1}$ to
500~km s$^{-1}$ from the center to the periphery of the emission-line
region. The low spatial resolution implies that we may be
underestimating the intrinsic velocity gradient by about a factor~2
(C. Collet, 2014, PhD thesis). The velocity gradient is well aligned
with the direction of the radio jets in TXS~1113$-$178
(Fig.~\ref{fig:maps2}), although we do not see extended gas. This
could be the consequence of the short exposure time for this source. 

\subsection{3C~257}
\label{ssec:ind3c257}

3C~257 at z$=$2.48 has extended line emission along an axis going from
the north-west to the south-east, extending over
3.4\arcsec$\times$2.0\arcsec\ (27.5~kpc$\times$15.1~kpc after
deconvolving with the size of the seeing disk of
0.8\arcsec$\times$0.7\arcsec). The continuum emission is extended with
a peak towards south and an extended fainter area towards north-west,
roughly aligned with the emission-line gas (contours in
Fig.~\ref{fig:maps}). The line emission is much brighter in the south-east,
and is clearly spatially offset from the continuum.

The integrated spectrum shows [OIII]$\lambda\lambda$4959,5007,
H$\beta$, and H$\alpha$, [NII]$\lambda$6583, and
[SII]$\lambda\lambda$6716,6731, where the two lines of the [SII]
doublet are blended. A single Gaussian component per line with
FWHM$=$1049$\pm$7 km s$^{-1}$ is sufficient to fit the line profiles
(the error is that of [OIII]$\lambda$5007). The spectrum is shown in
Fig.~\ref{fig:intspec} and all line properties are summarized in
Table~\ref{tab:emlines}.

The velocity field of 3C~257 is irregular (Fig.~\ref{fig:maps}), with
a velocity maximum south-west from the center, and declining
velocities towards either end of the emission-line region. The total
velocity offset is 400~km~s$^{-1}$. FWHMs are between 500 and 1300 km
s$^{-1}$. The highest FWHMs are reached in the eastern part of the
region, partially coinciding with the most strongly redshifted gas,
but extending over a larger region. The region of highest velocities and
FWHMs is almost, but not quite, associated with the relatively compact,
double-lobed radio source (Fig.~\ref{fig:maps2}), that is embedded in
the gas. The highest velocities are reached just south of the two
jets.

\subsection{USS~1243+036}
\label{ssec:induss1243}

USS~1243$+$036 at z$=$2.36 is one of the galaxies with very extended
line emission in our sample. Overall, we detect H$\beta$ and
[OIII]$\lambda\lambda$4959,5007, as well as H$\alpha$,
[NII]$\lambda$6583, and [SII]$\lambda\lambda$6716,6731. The integrated
spectrum is shown in Fig.~\ref{fig:intspec} and the line properties
are listed in Table~\ref{tab:emlines}.

The projected size of the emission-line region is
5.3\arcsec$\times$0.9\arcsec\ (corresponding to 38~kpc$\times$4~kpc
after deconvolution with the seeing disk), and the emission-line
morphology is very irregular. The nucleus as probed by the continuum
emission is roughly centered within a resolved inner emission line
region, but surrounded by two more extended emission-line blobs
towards the north-west and south-east, which are very different from
each other.  An extended plume is found in the south-east, whereas in
the north-west we only find a small, nearly unresolved region of
bright line emission. The region where the jet is escaping from the
emission-line region \citep[first noted by][]{vanojik96} shows
particularly large velocity dispersions and a strong blueshift, as
expected for regions of strong interactions between jet and gas.

To ensure that we are not missing flux in this region owing
to the small field of view of SINFONI of only
8\arcsec$\times$8\arcsec, we produced a mosaic from two pointings used
for this galaxy and obtained in two different runs
(Table~\ref{tab:sample}), finding that our first observation
did not miss any of the emission-line gas.

Velocities and FWHM line widths in USS1243$+$036 have an exceptionally
large dynamic range, which is why we present two sets of
[OIII]$\lambda$5007 line maps in Fig.~\ref{fig:maps2}, which only
differ in their dynamic range for the line widths and velocity
gradients. The maps show the mosaic from both pointings. The velocity
gradient in the extended gas (lower panel of Fig.~\ref{fig:maps2}) is
$\sim$500~km~s$^{-1}$ between the south-eastern and north-western lobe, with a
monotonic velocity increase from the north-west to the south-east. The
south-eastern lobe has broader lines, FWHM$=$700$-$900 km s$^{-1}$
than the north-western lobe, which has FWHM$\sim$ 300 km
s$^{-1}$. Near the continuum location, velocities are up to 2500 km
s$^{-1}$, and line widths are up to
FWHM$=$900 km s$^{-1}$.

The right panel of Fig.~\ref{fig:maps2} shows that the southern radio
hotspot is associated with the most strongly blueshifted gas in the
southern emission-line region, and the region of broadest line
widths. The radio plasma extends further outside the emission-line
region. \citet{vanojik96} already identified this area as the site
where the jet is being deflected. The northern radio lobe falls right
next to the northern emission-line region.

The bright continuum emitter west-south-west from the galaxy has a
featureless continuum consistent with a black body. We suspect it is a
foreground star.

\begin{figure}
\includegraphics[width=0.45\textwidth]{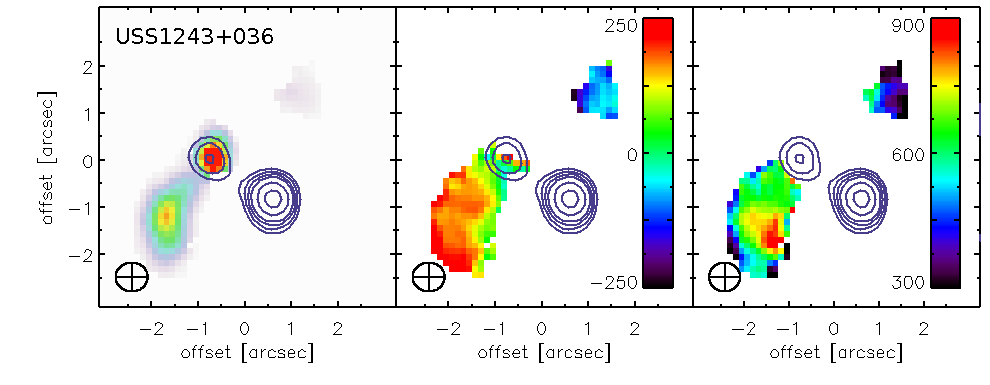}
\caption{
Maps of USS1243$+$036 highlighting the kinematics within the extended
emission-line gas. Contours show the rest-frame optical continuum
morphology. The very high-velocity gas in the central galaxy is not
shown (see Fig.~\ref{fig:maps} for a figure showing the full dynamic
range in USS1243$+$036.}
\label{fig:maps1243}
\end{figure}

\subsection{MG~1251$+$1104}
\label{ssec:indmg1251}

MG~1251$+$1104 at z$=$2.32 is a rather compact line emitter, for which
we have only shallow data with 2400~s of on-source exposure time with
an average seeing of FWHM$=$0.8\arcsec$\times$0.8\arcsec. The FWHM
size of the line image is 0.9\arcsec, observed at a seeing of
FWHM=0.8\arcsec. This corresponds to an upper limit on the size of
7.4~kpc. In the integrated spectrum we detect [OIII]$\lambda$5007,
H$\alpha$, and [NII]$\lambda$6583. The latter two lines are blended
owing to the broad line widths of FWHM$=$914$\pm$32 km s$^{-1}$.

Fig.~\ref{fig:maps} shows the morphology, velocity offsets and
FWHMs. We caution that the range in velocity and FWHMs that we can
detect in this source are likely very small compared to the intrinsic
values because of the compactness of the source.  While we do not
detect a large velocity offset, to the north-west and south-east from
the nucleus there appears to be a zone of more redshifted line
emission, with a total velocity offset of 200 km s$^{-1}$ relative to
the most blueshifted gas. Line widths are up to 1300 km s$^{-1}$ near
the center and decrease to 500~km s$^{-1}$ at the periphery.  We
cannot exclude that deeper observations would have revealed fainter
extended line emission. For MG~1251$+$1104, we do not have a
high-resolution radio image available. \citet{debreuck00} list it as a
compact source with LAS$<$1.2\arcsec, which corresponds roughly to the
size of the emission-line region.

\subsection{MRC~1324$-$262}
\label{ssec:indmrc1324}
MRC~1324$-$262 at z$=$2.28 is another source for which we only have
shallow data with an on-source exposure time of 2400~s and seeing of
1.1\arcsec$\times$1.1\arcsec. The [OIII]$\lambda$5007 line is bright,
well detected, and fairly broad, with FWHM$=$1253$\pm$36~km
s$^{-1}$. We also detect H$\alpha$, H$\beta$, and
[SII]$\lambda\lambda$6716,6731. The lines of the [SII] doublet are
blended due to their large intrinsic line widths and low
signal-to-noise ratio of the data. The integrated spectrum is shown in
Fig.~\ref{fig:intspec} and individual line properties are listed in
Table~\ref{tab:emlines}.

The source appears almost compact in this data set, perhaps with a
slight extension in east-western direction. The continuum is not
detected (Fig.~\ref{fig:maps}). We do not detect any distinctive
features in the velocity map, and given the large seeing and shallow
data set, irregularities in the map of FWHM line widths could be
dominated by noise. The radio maps shown in Fig.~\ref{fig:maps2} show
a small double source, where each radio hotspot is near the outer edge
of the emission line region.

\subsection{TN~J1338-1942}
\label{ssec:indtnj1338}

The second-highest redshift galaxy in our sample at
z$=4.1$. Unfortunately, TN~J1338$-$1942 falls at a somewhat difficult
redshift for NIR observations, where the bright
[OIII]$\lambda\lambda$4959,5007 and H$\alpha$ lines are at wavelengths
redward of the K-band. [OII]$\lambda$3737 falls into the spectral
region of low atmospheric transmission between the H and K band and
[NeIII] falls into the blue part of the K-band at $\lambda
\sim$2$\mu$m.

We caution that due to the faintness of the lines, the uncertainties
of relative velocities and line widths are high.  We therefore
extracted two spectra centered on the relative bright emission towards
the north and south of the emission-line image in
Fig.~\ref{fig:maps3}. These spectra are shown in
Fig.~\ref{fig:intspec}. We identify [OII], and [NeIII]$\lambda$3869 in
both regions, and [NeIII]$\lambda$3938 in the northern spectrum.

In spite of these challenges we detect TN~J1338$-$1942 in
[NeIII]$\lambda\lambda$3869,3968 and in [OII]$\lambda$3727. The
brightest line is [NeIII]$\lambda$3869, which we also resolve
spatially (Fig.~\ref{fig:maps2}). Line emission extends over
$\sim$1.5\arcsec\ and matches the size and orientation of the
Ly$\alpha$ and continuum emission of \citet{zirm05}, which is embedded
in a $>$100~kpc, highly asymmetric Ly$\alpha$ halo
\citep{venemans02}. The S/N of the line emission is very low, but
suggests a velocity gradient of-order 500 km s$^{-1}$, and very faint
line emission with a velocity offset of 500 km s$^{-1}$ towards
west. At this position, \citet{zirm05} previously found a wedge-shaped
feature in their {\it HST} narrow-band Ly$\alpha$ image. The radio
morphology of TN~J1338$-$1942 is compact. The nucleus is found just
east of this wedge, and conincides with the area of highest FWHMs.

\begin{figure}
\centering
\includegraphics[width=0.45\textwidth]{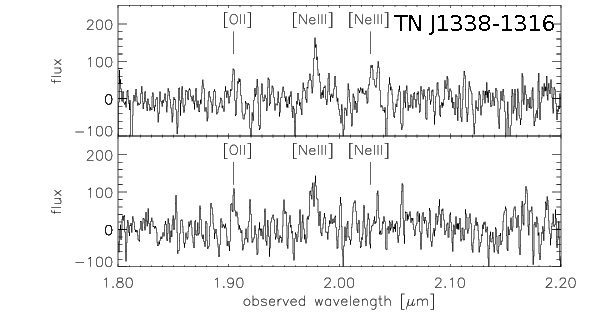}\\
\caption{Integrated spectrum of TN~J1338-1942 showing [OII], and
  [NeIII]$\lambda\lambda$3860,3068.}
\label{fig:1338spec}
\end{figure}

\subsection{USS~1410$-$001}
\label{ssec:induss1410} 

USS~1410$-$011 at z$=2.36$ is another source with a particularly large
emission-line region (Fig.~\ref{fig:maps}) extending over an area of
5.6\arcsec$\times$2\arcsec\ along the major and minor axis,
respectively (corresponding to 41~kpc$\times$14~kpc after deconvolving
with the size of the seeing disk).  [OIII]$\lambda\lambda$4959,5007
and H$\alpha$ line emission extends nearly along an axis going from
south to north with a monotonically increasing
velocity gradient of $\Delta v=$685~km~s$^{-1}$. Line widths are
FWHM$=400-900$ km s$^{-1}$, except in a small region to the north-east of
the nucleus where FWHMs up to 1300~km s$^{-1}$ are reached.  All lines
are well fit with single Gaussian components. The continuum is compact
and approximately centered on the line-emitting clouds.

The [NII]$\lambda6583$/H$\alpha$ ratio is fairly low,
[NII]$\lambda6583$/H$\alpha=0.17\pm0.02$. We also detect the
[SII]$\lambda\lambda$6716,6731 doublet at good S/N and identify each
line of the doublet individually, but no [OI]$\lambda$6300. The
3$\sigma$ upper limit on the [OI]$\lambda$6300 flux is
$2.4\times10^{-16}$ erg s$^{-1}$ cm$^{-2}$ assuming a line width of
FWHM$=$500 km s$^{-1}$. [OI]$\lambda$6300 coincides with a bright night sky
line residual. The [OIII]$\lambda$5007 line in the integrated spectrum
shows a faint blue wing.

\begin{figure}
\includegraphics[width=0.45\textwidth]{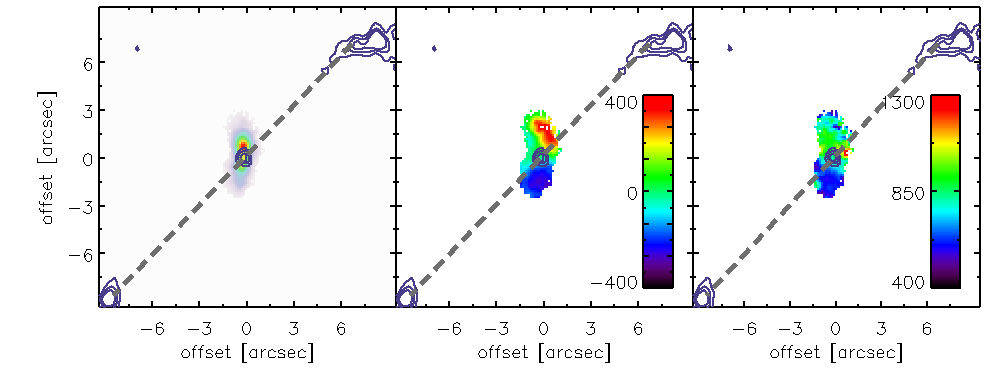}\\
\caption{
{\it left to right:} [OIII]
  morphologies, velocity offsets and FWHM line widths in USS1410$-$011
  in comparison with the radio morphology. The northern radio source
  shows a narrow, faint morphology along an axis that is roughly an
  extension with the most redshited gas in the northern emission-line
  region, before turning towards east at the hotspot.}
\label{fig:mapuss1410big}
\end{figure}

USS~1410$-$011 has a very large radio source with LAS$=25.2$\arcsec\,
significantly more extended than the emission-line gas
(Fig.~\ref{fig:mapuss1410big}). Apart from the compact radio core, we
do not detect radio continuum associated with the emission-line
gas. The northern radio source shows extended faint emission along an
axis that is associated with the region of the most redshifted line
emission in the northern emission-line region of USS~1410$-$011.

\subsection{MRC~1558$-$003}
\label{ssec:indmrc1558}
MRC~1558$-$003 at z$=$2.53 is one of the 6~galaxies of
\citet{nesvadba11a} which have broad nuclear H$\alpha$ line emission
heralding a direct sight line into the AGN. We removed the nuclear
point source before fitting the extended line emission
(\S\ref{ssec:blrremoval}). We find that the extended emission-line
nebulosity is roughly centered on the nucleus with a size of
5.1\arcsec$\times$2.8\arcsec\ (41~kpc$\times$22~kpc at z$=$2.53 and
after deconvolving with the seeing disk) along the major and minor
axis with complex kinematics. Relative velocities peak in a region
south from the nucleus with 500 km s$^{-1}$, blueshifts with up to
$-500$ km s$^{-1}$ are found in the periphery of the emission-line
gas. The largest widths are found along a ridge south of the nucleus
with FWHM=1200 km s$^{-1}$. Most line emission has widths of about
FWHM$=$700$-$900~km s$^{-1}$.

MRC~1558$-$003 is also extended in the radio, with
LAS$=7.7$\arcsec. The right panel of Fig.~\ref{fig:maps3} shows a
faint extension from the radio core towards the western lobe, which is
associated with the most strongly redshifted gas, although this does
not stand out through particularly broad FWHMs in this case. We
caution that this could also be a consequence of the removal of
broad-line emission from the AGN.

\subsection{USS~1707+105 (4C$+$10.48)}
\label{ssec:induss1707}
USS~1707+105 at z$=$2.35 is another source with very asymmetric line
emission. We detected well extended line emission
(3.5\arcsec$\times$1.6\arcsec), corresponding to a deconvolved size of
25~kpc$\times$5~kpc), extending from the nucleus to the north-east. We
did not detect continuum emission in our data, but the size of the
emission-line region is sufficiently large that we can safely place
the nucleus near the southern edge of the extended emission-line
region.  No line emission is detected in the south-west, although this
area is well covered by our data cube, with the same depth as the
northern emission-line region. The northern region is nearly straight
out to a distance of 20~kpc from the nuclear region, and then makes a
sudden bent towards the south-east.

The velocity field is irregular with several velocity maxima and
minima. The largest velocity gradient of 800~km~s$^{-1}$ is found in a
small region 1.5~kpc from the nucleus. FWHMs are
greatest in the near-nuclear region (1900~km s$^{-1}$),
and are $800-1200$~km~s$^{-1}$ in the region associated with the large
velocity gradient, and about $400-500$~km s$^{-1}$ in the emission-line
region beyond the bent.

On larger scales, the velocity offset is 490 km s$^{-1}$. We only
detect [OIII]$\lambda$4959,5007, H$\alpha$, [NII]$\lambda$6583, and
H$\beta$. [OI]$\lambda$6300 and [SII]$\lambda\lambda$6716,6731 are too
faint and in bad regions of the night sky.

\subsection{3C~362 (MG~1747$+$182)}
\label{ssec:indmg1747}
The warm ionized gas in MG~1747$+$182 at z$=$2.29 extends over 
  4.0\arcsec$\times$2.2\arcsec along the major and minor axis,
respectively (corresponding to a deconvolved size of 31$\times$15~kpc
at z$=$2.29). The emission-line gas extends along a north-east to
south-west axis. The integrated spectrum (Fig.~\ref{fig:intspec})
shows [OIII]$\lambda\lambda$4959,5007, H$\beta$, H$\alpha$,
[NII]$\lambda$6583, and [SII]$\lambda\lambda$6716,6731. The two
components of the [SII] doublet are blended because of the large line
width of 980~km s$^{-1}$ in the integrated spectrum. H$\alpha$ and
[NII]$\lambda\lambda$6548,6583 are also strongly
blended. Table~\ref{tab:emlines} gives the line properties derived
from a single-Gaussian fit (which is adequate for the lines that are
not blended) and by assuming that all lines have the same FWHM
measured from [OIII]$\lambda$5007.

The total velocity gradient of [OIII]$\lambda$5007 in MG~1747$+$182 is
about 600~km~s$^{-1}$, and the velocity field is not very regular. We
find a redshifted peak at about 8~kpc from the center of the
emission-line region, and velocity minima are found near the
north-eastern and south-western edge of line emission with velocities
of about -300~km s$^{-1}$ each. The line widths are highest in the
south-west (FWHM$\sim$1500~km s$^{-1}$), and about 1000~km s$^{-1}$ in
other parts of the gas, with two dips (FWHM$\sim$500 km s$^{-1}$)
north and south from the center. We do not detect the continuum from
this source. The source has two radio lobes that are detected outside
of the emission-line gas, delineating a jet axis that is well aligned
with the major axis of the emission-line gas (Fig.~\ref{fig:maps3}).

\subsection{MP~J1758$-$6738}
\label{ssec:indmp1758} 
The emission-line morphology of MP~J1758-6738 at z$=$2.03 is dominated
by a resolved, nearly circular, high-surface brightness region with a
diameter of 2.4\arcsec\ centered on the nucleus (which we associated
with the continuum source), and a fainter extension towards east
(Fig.~\ref{fig:maps3}). The velocity offset between these two regions
is about 800~km s$^{-1}$, and velocities are fairly uniform (with
offsets $\le$100 km s$^{-1}$) within each region. FWHM line widths are
of-order 700-800 km s$^{-1}$ throughout the central regions of the
source, and narrower (500$-$600~km s$^{-1}$) near the very
periphery of the faint emission-line region.

At z$=$2.03, H$\beta$ and [OIII]$\lambda$4959 fall outside the H-band,
and we only detect [OIII]$\lambda$5007, H$\alpha$, [NII]$\lambda$6583,
and [SII]$\lambda\lambda$6716,6731 (Fig.~\ref{fig:maps3}). With
FWHM$\sim$600~km s$^{-1}$, the H$\alpha$ and [NII]$\lambda$6583 lines
and the two lines of the [SII] doublet are blended, but can
nonetheless be individually identified. 

We only have a low-resolution radio map of this source with a
24\arcsec\ beam (Table~\ref{tab:radioobs}), which shows two hotspots well
outside the emission-line gas, and roughly aligned with the major axis
of the gas.

\subsection{TN~J2007$-$1316}
\label{ssec:indnj2007} 
TN~J2007$-$1316 at z$=$3.8 is amongst our highest-redshift
sources. Our integrated spectrum covers [OII]$\lambda$3727,
[OIII]$\lambda\lambda\lambda$4363,4959,5007, and H$\beta$
(Fig.~\ref{fig:intspec}). The results of individual line fits are
given in Table~\ref{tab:emlines}.

Line emission is compact at a spatial resolution of
0.9\arcsec$\times$0.8\arcsec. We find a small velocity gradient of
100~km s$^{-1}$, which may underestimate the intrinsic velocity
gradient by about a factor~2 due to beam-smearing effects (C.~Collet,
2014, PhD~thesis.) The line width in the integrated spectrum,
FWHM$=$950~km~s$^{-1}$ corresponds to the FWHM found also in the map
of Fig.~\ref{fig:maps}. The radio source shows two extended lobes
with a relative distance of 3.3\arcsec. The northern side shows a
faint, linear, narrow extention between the periphery of the
emission-line gas an the radio hotspot. The axis of the radio source
is well aligned with the major axis of the emission-line gas, and
nearly perpendicular to the small velocity gradient seen in the
gas. TN~J2007$-$1316 is one of the galaxies discussed in
\citet{drouart14} and \citet{rocca13} with a high star formation rate.

\subsection{MRC~2025-218}
\label{ssec:indmrc2025}

MRC~2025-218 at z=2.63 has a bright nuclear point source and broad
H$\alpha$ line emission originating from the AGN. \citet{humphrey08}
found a deep neutral absorber on top of the Ly$\alpha$ line emission,
which appears to be kinematically detached from the extended
emission-line region, and probably probes a massive reservoir of
either infalling our outflowing gas. We used the approach described in
\S\ref{ssec:blrremoval} to remove the nuclear component before fitting
the extended line emission. The remaining emission-line region is
extended with a size of 4.75\arcsec$\times$2.25\arcsec, corresponding
to a deconvolved size of 37~kpc$\times$16~kpc. The source has two
bright radio hotspots just outside the emission-line gas, and an axis
between them which aligns well with the major axis of the emission
line gas. The radio core is not detected.  The nucleus traced by the
bright point souce associated with the broad H$\alpha$ line is
slightly south from the center of the emission-line region. The gas
kinematics are irregular, with a total velocity offset of 900~km
s$^{-1}$. The highest redshifts are found about 0.5\arcsec\ north from
the nucleus. The lowest redshifts extend from the nucleus towards
south.
The FWHM line widths are between 700 and 1600~km s$^{-1}$, and the
highest velocities are found in two regions, one centered on the
nucleus,  the
other near the periphery of the northern emission-line gas. These
widths are greater than those of the H$\alpha$ broad-line region,
8024~km s$^{-1}$ \citep[][]{nesvadba11a}, so that we do not think that
they are residuals from the nuclear broad-line subtraction.

The integrated spectrum of MRC~2025-218 is shown in
Fig.~\ref{fig:intspec}. We identify [OIII]$\lambda\lambda$4959,5007,
H$\alpha$, [NII]$\lambda$6583, and [OI]$\lambda$6300 with widths
between 500 km s$^{-1}$ (for [OI]$\lambda$6300) and 1100~km s$^{-1}$
(for [OIII]$\lambda\lambda$4959,5007 and [NII]$\lambda$6583). The low
width of the [OI] line could be due to the faintness of the line,
which is detected only at 4$\sigma$. The
[SII]$\lambda\lambda$6716,6731 doublet falls beyond 2.4~$\mu$m at
z$=$2.63, and is therefore not covered by our data set.

\subsection{MG~2037$-$0011}
\label{ssec:indmg2037} 
MG~$2037-0011$ is a compact radio source at z$=$1.51
\citep[LAS$<$0.2\arcsec][]{}, which we observed with the J and H-band
gratings at R$=$2000 and R$=$3000, respectively. 
The integrated
spectrum (Fig.~\ref{fig:intspec}) is dominated by bright and broad
H$\alpha$ and [OIII]$\lambda\lambda$4959,5007 line emission, and we
also detect H$\beta$ and [SII]$\lambda\lambda$6716,6731. H$\alpha$ is
heavily blended with [NII]$\lambda$6583, and the two components of the
[SII]$\lambda\lambda$6716,6731 doublet with each other. The line width
derived from the [OIII]$\lambda\lambda$4959,5007 doublet in the
integrated spectrum is FWHM$=$802$\pm$25~km s$^{-1}$, and we obtain
adequate fits when imposing the same width on the blended lines.

The emission-line morphology of [OIII]$\lambda$5007 is compact, but we
do see a small velocity gradient of 300~km s$^{-1}$ going from
north-east to south-west. This gradient may be smaller by about a
factor~2 than the intrinsic gradient because of beam-smearing effects
(C.~Collet, 2014, PhD~thesis). The line widths measured in the map are
between 1200-1500 km s$^{-1}$.

\subsection{MRC~2048-272}
\label{ssec:indmrc2048}  
MRC~$2048-272$ at z$=$2.06 has a relatively powerful ($\log{P_{\rm
    500}}=$28.7 W Hz$^{-1}$) radio source with a size of 6.7\arcsec.
We did not detect any line emission, presumably because of deep
telluric absorption features at the redshifted wavelength of
[OIII]$\lambda$5007 and H$\alpha$. H$\beta$ and [OIII]$\lambda$4959
fall on top of bright night sky lines. We will therefore not discuss
this galaxy any further.

\begin{figure}
\centering
\includegraphics[width=0.45\textwidth]{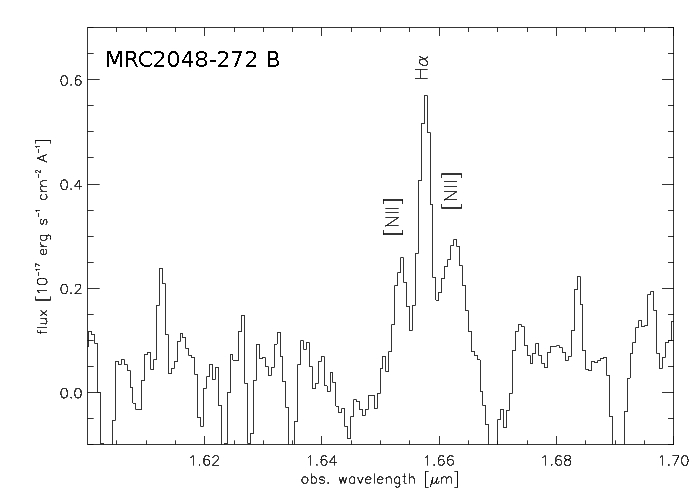}\\
\caption{Integrated spectrum of the intervening z$=$1.52 galaxy found near MRC~2048-272.}
\label{fig:MRC2048spec}
\end{figure}

\begin{figure}
\centering
\includegraphics[width=0.45\textwidth]{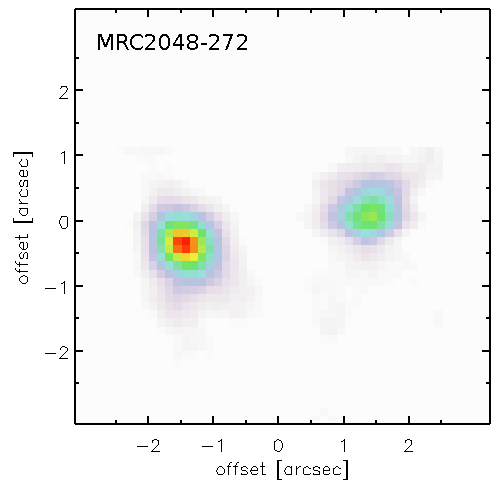}\\
\caption{Continuum image of MRC~2048$-$272 and the intervening galaxy at z$=$1.5.}
\label{fig:MRC2048maps}
\end{figure}

Interesting however is that a second galaxy at redshift z$=$1.52 falls
into the field of view (see the continuum image in
Fig.~\ref{fig:MRC2048maps}), for which we detect H$\alpha$ and
[NII]$\lambda\lambda$6548,6583 in the H-band (Tab.~\ref{tab:emlines}
and Fig.~\ref{fig:MRC2048spec}). Obviously, the large redshift
difference rules out that the two galaxies are physically
related. This second galaxy appears to be the source of the bright
far-infrared emission detected with Herschel \citep[][]{drouart14}. 

\subsection{MRC~2104-242}
\label{ssec:indmrc2104} 

MRC2104$-$242 at z$=$2.49 has one of the most extended radio sources
in our sample, with a size of 21.8\arcsec, corresponding to 176.6~kpc
at z$=$2.49. We observed this galaxy during run 079.A-0617, in parts
with adaptive optics, taking advantage of a nearby bright star, which
we used for the tip-tilt correction. We obtained data cubes in the H
and K-band individually, detecting [OIII]$\lambda\lambda$4959,5007,
H$\beta$, H$\alpha$, [NII]$\lambda\lambda$6548,6583, and the
[SII]$\lambda\lambda$6716,6731 doublet.  At a line width of
FWHM$=$633~km s$^{-1}$, H$\alpha$ is partially blended with
[NII]$\lambda\lambda$6548,6583, and the two lines of the
[SII]$\lambda\lambda$6716,6731 doublet with each
other. Single-component fits were sufficient to obtain adequate line
fits (Fig.~\ref{fig:intspec}). Fit results are listed in
Table~\ref{tab:emlines}.

The [OIII]$\lambda$5007 line image in Fig.~\ref{fig:maps} shows two
gas blobs, and only the southern blob is associated with continuum
emission. The northern blob has however a very high
[OIII]$\lambda$5007/H$\alpha$ ratio, which is not consistent with gas
heated by star formation and suggests that this source is a gas cloud
that is illuminated by the AGN of MRC~2104-242. We find a regular
velocity gradient in the southern cloud with a total velocity offset
of 350~km s$^{-1}$, and little variation in velocity in the northern
blob. The FWHM line width are relatively narrow, about 500~km s$^{-1}$
in both clouds, with a broader region with FWHM up to 900~km s$^{-1}$
inbetween. The major axis of the overall emission-line region aligns
well with the direction of the radio jets in the southern blob. The
axis towards the northern hotspot intercepts the periphery of the
northern cloud. The observed kinematics are consistent with those seen
in Ly$\alpha$ with VIMOS by \citet{villar06}, in particular the two
bright components are also seen, and with a similar velocity offset.

\subsection{4C~23.56}
\label{ssec:ind4c2356}

4C~23.56 at z$=$2.48 is one of our most extended sources in the radio,
with a size of 47\arcsec, corresponding to 380~kpc. It is also one of
the most powerful radio sources in our sample, with a radio power of
$\log{P_{\rm 500}} = 28.9$ W Hz$^{-1}$ at 500~MHz in the rest-frame. The
direction towards the radio hotspots is shown as a dashed line in the
right panel of Fig.~\ref{fig:maps4}. Comparing radio and X-ray
observations, \citet{blundell11} argued that there has been a previous
episode of radio activity in this source.

With SINFONI we detect [OIII]$\lambda\lambda$4959,5007, H$\alpha$,
H$\beta$, and [NII]$\lambda$6583 in this source. Line widths in the
integrated spectrum are between FWHM$=$500 and 600 km s$^{-1}$, and
all lines are adequately fit with a single Gaussian component.

4C~23.56 was one of the first HzRGs where a biconal emission-line
nebula was detected \citep[][]{knopp97}. Line emission in our SINFONI
cube is very extended in 4C~23.56, mostly along, but also
perpendicular to the axis of the radio jet, and roughly centered on
the continuum source, which we use as tracer of the nuclear
regions. We find a velocity gradient of 1200~km s$^{-1}$ with the
velocity monotonically increasing from the south-west to the
north-east. The line widths are FWHM$=$700$-$1000 km s$^{-1}$ through
most of the source, with lower widths of FWHM=400-500 km s$^{-1}$ in
the periphery, including a faint extension towards north-east.

\subsection{NVSS~J2135$-$3337}
\label{ssec:nvss2135}

NVSS~J2135$-$3337 at z$=$2.52 was taken from the catalog of
\citet{broderick07a} and \citet{bryant09} and is one of our galaxies
with fairly moderate radio luminosity, $\log{P_{\rm 500}}=27.9$~W Hz$^{-1}$
\citep{miley08}. It has a bright, R$=$14.5~mag star at a distance of
only 16.9\arcsec\ which allows observations with adaptive optics. At a
250~mas$\times$125~mas pixel scale this enabled us to reach a spatial resolution of
FWHM$=0.4$\arcsec$\times$0.4\arcsec. Ionized line emission in
NVSS2135$-$3337 is exended over 2.5\arcsec$\times$1.4\arcsec,
corresponding to 16.3~kpc$\times$10.4~kpc after deconvolution with the
size of the point spread function. We observed NVSS~J2135$-$3337
individually in the H and K band at R$=$3000 and R$=$4000,
respectively. The deep AO data were obtained in the K-band through
program 381.A-0541.

Fig.~\ref{fig:maps4} shows the gas morphology and kinematics of
NVSS~J2135$-$3337, which was derived from H$\alpha$ in this case. The
line emission is dominated by a centrally peaked component with a
small extension towards the north-west. The total velocity offset is
600 km s$^{-1}$ along an axis that is going from the south-east to the
north-west of the nucleus. The line widths peak near the nucleus with
FWHM=1200 km s$^{-1}$, and with much lower widths of FWHM$=200-300$~km
s$^{-1}$ in the far periphery and in particular the faint northern
extension. The radio source is compact and falls inbetween the regions
with the most strongly blueshifted and redshifted gas, near the peak
in FWHM line width.

\subsection{MG~J2144$+$1928}
\label{ssec:indmg2144}

The gas morphology in MG~2144$+$1928 at z$=$3.59 is very unusual
compared to the rest of the sample. [OIII]$\lambda$5007 line emission
extends over 9.1\arcsec\ along the major axis, with a very small minor
axis of only 2.2\arcsec. This corresponds to a size of
66~kpc$\times$16~kpc. Fig.~\ref{fig:maps4} shows a relatively compact,
but elongated central emission-line region, centered roughly on the
continuum peak. Towards north and south, but offset from the central
nebulosity, we find two gas plumes that are both not directly
connected to the central nebulosity, but have very similar
velocities. The line widths are also fairly uniform throughout the
source, with FWHM$\sim$800 km s$^{-1}$. Two unresolved radio hotspots are
found right at the far end of each emission-line gas plume
(Fig.~\ref{fig:maps4}).

\subsection{MRC~2224$-$273}
\label{ssec:indmrc2224}

MRC~2224$-$273 is a relatively small source at z$=1.7$, for which we
only have shallow data. We detect line emission centered on the
continuum peak with a size of 1.6\arcsec$\times$1.4\arcsec, corresponding to a
deconvolved size of 11~kpc$\times$8~kpc. The integrated spectrum shows
H$\alpha$ and [NII]$\lambda$6583 (which are heavily blended at an
integrated FWHM$=$1343~km s$^{-1}$), the [SII]$\lambda\lambda$6716,6731
doublet (the two components of the doublet are also strongly blended),
and [OI]$\lambda$6300. The radio source is compact with LAS$<$0.4\arcsec.

\subsection{TN~J2254+1857}
\label{ssec:indtnj2254}

TN~J2254$+$1857 at z$=$2.15 is one of our lower-power radio sources,
with $\log{P_{\rm 500}}=27.8$ W Hz$^{-1}$. [OIII]$\lambda$5007 line
emission is extended over 2.1\arcsec$\times$1.3\arcsec, corresponding
to a deconvolved size of 16.0~kpc$\times$9.6~kpc. Relative velocities
decrease monotonically from the south-west to the north-east with a
total velocity offset of 500~km s$^{-1}$. FWHM line widths are very
broad near the continuum peak, with FWHM$=1500-2000$ km s$^{-1}$,
and around $600-800$~km s$^{-1}$ in the periphery. We show the
emission-line maps in Fig.~\ref{fig:maps}. The integrated spectrum
shown in Fig.~\ref{fig:intspec} shows broad (FWHM$=863$~km s$^{-1}$)
lines of [OIII]$\lambda\lambda$4959,5007,
[SII]$\lambda\lambda$6716,6731, and [OI]$\lambda$6300. Properties of
individual lines are listed in Table~\ref{tab:emlines}. 

The radio map shown in the right panel of Fig.~\ref{fig:maps5} shows a
small double. The axis between the two radio hotspots is well aligned
with the major axis of the emission-line region and with the velocity
gradient. The two radio hotspots are right at the far end of the
emission line regions.

\subsection{MG~2308$+$0336}
\label{ssec:indmg2308} 

MG~2308$+$0336 at z$=$2.46 has a resolved, but relatively small radio
source with a size of 3.6\arcsec\ (29~kpc at z$=$2.46). Line emission
shown in Fig.~\ref{fig:maps5} fills the area between the two radio
lobes (right panel of Fig.~\ref{fig:maps5}). We did not detect the optical
continuum in this source, but the emission-line surface brightness
profile is very regular with a peak in the middle between the two
radio lobes. We assume that this is approximately the position of the
nucleus in MG~2308$+$0336. 

The integrated spectrum of MG~J2308$+$0336 shows H$\beta$,
[OIII]$\lambda\lambda$4959,5007, [OI], H$\alpha$, [NII]$\lambda$6583,
and the [SII]$\lambda\lambda$6716,6731 doublet
(Fig.~\ref{fig:intspec}). Line properties are listed in
Table~\ref{tab:emlines}. All lines were adequately fit with a single
Gaussian component of a common width FWHM$=$750~km s$^{-1}$.

The kinematic maps show a monotonic velocity gradient of 800~km
s$^{-1}$ increasing from the south-south-east to the
north-north-west. The FWHMs show that the lowest line widths are near
the center, with FWHM$\sim$500$-$600 km s$^{-1}$, and a broadening at
either end of the emission line region, near the radio lobes. In the
north, FWHMs are up to 1000 km s$^{-1}$, and 1200 km s$^{-1}$ in the
south.
\newpage
\onecolumn

\addtocounter{table}{5}
\longtab{5}{
\begin{longtable}{llccccc}
\caption{\label{tab:emlines}
Integrated spectral properties}\\
\hline\hline
Source & Line ID   & $\lambda_0$ & $\lambda_{\rm obs}$  & redshift & FWHM     & flux  \\
      &            & $[\AA]$     & $[\AA]$         &           & $[$km s$^{-1}]$ & $[10^{-16}$ erg s$^{-1}$ cm$^{-2}]$\\
\hline\hline
\endfirsthead
\caption{Integrated spectral properties (continued).}\\
\hline\hline
Source & Line ID   & $\lambda_0$ & $\lambda_{\rm obs}$  & redshift & FWHM     & flux  \\
      &            & $[\AA]$     & $[\AA]$         &           & $[$km s$^{-1}]$ & $[10^{-16}$ erg s$^{-1}$ cm$^{-2}]$\\
\hline\hline
\endhead
\hline
\endfoot
MRC~0114$-$211 
&$[$OIII$]$& 5007 & 12093.3$\pm$0.1 & 1.41528$\pm$0.00002 & 673$\pm$6  & 22.6$\pm$0.4 \\ 
&H$\beta$  & 4861 & 11741.6$\pm$0.8 & 1.41548$\pm$0.00015 & 673$\pm$46 & 2.92$\pm$ 0.4\\ 
&H$\alpha$ & 6563 & 15851.3$\pm$0.1 & 1.41526$\pm$0.00001 & 673$\pm$4  & 47.4$\pm$0.5 \\  
&$[$NII$]$ & 6583 & 15900.1$\pm$0.1 & 1.41533$\pm$0.00002 & 673$\pm$5  & 35.6$\pm$0.5 \\  
&$[$SII$]$ & 6716 & 16221.3$\pm$0.2 & 1.41533$\pm$0.00002 & 673$\pm$7  & 20.2$\pm$0.4 \\  
&$[$SII$]$ & 6731 & 16257.6$\pm$0.2 & 1.41533$\pm$0.00003 & 673$\pm$8  & 16.2$\pm$0.3 \\  
&$[$OI$]$  & 6300 & 15216.6$\pm$0.1 & 1.41533$\pm$0.00002 & 673$\pm$6  & 19.0$\pm$0.2 \\  
\hline
BLR~0128$-$264 
&$[$OIII$]$& 5007 & 16756.3$\pm$0.5 & 2.34658$\pm$0.00009 & 1136$\pm$20  & 34.5$\pm$1.1 \\ 
&H$\beta$  & 4861 & 16269.1$\pm$3.1 & 2.34686$\pm$0.00064 & 1136$\pm$140 &  4.9$\pm$1.1 \\
&H$\alpha$ & 6563 & 21963.4$\pm$1.0 & 2.34655$\pm$0.00016 & 1136$\pm$33  & 12.1$\pm$0.7 \\
&$[$NII$]$ & 6583 & 22031.0$\pm$1.8 & 2.34665$\pm$0.00027 & 1136$\pm$58  &  7.0$\pm$0.7 \\
&$[$SII$]$ & 6716 & 22476.1$\pm$13.7& 2.34665$\pm$0.00204 & 1136$\pm$90  &  1.0$\pm$0.8 \\
&$[$SII$]$ & 6731 & 22526.3$\pm$5.0 & 2.34665$\pm$0.00074 & 1136$\pm$177 &  2.1$\pm$0.6 \\
\hline
MRC~0156$-$252 
&[OIII]    & 5007 & 15112.3$\pm$0.2 & 2.01823$\pm$0.00004 &  621$\pm$ 9. & 208.9$\pm$5.7\\
&H$\beta$  & 4861 & 14672.8$\pm$3.9 & 2.01848$\pm$0.00080 &  621$\pm$ 207&  12.6$\pm$6.8\\
&H$\alpha$ & 6563 & 19808.4$\pm$0.3 & 2.01820$\pm$0.00004 &  621$\pm$  10&  47.6$\pm$1.4\\
&$[$NII$]$ & 6583 & 19869.4$\pm$0.4 & 2.01829$\pm$0.0000  &  621$\pm$ 15 &  31.1$\pm$1.4\\
 \hline
USS~0211$-$122
& $[$OIII$]_s$ & 5007 & 16717.2$\pm$0.1 & 2.3387$\pm$0.0001 &  434$\pm$2  &  61.0$\pm$0.4\\
& $[$OIII$]_n$ & 5007 & 16694.5$\pm$0.1 & 2.3342$\pm$0.0001 &  275$\pm$1  &  45.1$\pm$0.4\\
& $[$OIII$]_b$ & 5007 & 16717.8$\pm$0.7 & 2.3388$\pm$0.0001 & 1380$\pm$29 &  16.0$\pm$0.7\\
& H$\beta$   & 4861 & 16231.1$\pm$0.6 & 2.3390$\pm$0.0001 &  434$\pm$20   &  5.6$\pm$0.4\\
&H$\alpha$    & 6563 & 21912.1$\pm$0.2 & 2.3387$\pm$0.0001 &  434$\pm$6   & 19.9$\pm$0.5\\
&$[$NII$]$   & 6583 & 21979.5$\pm$0.8 & 2.3388$\pm$0.0001 &  434$\pm$21   &  5.5$\pm$0.5\\
&$[$SII$]$   & 6716 & 22423.6$\pm$2.5 & 2.3388$\pm$0.0004 &  434$\pm$67   &  1.6$\pm$0.5\\
&$[$SII$]$   & 6731 & 22473.7$\pm$3.3 & 2.3388$\pm$0.0005 &  434$\pm$86   &  1.3$\pm$0.5\\
\hline
MRC~0251$-$273 
&$[$OIII$]$ & 5007 & 20859.2$\pm$0.3 & 3.16600$\pm$0.00005 & 681$\pm$9   & 39.9$\pm$0.9\\
&H$\beta$   & 4861 & 20252.6$\pm$4   & 3.16634$\pm$0.00084 & 681$\pm$145 &  2.6$\pm$1.0\\ 
&$[$OII$]$  & 3727 & 15522.8$\pm$63  & 3.16497$\pm$0.00150 & 681$\pm$287 &  3.2$\pm$3.1\\ 
&$[$OII$]$  & 3729 & 15534.5$\pm$6   & 3.16586$\pm$0.00150 & 681$\pm$287 &  3.2$\pm$3.1\\ 
\hline
RC~J0311$+$0507
&$[$OII$]_n$& 3727 & 20536.0$\pm$0.5 & 4.5100$\pm$       & 112$\pm$19   & 12.2$\pm$1.6\\
&$[$OII$]_n$& 3729 & 20557.2$\pm$0.6 & 4.5040$\pm$       & 112$\pm$20   & 11.7$\pm$1.6\\
&$[$OII$]_b$& 3728 & 20518.9$\pm$5   & 4.5128$\pm$       & 1407$\pm$195 & 62.1$\pm$6.5\\
\hline
MP~J0340$-$6507 
& $[$OIII$]$& 5007 & 16456.4$\pm$0.1 & 2.28667$\pm$0.00002 &  829$\pm$ 5  & 94.3$\pm$ 1.1\\ 
&H$\beta$   & 4861 & 15977.8$\pm$0.8 & 2.28694$\pm$0.00016 &  829$\pm$36  & 14.9$\pm$ 1.2\\ 
&H$\alpha$  & 6563 & 21570.2$\pm$0.4 & 2.28664$\pm$0.00006 &  829$\pm$ 14 & 33.8$\pm$ 1.0\\ 
&$[$NII$]$  & 6583 & 21636.6$\pm$0.9 & 2.28674$\pm$0.00013 &  829$\pm$ 29 & 15.8$\pm$ 1.0\\ 
&$[$SII$]$  & 6716 & 22073.7$\pm$6.0 & 2.28674$\pm$0.00089 &  829$\pm$226 &  3.5$\pm$ 1.6\\ 
&$[$SII$]$  & 6731 & 22123.0$\pm$3.0 & 2.28674$\pm$0.00044 &  829$\pm$103 &  6.6$\pm$ 1.5 \\
\hline
PKS~0529$-$549 
& $[$OIII$]$& 5007 & 17896.46$\pm$0.04& 2.57429$\pm$0.00001&693$\pm$2  & 72.4$\pm$0.3\\ 
 &H$\beta$  & 4861 & 17376.0$\pm$0.3  & 2.57458$\pm$0.00007&693$\pm$13 &  9.9$\pm$0.3\\  
 &H$\alpha$ & 6563 & 23457.8$\pm$0.2  & 2.57425$\pm$0.00003&693$\pm$ 5 & 40.6$\pm$0.6\\  
 &$[$NII$]$ & 6583 & 23530.0$\pm$0.3  & 2.57436$\pm$0.00005&693$\pm$ 9 & 21.4$\pm$0.6\\  
 &$[$OI$]$  & 6300 & 22518.5$\pm$0.9  & 2.57436$\pm$0.00014&693$\pm$27 &  8.0$\pm$0.6\\  
 &$[$OI$]$  & 6363 & 22756.9$\pm$3    & 2.57644$\pm$0.00048&683$\pm$94 &  2.4$\pm$0.6\\  
 &H$\gamma$ & 4340 & 15508.0$\pm$0.3  & 2.57327$\pm$0.00007&763$\pm$14 &  5.2$\pm$0.2\\  
\hline
5C~07.269 
& $[$OIII$]$ & 5007 & 16148$\pm$5  &  2.22505$\pm$0.0010  & 1224$\pm$245 & 1.7$\pm$0.6\\
& H$\alpha$  & 6563 & 21157$\pm$4  &  2.22374$\pm$0.0007  & 2067$\pm$160 & 3.1$\pm$0.4\\
& $[$NII$]$  & 6583 & 21212$\pm$18 &  2.22219$\pm$0.0027  & 2475$\pm$754 & 1.15$\pm$0.6\\
& $[$SII$]$  & 6724 & 21676$\pm$22 &  2.22394$\pm$0.0033  & 2425$\pm$928 & 1.3$\pm$0.6\\
\hline
TN~J1112$-$2948 
& $[$OIII$]$& 5007 & 20480.7$\pm$0.2 & 3.09042$\pm$0.00004 & 393$\pm$  6 & 16.4$\pm$0.4 \\
& H$\beta$  & 4861 & 19885.2$\pm$1.9 & 3.09075$\pm$0.00039 & 393$\pm$ 64 &  1.5$\pm$0.4 \\
&$[$OII$]$  & 3727 & 15241.2$\pm$1.9 & 3.08940$\pm$0.00052 & 393$\pm$ 85 &  1.4$\pm$0.5 \\
&$[$OII$]$  & 3729 & 15252.7$\pm$3.5 & 3.09028$\pm$0.00093 & 393$\pm$157 &  0.8$\pm$1.0 \\ 
\hline
TXS~1113$-$178 
&$[$OIII$]$ & 5007 & 16222.1$\pm$0.2& 2.23989$\pm$0.00002& 632$\pm$5 & 24.8$\pm$0.3\\ 
&H$\beta$   & 4861 & 15750.4$\pm$1.2& 2.24015$\pm$0.00024& 632$\pm$48&  2.4$\pm$0.3\\
&H$\alpha$  & 6563 & 21263.2$\pm$0.5& 2.23985$\pm$0.00008& 632$\pm$16&  8.3$\pm$0.4\\
&$[$NII$]$  & 6583 & 21328.6$\pm$1.0& 2.23995$\pm$0.00015& 632$\pm$29&  4.5$\pm$0.4\\
\hline
3C~257 
&$[$OIII$]$ & 5007& 17432.1$\pm$0.2& 2.48154$\pm$0.00003& 1049$\pm$ 7& 87.2$\pm$1.0 \\
&H$\beta$   & 4861& 16925.1$\pm$1.5& 2.48182$\pm$0.00032& 1049$\pm$64&  9.8$\pm$1.1 \\
&H$\alpha$  & 6563& 22849.1$\pm$1.2& 2.48150$\pm$0.00019& 1049$\pm$38& 35.2$\pm$2.4 \\
&$[$NII$]$  & 6583& 22919.4$\pm$1.8& 2.48161$\pm$0.00027& 1049$\pm$54& 24.9$\pm$2.4 \\
&$[$SII$]$  & 6716& 23382.5$\pm$6.7& 2.48161$\pm$0.00100& 1049$\pm$223& 6.7$\pm$2.5 \\
&$[$SII$]$  & 6731& 23434.7$\pm$9.3& 2.48161$\pm$0.00138& 1049$\pm$329& 5.1$\pm$2.7 \\
\hline
USS~1243$+$036 
&$[$OIII$]$ & 5007 & 22860.6$\pm$0.2 & 3.56573$\pm$0.00005 & 670$\pm$ 7 & 28.7$\pm$ 0.5 \\
&H$\beta$   & 4861 & 22195.8$\pm$2.9 & 3.56610$\pm$0.00060 & 670$\pm$87 &  2.8$\pm$ 0.7 \\
&$[$OII$]$  & 3727 & 17012.3$\pm$1.6 & 3.56460$\pm$0.00042 & 670$\pm$62 &  5.6$\pm$ 0.9 \\
&$[$OII$]$  & 3729 & 17025.0$\pm$1.6 & 3.56558$\pm$0.00042 & 670$\pm$62 &  5.6$\pm$ 0.9 \\ 
\hline
MG~1251$+$1104 
&$[$OIII$]$ & 5007 & 16631.9$\pm$0.7 & 2.3217$\pm$0.0002 & 914$\pm$32  & 109$\pm$7 \\ 
&H$\alpha$  & 6563 & 21800.3$\pm$2.4 & 2.3217$\pm$0.0004 & 914$\pm$80  &  58$\pm$9 \\
&$[$NII$]$  & 6583 & 21867.4$\pm$5.8 & 2.3218$\pm$0.0009 & 914$\pm$208 &  26$\pm$10\\
 \hline
MRC~1324$-$262 
&$[$OIII$]$ & 5007 & 16438.1$\pm$ 0.9 & 2.2830$\pm$0.0002 & 1253$\pm$ 36  & 38.7$\pm$ 2.1 \\
&H$\beta$   & 4861 & 15970.0$\pm$ 6.0 & 2.2853$\pm$0.0012 & 1022$\pm$ 283 &  5.0343$\pm$ 2.4 \\
&H$\alpha$  & 6563 & 21540.5$\pm$ 1.1 & 2.2821$\pm$0.0002 &  594$\pm$  37 & 15.0559$\pm$ 1.7 \\
&$[$NII$]$  & 6583 & 21604.5$\pm$ 2.5 & 2.2819$\pm$0.0004 &  815$\pm$  80 & 10.6862$\pm$ 1.9 \\ 
\hline
TN~J1338$-$1942
&$[$NeIII$]$& 3967.5 & 20293.3$\pm$4.9 & 4.115$\pm$0.001 & 842$\pm$118 & 6.2$\pm$1.0 \\
&$[$NeIII$]$& 3868.8 & 19768.5$\pm$5.2 & 4.110$\pm$0.001 & 842$\pm$118 & 11.2$\pm$1.6 \\
&$[$OII$]$  & 3727   & 19047.0$\pm$5.1 & 4.111$\pm$0.001 & 725$\pm$141 & 4.7$\pm$ 1.1 \\
\hline
USS~1410$-$001 
&$[$OIII$]$ & 5007 & 16843.2$\pm$ 0.1 & 2.36394$\pm$ 0.00002& 710$\pm$3  & 144$\pm$ 1.2 \\
&H$\beta$   & 4861 & 16356.5$\pm$ 1.1 & 2.36486$\pm$ 0.00023& 678$\pm$46 &  9$\pm$  1.1 \\
&H$\alpha$  & 6563 & 22075.8$\pm$ 0.2 & 2.36369$\pm$ 0.00003& 711$\pm$6  &  46$\pm$ 0.8 \\
&$[$NII$]$  & 6583 & 22147.0$\pm$ 1.3 & 2.36428$\pm$ 0.00019& 748$\pm$40 &  8$\pm$  0.8 \\
&$[$SII$]$  & 6724 & 22588.5$\pm$ 2.5 & 2.36340$\pm$ 0.00038& 824$\pm$84 &  7$\pm$  1.0 \\
\hline
MRC~1558$-$003 
&$[$OIII$]$ & 5007 & 17683.4$\pm$0.1 & 2.53173$\pm$0.00003 & 695$\pm$ 6  & 77.6$\pm$ 1.2\\ 
&H$\beta$   & 4861 & 17169.1$\pm$1.2 & 2.53202$\pm$0.00024 & 695$\pm$ 47 &  8.5$\pm$ 1.0\\
&H$\alpha$  & 6563 & 23178.5$\pm$0.4 & 2.53169$\pm$0.00007 & 695$\pm$ 13 & 34.9$\pm$ 1.2\\
&$[$NII$]$  & 6583 & 23249.8$\pm$0.9 & 2.53180$\pm$0.00013 & 695$\pm$ 25 & 17.9$\pm$ 1.2\\ 
&$[$SII$]$  & 6716 & 23719.6$\pm$5.2 & 2.53180$\pm$0.00077 & 695$\pm$164 &  3.5$\pm$ 1.4\\ 
&$[$SII$]$  & 6731 & 23772.5$\pm$2.6 & 2.53180$\pm$0.00039 & 695$\pm$ 77 &  6.5$\pm$ 1.3\\ 
&$[$OI$]$   & 6300 & 22250.3$\pm$2.4 & 2.53180$\pm$0.00038 & 695$\pm$ 76 &  4.7$\pm$ 0.9\\ 
\hline
USS~1707$+$105 
&$[$OIII$]$ & 5007.00 & 16771.7  $\pm$0.2 & 2.3497$\pm$ 0.00004 & 698$\pm$ 8 & 23.7$\pm$0.5 \\
&H$\beta$   & 4861.00 & 16280.   $\pm$1.3 & 2.3491$\pm$ 0.00026 & 297$\pm$46 & 1.4$\pm$0.4  \\
&H$\alpha$  & 6563.00 & 21991.7  $\pm$0.8 & 2.3509$\pm$ 0.00012 & 641$\pm$23 & 7.0$\pm$0.5  \\
&$[$NII$]$  & 6583.00 & 22062.   $\pm$2.2 & 2.3513$\pm$ 0.00033 & 494$\pm$65 & 1.8$\pm$0.4  \\
\hline
3C~362 
&$[$OIII$]$ & 5007 & 16449.5$\pm$0.2 & 2.28531$\pm$0.00003&982$\pm$6   & 77.2$\pm$0.9\\ 
&H$\beta$   & 4861 & 15971.2$\pm$4.2 & 2.28557$\pm$0.0009 &982$\pm$200 & 3.1$\pm$1.0\\
&H$\alpha$  & 6563 & 21561.2$\pm$0.4 & 2.28527$\pm$0.00007&982$\pm$15  & 32.8$\pm$0.9\\
&$[$NII$]$  & 6583 & 21627.6$\pm$1.4 & 2.28537$\pm$0.0002 &982$\pm$47  & 10.7$\pm$0.9\\
&$[$SII$]$  & 6716 & 22064.6$\pm$3.0 & 2.28537$\pm$0.0005 &982$\pm$100 &  7.0$\pm$1.3\\ 
\hline
MP~1758$-$6738 
&$[$OIII$]$ & 5007 & 15178.5$\pm$0.2 & 2.03145$\pm$0.00003& 573$\pm$7   & 36.6$\pm$0.8 \\
&H$\alpha$  & 6563 & 19893.0$\pm$0.8 & 2.0311$\pm$0.0001  & 508$\pm$24  & 12$\pm$1.    \\
&$[$NII$]$  & 6583 & 19949.3$\pm$1.3 & 2.0304$\pm$0.0002  & 601$\pm$41  & 9.$\pm$1.    \\
& $[$SII$]$ & 6716 & 20359.6$\pm$1.5&2.0315$\pm$0.0002&572$\pm$49  & 4.5$\pm$0.7  \\
& $[$SII$]$ & 6731 & 20405.1$\pm$1.6&2.0315$\pm$0.0003&572$\pm$53  & 4.2$\pm$0.7  \\
\hline
TN~J2007$-$1316 
&$[$OIII$]$ & 5007 & 24260.6$\pm$0.3 & 3.84534$\pm$0.0006 & 950$\pm$10 & 28.9$\pm$0.5 \\ 
&H$\beta$   & 4861 & 23555.1$\pm$1.6 & 3.84573$\pm$0.0009 & 950$\pm$50 & 5.5$\pm$0.5  \\ 
&$[$OII$]$  & 3727 & 18067.6$\pm$0.9 & 3.84523$\pm$0.0003 & 950$\pm$35 & 4.3$\pm$0.3  \\
\hline
MRC~2025$-$218 
&$[$OIII$]$ & 5007 & 18169.8$\pm$0.2 & 2.62889$\pm$0.00004 & 1037$\pm$7 & 79.25$\pm$1.0 \\
&H$\alpha$  & 6563 & 23817.4$\pm$1.4 & 2.62904$\pm$0.00022 &  819$\pm$43& 17.15$\pm$1.6 \\
&$[$NII$]$  & 6583 & 23885.0$\pm$2.5 & 2.62829$\pm$0.00038 & 1096$\pm$77& 14.77$\pm$1.9 \\
&$[$OI$]$   & 6300 & 22860.6$\pm$2.3 & 2.62867$\pm$0.00037 &  479$\pm$71&  2.44$\pm$0.6 \\
\hline
MG~2037$-$0011 
&$[$OIII$]$ & 5007 & 12571.5$\pm$ 0.5 & 1.51078$\pm$ 0.00009 & 802$\pm$25  & 5.0$\pm$0.3  \\
&H$\alpha$  & 6563 & 16478.1$\pm$ 0.9 & 1.51076$\pm$ 0.00014 & 802$\pm$39  & 5.4$\pm$0.5 \\ 
&$[$NII$]$  & 6583 & 16528.8$\pm$ 0.9 & 1.51083$\pm$ 0.00014 & 802$\pm$38  & 5.5$\pm$0.5 \\ 
&$[$SII$]$  & 6716 & 16862.8$\pm$ 2.4 & 1.51083$\pm$ 0.00036 & 802$\pm$99  & 1.8$\pm$0.4 \\ 
&$[$SII$]$  & 6731 & 16900.4$\pm$ 2.5 & 1.51083$\pm$ 0.00037 & 802$\pm$102 & 1.8$\pm$0.4 \\ 
\hline
MRC~2048$-$272
&H$\alpha$  & 6563 & 16576.7$\pm$1.0 & 1.52579$\pm$0.0002 & 468$\pm$40  &1.5$\pm$2 \\
&$[$NII$]$  & 6583 & 16626.1$\pm$1.7 & 1.52561$\pm$0.0003 & 466$\pm$70  &0.9$\pm$2 \\
&$[$NII$]$  & 6548 & 16534.2$\pm$2.9 & 1.52507$\pm$0.0004 & 470$\pm$119 &0.6$\pm$3 \\ 
\hline
MRC~2104$-$242 
&$[$OIII$]$ & 5007 & 17482.5$\pm$ 0.1 & 2.4916$\pm$ 0.00002 & 633$\pm$   4 & 37.7$\pm$ 0.4 \\
&H$\beta$   & 4861 & 16975.2$\pm$ 0.8 & 2.4921$\pm$ 0.00017 & 491$\pm$  32 &  2.6$\pm$ 0.3 \\
&H$\alpha$  & 6563 & 22915.3$\pm$ 0.3 & 2.4916$\pm$ 0.00004 & 633$\pm$   8 & 22.0$\pm$ 0.5 \\
&$[$NII$]$  & 6583 & 22985.8$\pm$ 0.9 & 2.4917$\pm$ 0.00013 & 633$\pm$  25 &  7.0$\pm$ 0.5 \\
&$[$SII$]$  & 6716 & 23450.2$\pm$ 3.4 & 2.4917$\pm$ 0.00051 & 633$\pm$ 105 &  2.2$\pm$ 0.5 \\
&$[$SII$]$  & 6731 & 23502.6$\pm$360  & 2.4917$\pm$ 0.05359 & 633$\pm$2211 &  1.6$\pm$0.5 \\
\hline
NVSS~J2135-3337
&H$\alpha$   & 6563 & 23052.6$\pm$0.5 & 2.51251$\pm$0.0001 & 637$\pm$15     & 3.2$\pm$0.13\\
&$[$NII$]$   & 6583 & 23123.5$\pm$0.5 & 2.51261$\pm$0.0001 & 637$\pm$16     & 2.9$\pm$0.13\\
&$[$SII$]$   & 6716 & 23590.7$\pm$2   & 2.51261$\pm$0.0003 & 637$\pm$52     &  0.9$\pm$0.13\\
&$[$SII$]$   & 6731 & 23643.4$\pm$2   & 2.51261$\pm$0.0003 & 637$\pm$62     &  0.8$\pm$0.13\\
&$[$OI$]$    & 6300 & 22129.5$\pm$3   & 2.51261$\pm$0.0006 & 1307$\pm$120   &  1.3$\pm$0.2\\
\hline
4C~23.56 
&$[$OIII$]$ & 5007 & 17440.6$\pm$0.3 & 2.48325$\pm$0.00006 & 629$\pm$ 12 & 104.7$\pm$3.7\\ 
&H$\beta$   & 4861 & 16928.5$\pm$3.4 & 2.48251$\pm$0.00069 & 517$\pm$139 &   8.1$\pm$3.7 \\
&H$\alpha$  & 6563 & 22862.6$\pm$0.8 & 2.48356$\pm$0.00012 & 575$\pm$ 22 &  28.8$\pm$2.1 \\
&$[$NII$]$  & 6583 & 22933.2$\pm$2.4 & 2.48369$\pm$0.00037 & 560$\pm$ 67 &   9.5$\pm$2.1\\
\hline
MG~J2144$+$1928 
&$[$OIII$]$ & 5007 &  23000.3$\pm$ 0.4 & 3.59362$\pm$0.00008 & 651$\pm$  12 & 31$\pm$1 \\
&H$\beta$   & 4861 &  22331.4$\pm$ 21.  & 3.59400$\pm$0.00428 & 651$\pm$ 842 & 1.7$\pm$2.5 \\
&$[$OII$]$  & 3727 &  17116.2$\pm$ 2.8  & 3.59248$\pm$0.00075 & 651$\pm$ 118 & 4.2$\pm$1.4 \\
&$[$OII$]$  & 3729 &  17129.0$\pm$ 2.5  & 3.59347$\pm$0.00066 & 651$\pm$ 102 & 4.9$\pm$1.3 \\
\hline
MRC~2224$-$273 
&H$\alpha$  & 6563 & 17605.6$\pm$0.9 & 1.68255$\pm$0.00014 & 1343$\pm$ 36 & 34.0$\pm$1.7 \\
&$[$NII$]$  & 6583 & 17659.8$\pm$1.7 & 1.68263$\pm$0.00025 & 1343$\pm$ 68 & 18.4$\pm$1.7 \\
&$[$SII$]$  & 6716 & 18016.7$\pm$6.4 & 1.68263$\pm$0.00095 & 1343$\pm$275 &  5.2$\pm$1.9 \\
&$[$SII$]$  & 6731 & 18056.8$\pm$3.2 & 1.68263$\pm$0.00048 & 1343$\pm$132 &  9.8$\pm$1.7 \\
&$[$OI$]$   & 6300 & 16900.6$\pm$2.7 & 1.68263$\pm$0.00043 & 1343$\pm$119 & 10.0$\pm$1.7 \\
\hline
TN~J2254$+$1857
&$[$OIII$]$  & 5007 & 15787.8$\pm$0.8 & 2.1531$\pm$0.0002 & 863$\pm$36  & 5.4$\pm$0.4\\
&H$\alpha$   & 6563 & 20693.9$\pm$0.9 & 2.1531$\pm$0.0001 & 863$\pm$32  & 6.3$\pm$0.4\\
&$[$NII$]$   & 6583 & 20757.6$\pm$1.1 & 2.1532$\pm$0.0002 & 863$\pm$39  & 5.2$\pm$0.4\\
&$[$SII$]$   & 6716 & 21176.9$\pm$1.7 & 2.1532$\pm$0.0003 & 863$\pm$58  & 2.6$\pm$0.3\\
&$[$SII$]$   & 6731 & 21224.2$\pm$1.8 & 2.1532$\pm$0.0003 & 863$\pm$59  & 2.5$\pm$0.3\\
&$[$OI$]$    & 6300 & 19865.2$\pm$2.3 & 2.1532$\pm$0.0004 & 863$\pm$83  & 2.1$\pm$0.4\\
&$[$OI$]$    & 3727 & 11751.8$\pm$0.2 & 2.1532$\pm$0.0001 & 1042$\pm$11 & 2.5$\pm$0.3\\
\hline
MG~2308$+$0336 
&$[$OIII$]$ & 5007 & 17306.1$\pm$0.2 & 2.45638$\pm$0.00004& 747$\pm$9 & 50.4$\pm$ 1.1 \\ 
&H$\beta$   & 4861 & 16802.8$\pm$1.5 & 2.45666$\pm$0.0003&  747$\pm$66&  6.1$\pm$ 1.0 \\ 
&H$\alpha$  & 6563 & 22684.0$\pm$0.3 & 2.45634$\pm$0.00006& 747$\pm$13& 22.7$\pm$ 0.7 \\  
&$[$NII$]$  & 6583 & 22753.8$\pm$1.0 & 2.45645$\pm$0.0002&  747$\pm$32&  9.2$\pm$ 0.7  \\
&$[$SII$]$  & 6716 & 23213.5$\pm$2.4 & 2.45645$\pm$0.0004&  747$\pm$79&  6.0$\pm$ 1.2  \\
&$[$SII$]$  & 6731 & 23265.3$\pm$2.1 & 2.45645$\pm$0.0003&  747$\pm$67&  7.0$\pm$ 1.2  \\
&$[$OI$]$   & 6300 & 21775.6$\pm$2.2 & 2.45645$\pm$0.0004&  747$\pm$79&  4.0$\pm$ 0.8  \\
\hline\hline
\end{longtable}}

\end{document}